\documentclass[aps,twocolumn,twoside,showpacs,preprintnumbers,nofootinbib,
  amsmath,amssymb,showkeys,letterpaper,10pt,final]{revtex4}

\usepackage{graphicx}
\usepackage{dcolumn}
\usepackage{bm}
\usepackage{pstricks}

\newcommand{\bc}{\begin{center}}
\newcommand{\ec}{\end{center}}
\newcommand{\bq}{\begin{quote}}
\newcommand{\eq}{\end{quote}}
\newcommand{\bi}{\begin{itemize}}
\newcommand{\ei}{\end{itemize}}
\newcommand{\be}{\begin{equation}}
\newcommand{\ee}{\end{equation}}
\newcommand{\bea}{\begin{eqnarray}}
\newcommand{\eea}{\end{eqnarray}}
\newcommand{\bt}{\begin{tabular}}
\newcommand{\et}{\end{tabular}}
\newcommand{\she}{{\tt SHERPA}}

\def\ie{i.e.\ }

\begin{document}


\title{Studying {\boldmath$W^+W^-$} production at the Fermilab
       Tevatron with SHERPA}
\thanks{Dedicated to the Memory of Gerhard~Soff, 1949 - 2004.}

\author{Tanju~Gleisberg}\email{tanju@theory.phy.tu-dresden.de}
\author{Frank~Krauss}\email{krauss@theory.phy.tu-dresden.de}
\author{Andreas~Sch{\"a}licke}\email{dreas@theory.phy.tu-dresden.de}
\author{Steffen~Schumann}\email{steffen@theory.phy.tu-dresden.de}
\author{Jan-Christopher~Winter}\email{winter@physik.tu-dresden.de}
\affiliation{Institute for Theoretical Physics, D-01062 Dresden, Germany}
\homepage{http://www.physik.tu-dresden.de/~krauss/hep/}

\date{\today}

\begin{abstract}
  \noindent
  The merging procedure of tree-level matrix elements with the subsequent 
  parton shower as implemented in \she\ will be studied for the example
  of $W$\/ boson pair production at the Fermilab Tevatron. Comparisons
  with fixed order calculations at leading and next-to-leading order in
  the strong coupling constant and with other Monte Carlo simulations
  validate once more the impact and the quality of the merging algorithm
  and its implementation. 
\end{abstract}


\maketitle

\section{Introduction\label{sec_in}}

\noindent
Studying the production of $W$\/ pairs at collider experiments offers a great 
possibility for tests of the gauge sector of the Standard Model, that has 
been extensively investigated by the LEP2 collaborations \cite{unknown:2003ih,
Barate:2000gi,Abbiendi:2000eg,Abdallah:2003zm,Achard:2004zw}. Tests in this
channel are quite sensitive, because there is a destructive interference of 
two contributions: a $t$-channel contribution, where both $W$ bosons couple 
to incoming fermions, and an $s$-channel contribution, where the $W$ bosons 
emerge through a triple gauge coupling, either $\gamma W^+W^-$ or $ZW^+W^-$. 
New physics beyond the Standard Model could easily manifest itself, either 
through new particles propagating in the $s$-channel, like, for instance, a 
$Z'$ particle in L-R symmetric models \cite{Pati:1974yy,Mohapatra:1974hk,
Mohapatra:1980yp,Eichten:1984eu}, or through anomalous triple gauge couplings, 
which could be loop-induced, mediated by heavy virtual particles
running in the loop. In \cite{Gaemers:1978hg,Hagiwara:1986vm,Bilenky:1993ms}
the most general form of an effective Lagrangian for such interactions
has been developed and discussed.
Such tests of anomalous triple gauge couplings have been performed both at 
LEP2 \cite{Abreu:1999vv,Heister:2001qt,Abbiendi:2003mk,Achard:2004ji} and at 
Tevatron, Run I \cite{Abe:1995jb,Abachi:1997xe,Abbott:1998gp,Abbott:1998jz} 
and at Run II \cite{Acosta:2005mu}. Both scenarios could clearly modify the 
total cross section or, at least, lead to different distributions of the final 
state particles. In addition, $W$\/ pairs, possibly in association with jets, 
represent a background to a number of relevant other processes, such as the 
production of top quarks, the production of a Higgs boson with a mass above 
roughly 135 GeV, or the production of supersymmetric particles, such
as charginos or neutralinos \cite{Abbott:1997rd,Abe:1998qm}.  
\\
Accordingly, there are a number of calculations and programs dealing with 
this process. At next-to-leading order (NLO) in the strong coupling constant, 
$W$\/ pair production has been calculated by \cite{Ohnemus:1991kk,
Frixione:1993yp,Ohnemus:1994ff}. In addition, a number of programs have been 
made available, allowing the user to implement phase space cuts and to
generate single events. First of all, there are fixed order
calculations. At leading order (LO), \ie at tree-level, 
they are usually performed through automated tools, called matrix element 
or parton level generators. Examples for such programs include {\tt COMPHEP} 
\cite{Pukhov:1999gg}, {\tt GRACE/GR@PPA} \cite{Ishikawa:1993qr,GRAPPA}, 
{\tt MADGRAPH/MADEVENT} \cite{Stelzer:1994ta,Maltoni:2002qb}, {\tt ALPGEN} 
\cite{Mangano:2002ea}, and {\tt AMEGIC++} \cite{Krauss:2001iv}. At
NLO, the program {\tt MCFM} \cite{Campbell:1999ah} provides cross
sections and distributions for this process.
\\
Apart from such fixed order calculations, multipurpose event 
generators such as {\tt PYTHIA} \cite{Sjostrand:2000wi,Sjostrand:2001yu} or 
{\tt HERWIG} \cite{Corcella:2000bw,Corcella:2002jc} play a major role in the 
experimental analyses of collider experiments. They proved to be extremely 
successful in describing global features of such processes, like, for
instance, the transverse momenta or rapidity distributions of the bosons.
They are usually based on exact tree-level matrix elements for the
production and decay of the boson pair, supplemented with a parton shower.
The latter takes proper care of multiple parton emission and resums
the corresponding leading and some of the subleading Sudakov logarithms.
\\
In view of the need for increasing precision, recently two approaches
have been developed that incorporate higher order corrections into the
framework of multipurpose event generators.
The first one, called {\tt MC@NLO}, provides a method to consistently
match NLO calculations for specific processes with the parton shower 
\cite{Frixione:2002ik,Frixione:2003ei}. The idea of this approach is
to organize the counter-terms necessary to cancel real and virtual infrared 
divergencies in such a way that the first emission of the parton shower is 
recovered. Of course, this method depends to some extent on the details of 
the parton shower, and it has some residual dependence on the process in 
question. So far, {\tt MC@NLO} has been implemented in conjunction
with {\tt HERWIG} \cite{Frixione:2004wy} for the following processes:
production of $W$\/ and $Z$\/ bosons, or pairs of these bosons
\cite{Frixione:2002ik}, production of the Higgs boson, production of
heavy quarks \cite{Frixione:2003ei}.
\\
An alternative approach is to consistently combine tree-level matrix
elements for different multiplicities of additional jets and to merge
them with the parton shower. This approach has been presented for the
first time for the case of $e^+e^-$ annihilations into jets
\cite{Catani:2001cc}; later it has been extended to hadronic
collisions \cite{Krauss:2002up} and it has been reformulated to a
merging procedure with a dipole shower in \cite{Lonnblad:2001iq}. The idea
underlying this method is to separate the kinematical range of parton emission
by a $k_\perp$-algorithm \cite{Catani:1991hj,Catani:1992zp,Catani:1993hr}
into a regime of jet production, covered by the appropriate matrix elements,
and a regime of jet evolution, covered by the respective shower. Then,
the matrix elements are reweighted through Sudakov form factors and hard 
emissions in the parton shower leading to a jet are vetoed such that there 
is only a residual dependence on the jet resolution cut. This method is one
of the cornerstones of the new event generator \she\ \cite{Gleisberg:2003xi};
it has been validated for the cases of $e^+e^-$ annihilations into jets
\cite{apacic2,Schaelicke:2005nv} and for the production of single
vector bosons at the Fermilab
Tevatron \cite{Krauss:2004bs} and the CERN LHC \cite{wz@LHC}.
\\
In this publication this series of studies will be continued with an
investigation of $W$\/ pair production at the Fermilab Tevatron, Run
II, where both $W$\/ bosons decay leptonically, \ie
$p\bar p\to W^+W^-+X\to e^+\mu^-\nu_e\bar\nu_\mu+X$\footnote{%
  Singly resonant diagrams contributing to the parton level processes
  of $p\bar p\to e^+\mu^-\nu_e\bar\nu_\mu+X$ have been included.}.
Input parameters used throughout this publication and the specifics,
how the \she\ runs have been obtained, are listed in the appendix, see
Apps.\ \ref{app_input} and \ref{app_cuts}.
After some consistency -- including scale variation -- checks of the 
merging algorithm in Sec.\ \ref{sec_check}, results obtained with \she\ 
will be confronted with those from an NLO calculation provided by {\tt
MCFM}, cf.\ Sec.\ \ref{sec_MCFM}. Then, in Sec.\ \ref{sec_MC} some
exemplary results of \she\ are compared with those obtained from
other event generators, in particular with those from {\tt PYTHIA} and
{\tt MC@NLO}. A summary closes this publication.

\section{Consistency checks\label{sec_check}}

\noindent
In this section some sanity checks of the merging algorithm for the
case of $W$\/ pair production are presented. For this, first, the
dependence of different observables on the key parameters of the
merging procedure, namely the internal matrix-element parton-shower
separation scale $Q_{\rm cut}$ and the highest multiplicity $n_{\rm
max}$ of included tree-level matrix elements, is examined.
Secondly, the sensitivity of the results with respect to changes in
the renormalization scale $\mu_R$ and the factorization scale $\mu_F$
will be discussed.
\\
All distributions shown in this section are inclusive results at the
hadron level, where restrictive jet and lepton cuts have been applied,
for details on the cuts cf.\ App.\ \ref{app_cuts}. In all cases, the
distributions are normalized to one using the respective total cross
section as delivered by the merging algorithm.

\subsection*{Impact of the phase space separation cut}

\noindent
First of all, the impact of varying the jet resolution cut $Q_{\rm
cut}$ is studied. \she\ results have been obtained with an inclusive
$2$jet production sample, \ie tree-level matrix elements up to two
additional QCD emissions have been combined and merged with the parton
shower. In all figures presented here the black solid line shows the
total inclusive result as obtained by \she\ for the respective
resolution cut $Q_{\rm cut}$. The reference curve drawn as a black
dashed line has been obtained as the mean of five different runs,
where the resolution cut has been gradually increased, $Q_{\rm
cut}=10, 15, 30, 50$ and $80$ GeV. The coloured curves represent the
contributions stemming from the different matrix-element final-state
multiplicities. Results are shown for three different resolution cuts,
namely $Q_{\rm cut}=15, 30$ and $80$ GeV. It should be noted that the
change of the rate predicted by the merging procedure under $Q_{\rm
cut}$ variation has been found to be very small, although it is a
leading order prediction only. Nevertheless, by varying the separation
cut between $10$ and $80$ GeV, the deviation of the total rate amounts
to $2.4\%$ only.
\begin{figure}[t!]
  \vspace{0mm}
  \begin{picture}(188,615)
    \put(0,410){
      \includegraphics[width=70mm]{%
        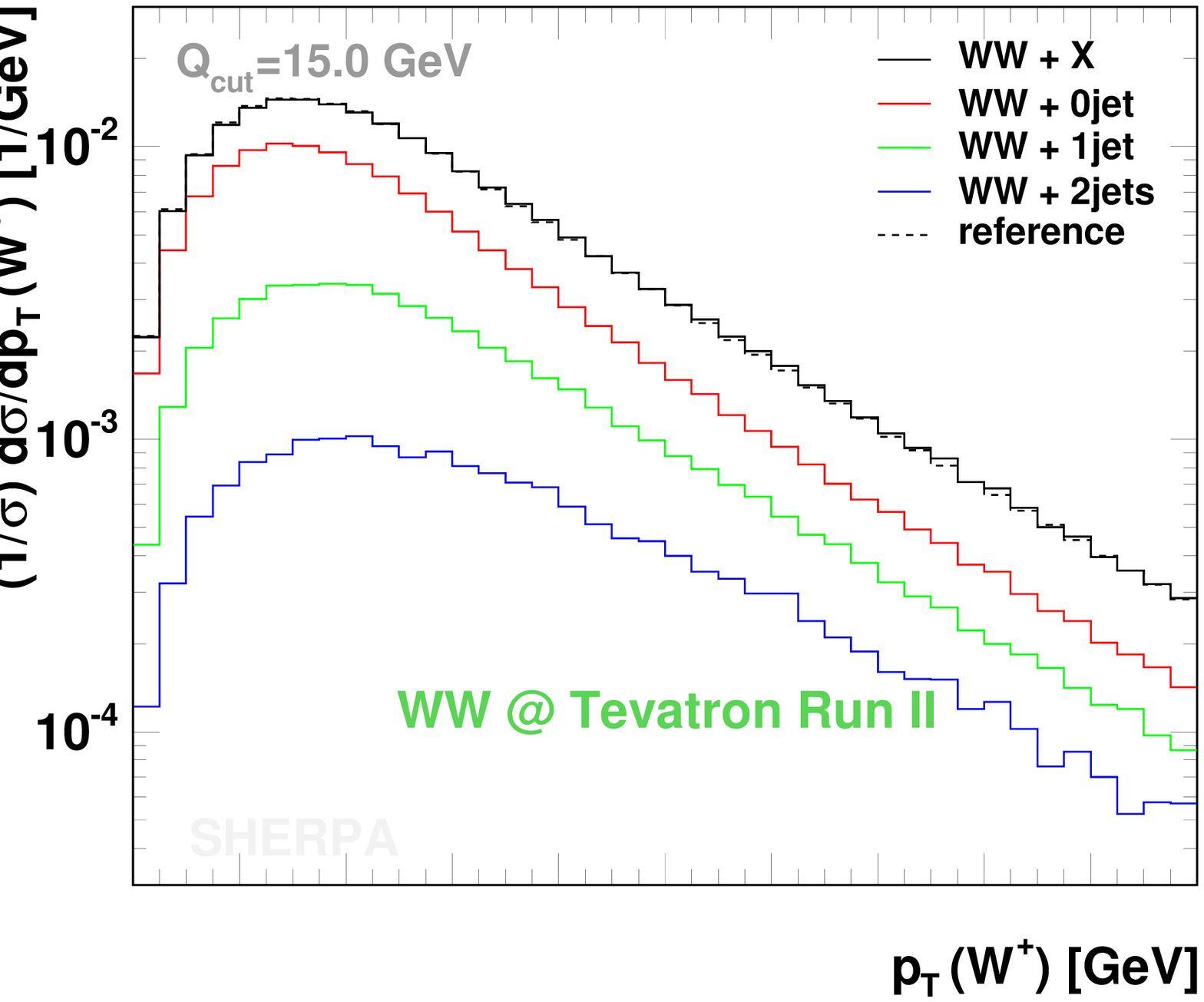}}
    \put(0,410){
      \includegraphics[width=70mm]{%
        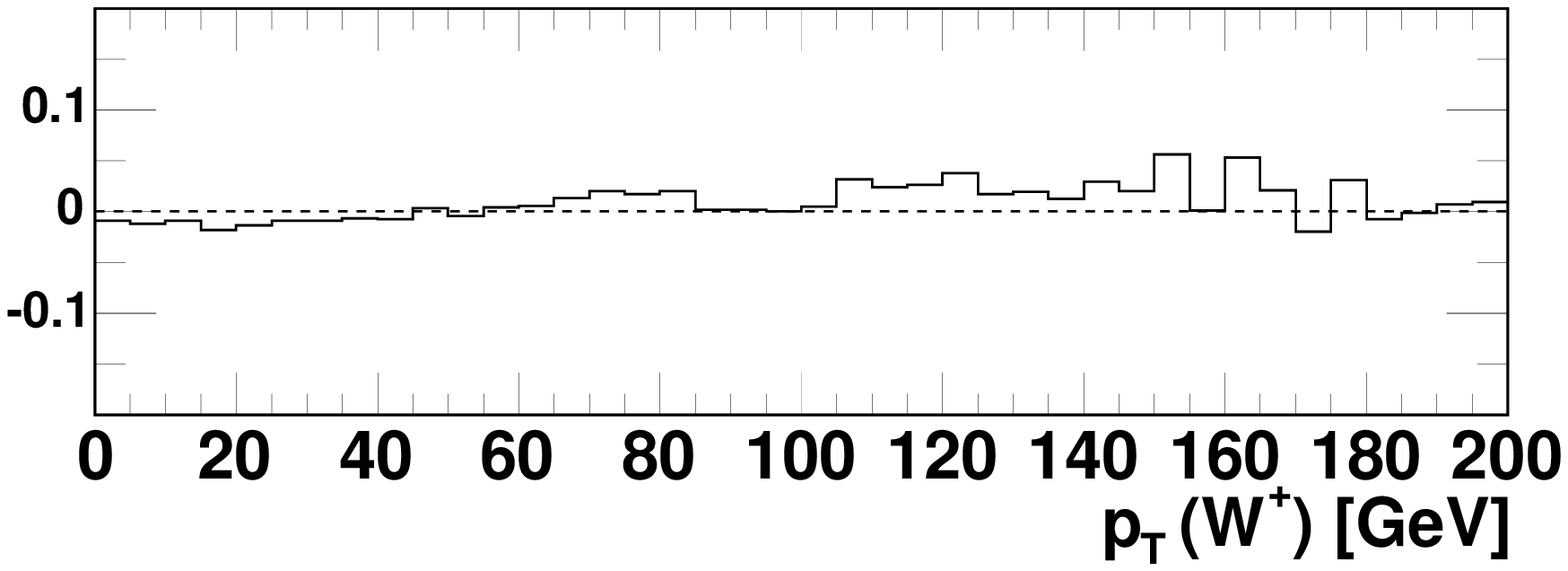}}
    \put(0,205){
      \includegraphics[width=70mm]{%
        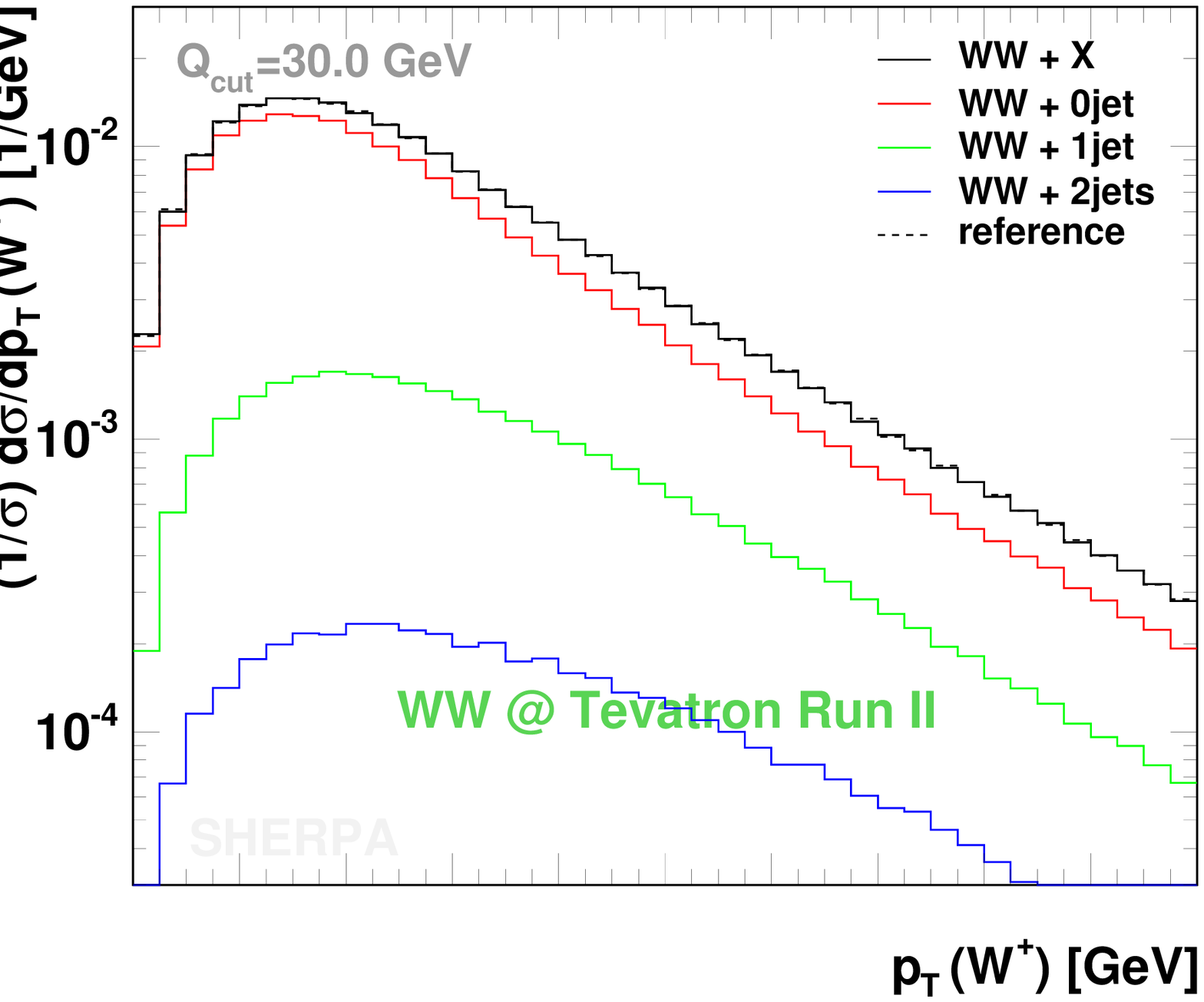}}
    \put(0,205){
      \includegraphics[width=70mm]{%
        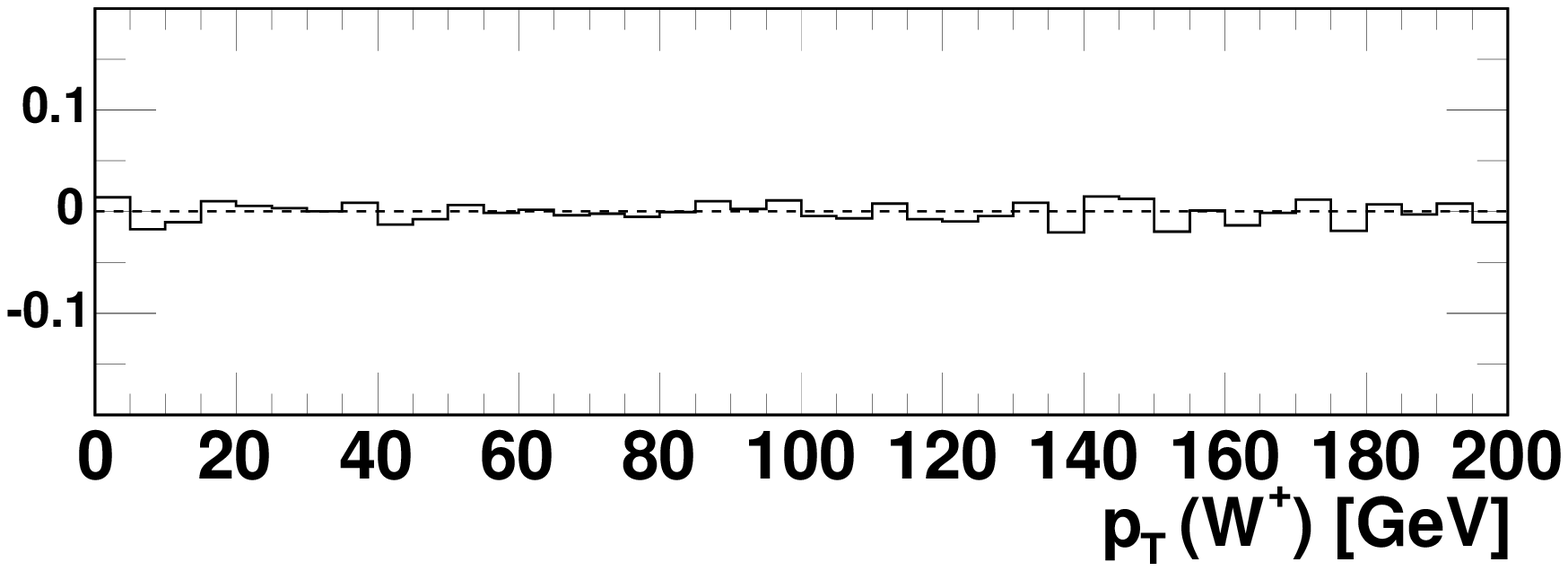}}
    \put(0,0){
      \includegraphics[width=70mm]{%
        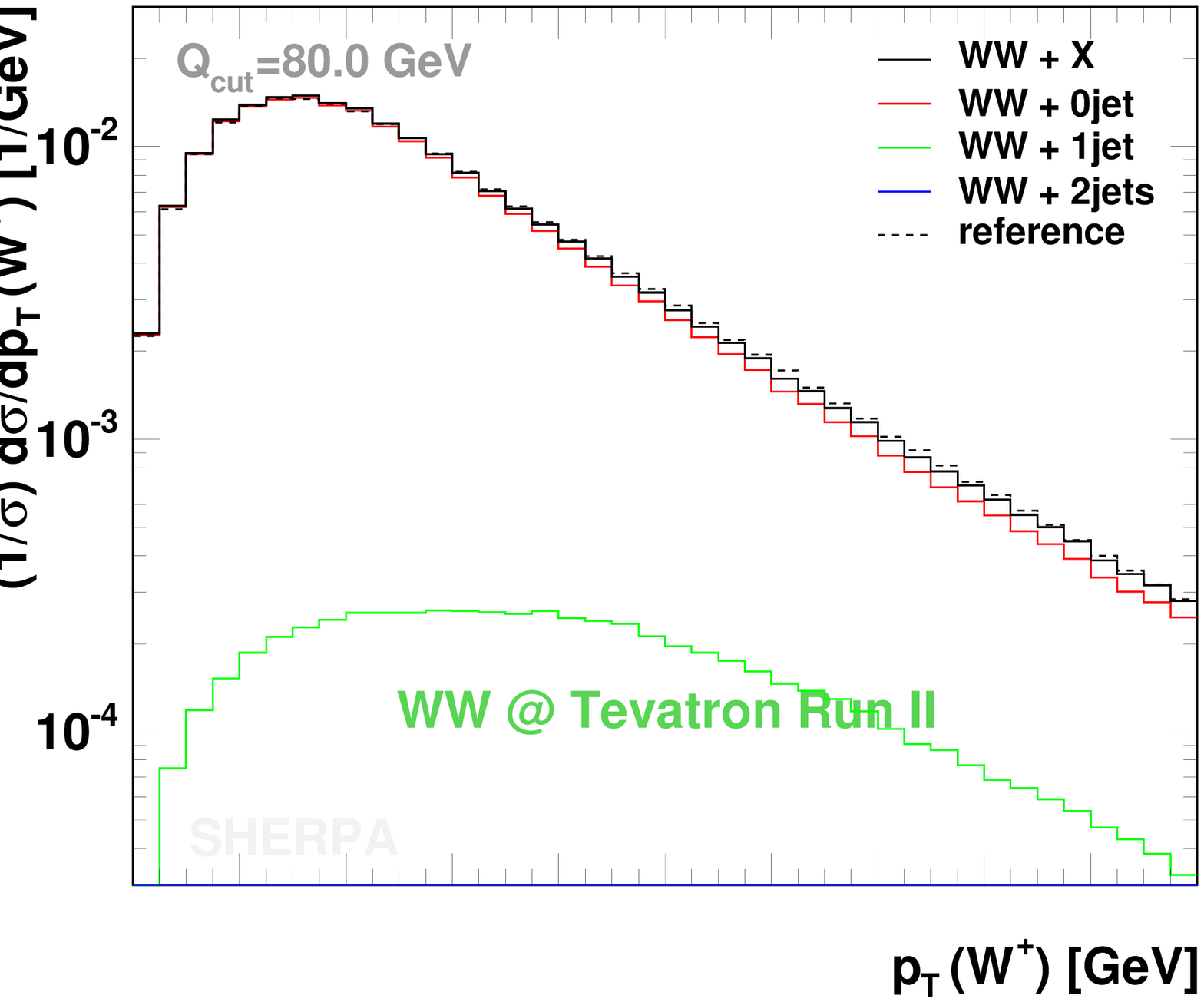}}
    \put(0,0){
      \includegraphics[width=70mm]{%
        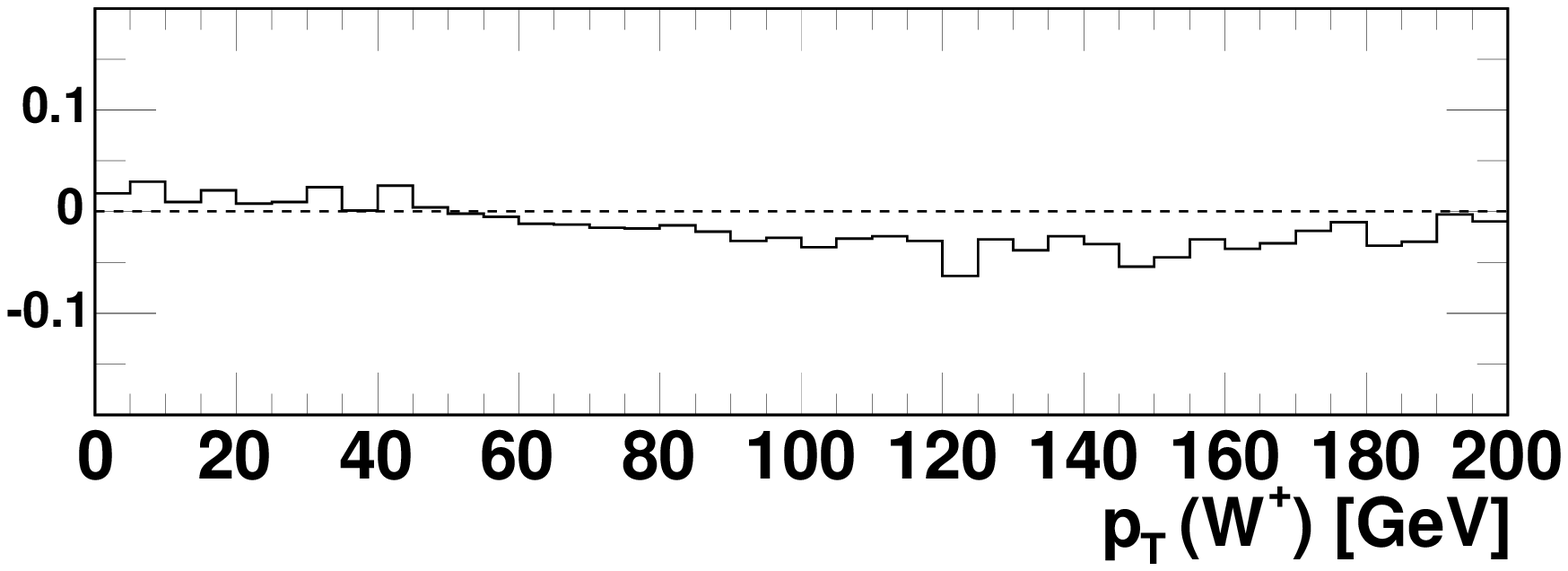}}
  \end{picture}
  \vspace{0mm}
  \caption{The $p_T$ distribution of the $W^+$ boson and its
    dependence on $Q_{\rm cut}$, chosen to be $15$, $30$ and $80$ GeV
    (from top to bottom). The black solid line shows the \she\
    prediction obtained with $n_{\rm max}=2$, the black dashed one is
    the reference obtained as the mean of different $Q_{\rm cut}$ runs
    and the coloured lines indicate the different multiplicity
    contributions. The lower part of the plots exhibits the normalized
    difference of the prediction with respect to the reference. Cuts
    and input parameters are specified in the appendices.}
  \label{ptWplus2_coychk}
  \vspace{-7mm}
\end{figure}
\begin{figure}[t!]
  \vspace{0mm}
  \begin{picture}(188,615)
    \put(0,410){
      \includegraphics[width=70mm]{%
        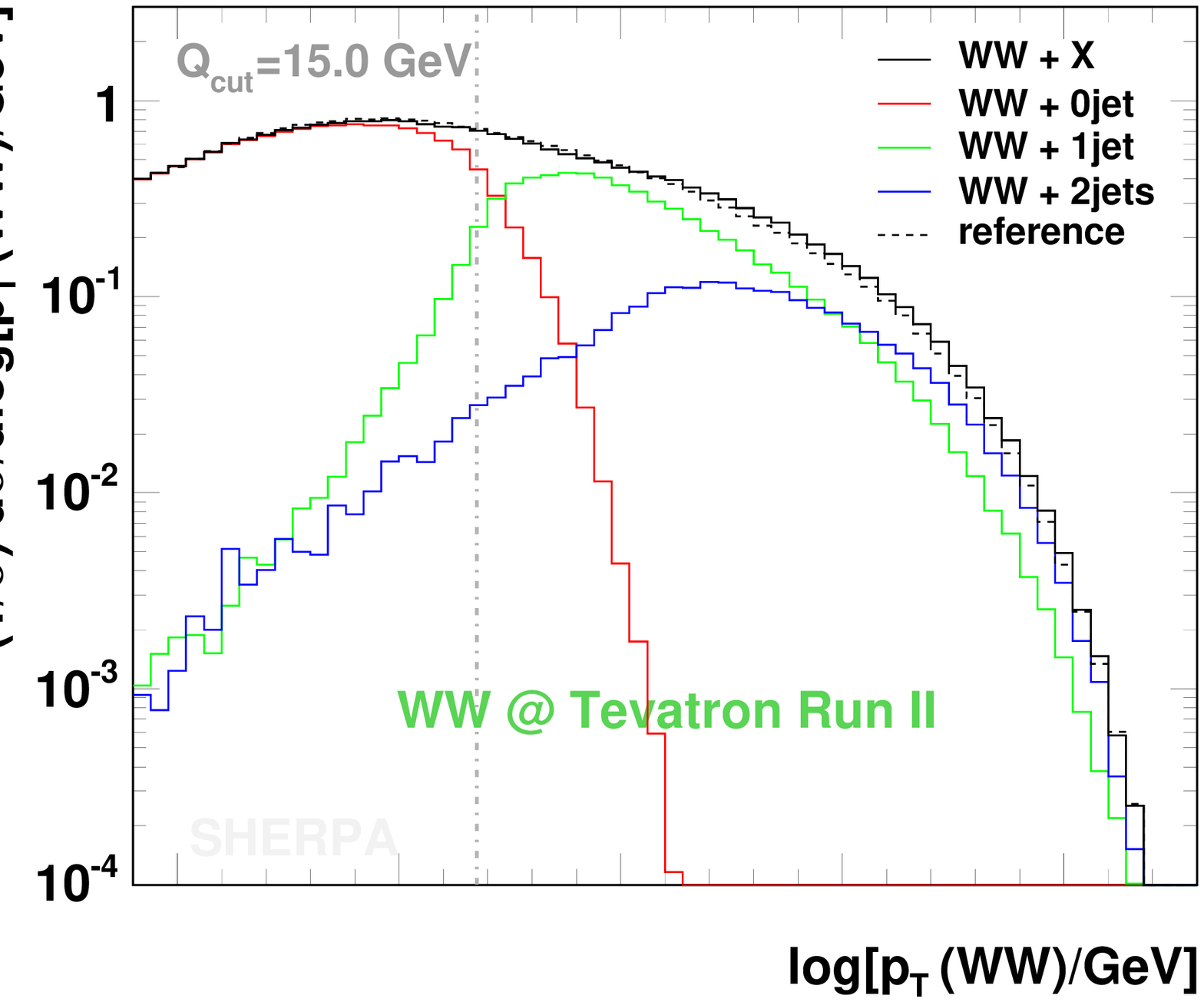}}
    \put(0,410){
      \includegraphics[width=70mm]{%
        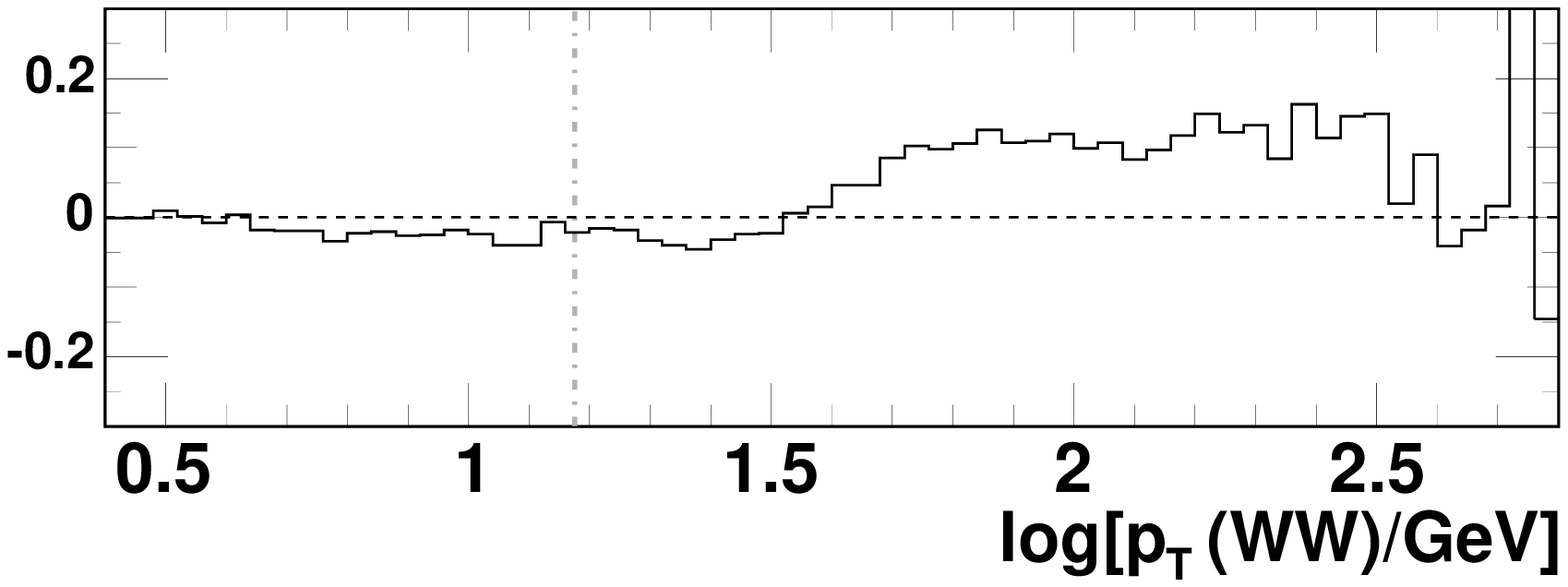}}
    \put(0,205){
      \includegraphics[width=70mm]{%
        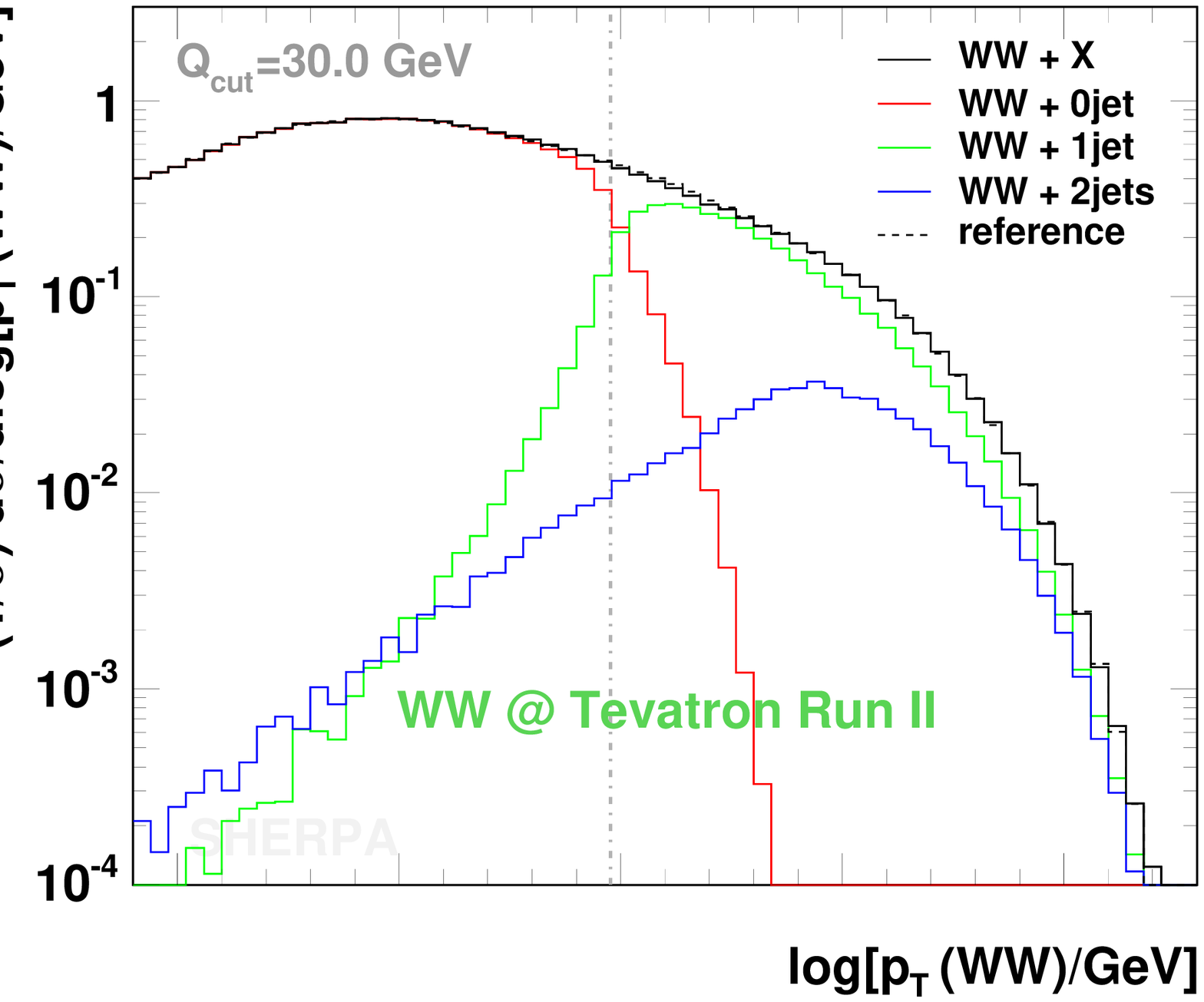}}
    \put(0,205){
      \includegraphics[width=70mm]{%
        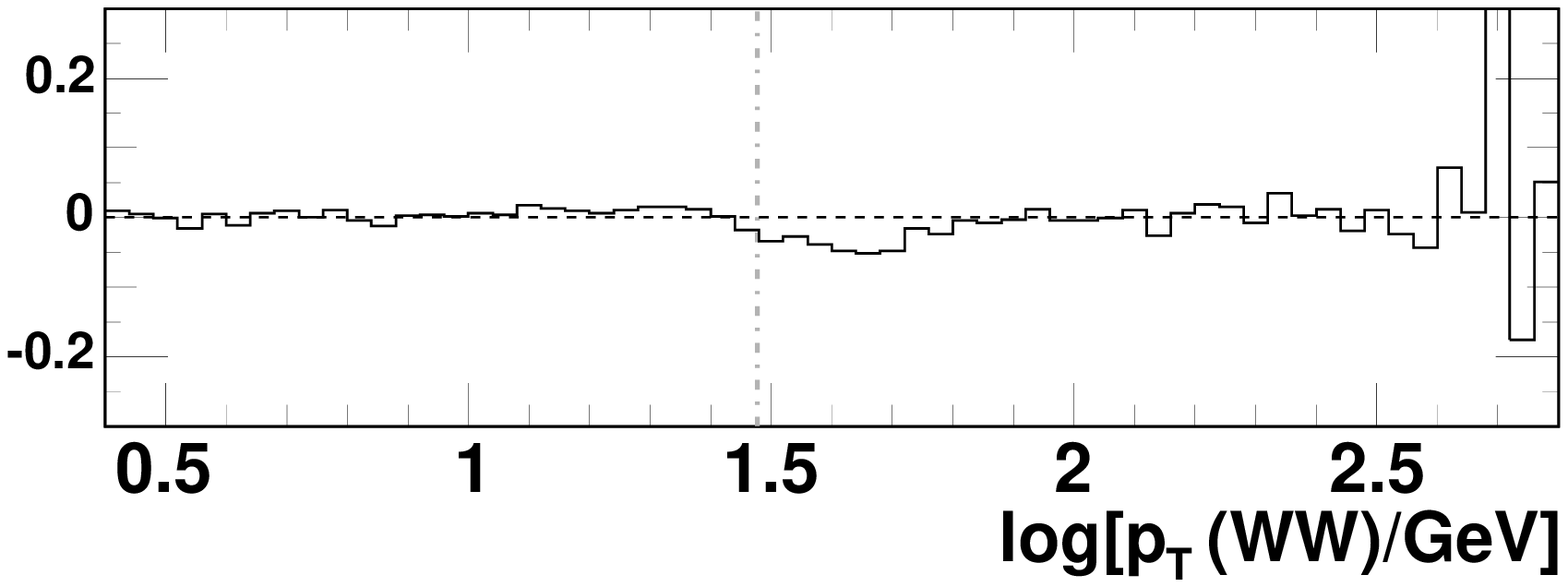}}
    \put(0,0){
      \includegraphics[width=70mm]{%
        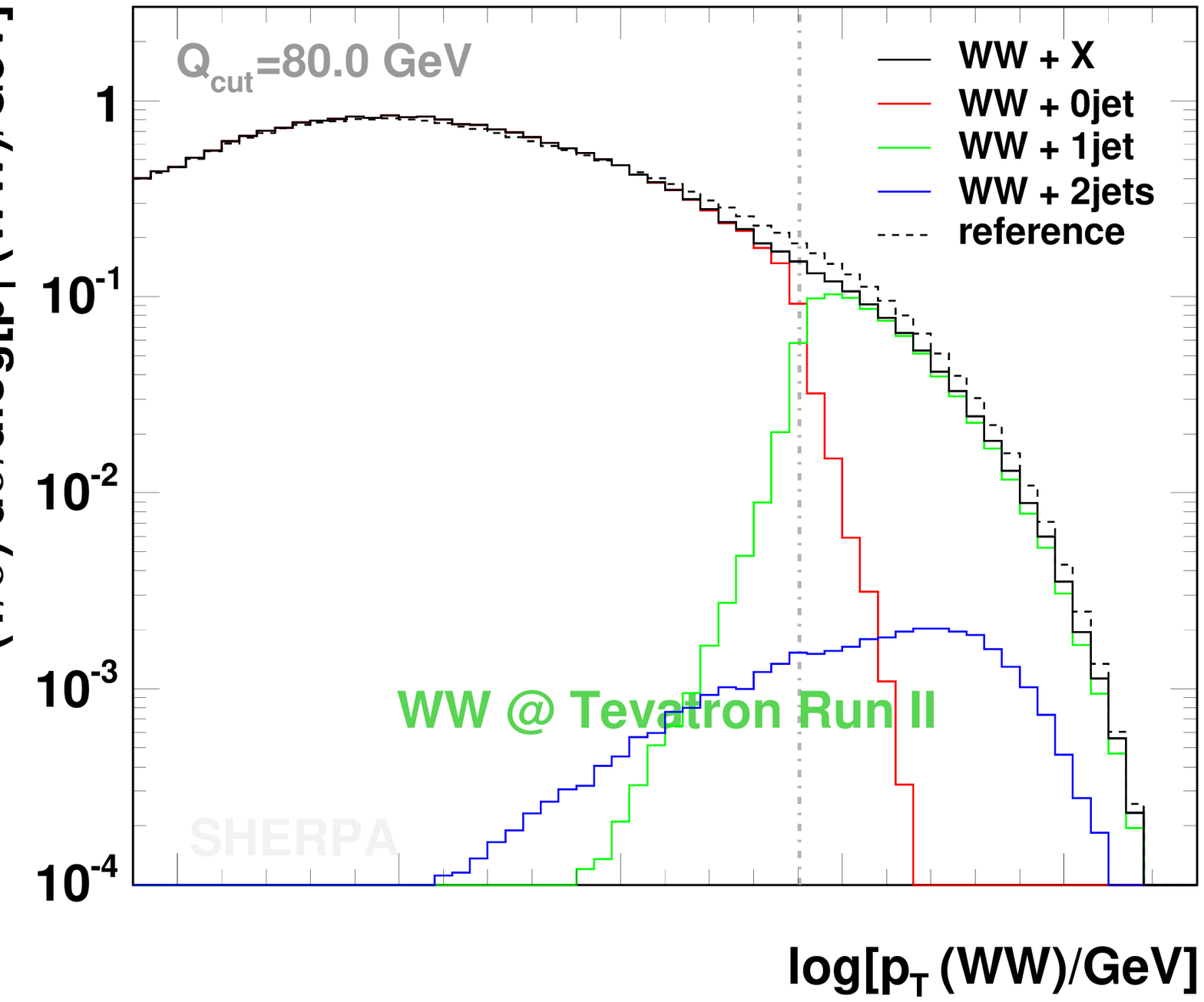}}
    \put(0,0){
      \includegraphics[width=70mm]{%
        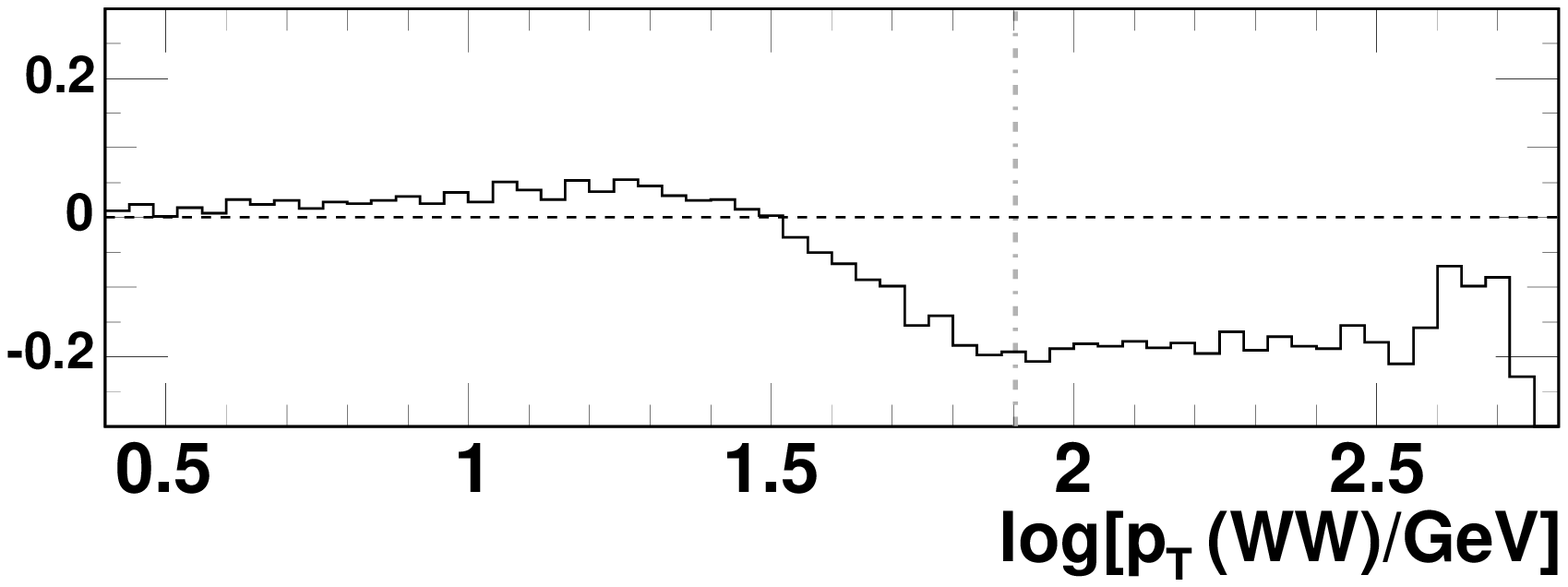}}
  \end{picture}
  \vspace{0mm}
  \caption{The $p_T$ distribution of the $W^+W^-$ system under merging
    scale variation. The cut indicated through a vertical
    dashed-dotted line has been chosen as $Q_{\rm cut}=15$, $30$ and
    $80$ GeV (from top to bottom). The black solid line shows the
    \she\ prediction obtained with $n_{\rm max}=2$, the black dashed
    one is the reference obtained as the mean of different $Q_{\rm
    cut}$ runs and the coloured lines indicate the different
    multiplicity contributions. The lower part of the plots exhibits
    the normalized difference of the prediction with respect to the
    reference. Cuts and input parameters are specified in the
    appendices.}
  \label{ptWW2_coychk}
  \vspace{-10mm}
\end{figure}
\begin{figure*}[t!]
  \vspace{0mm}
  \begin{picture}(444,480)
    \put(300,320){
      \includegraphics[width=54mm]{%
        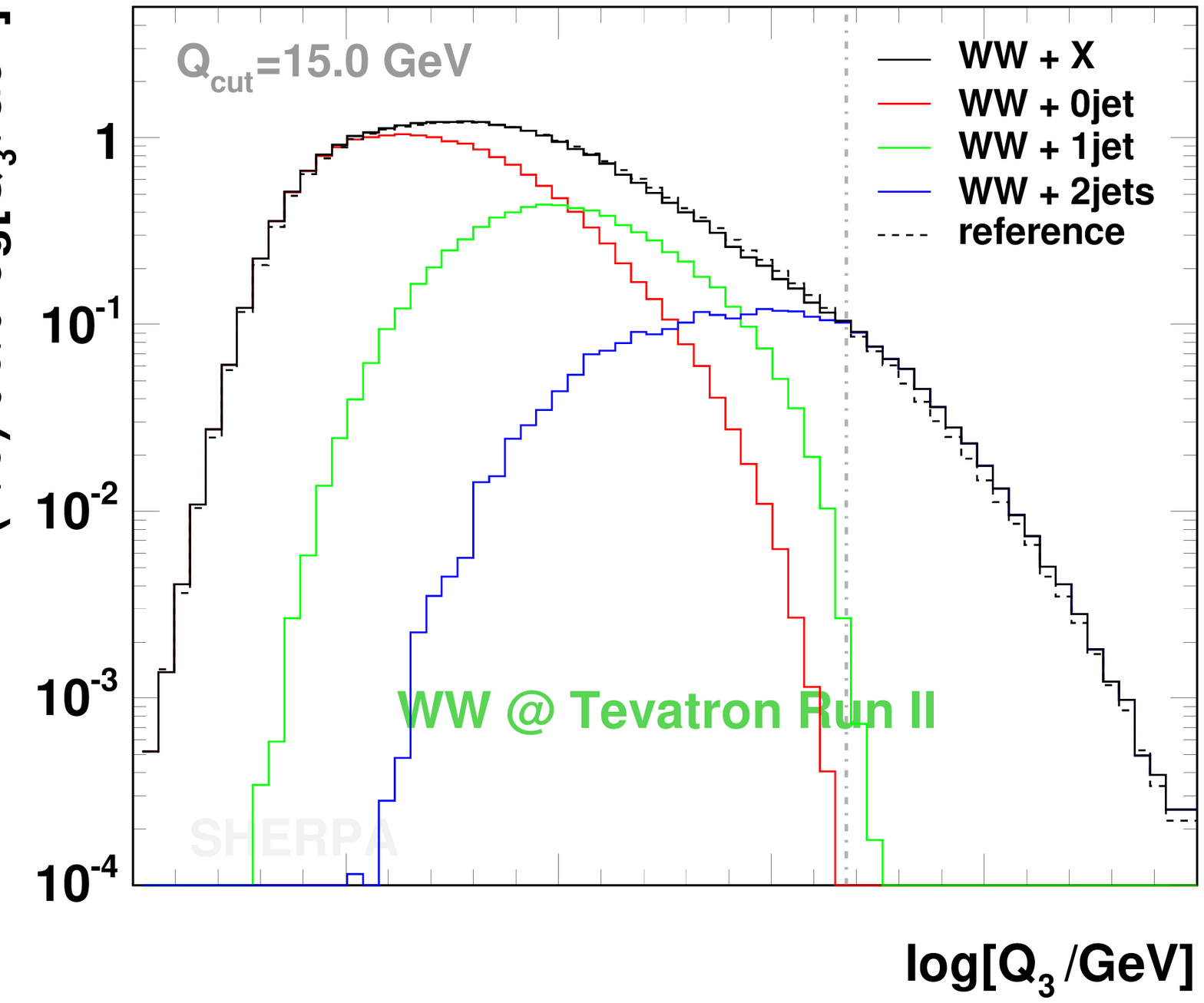}}
    \put(300,320){
      \includegraphics[width=54mm]{%
        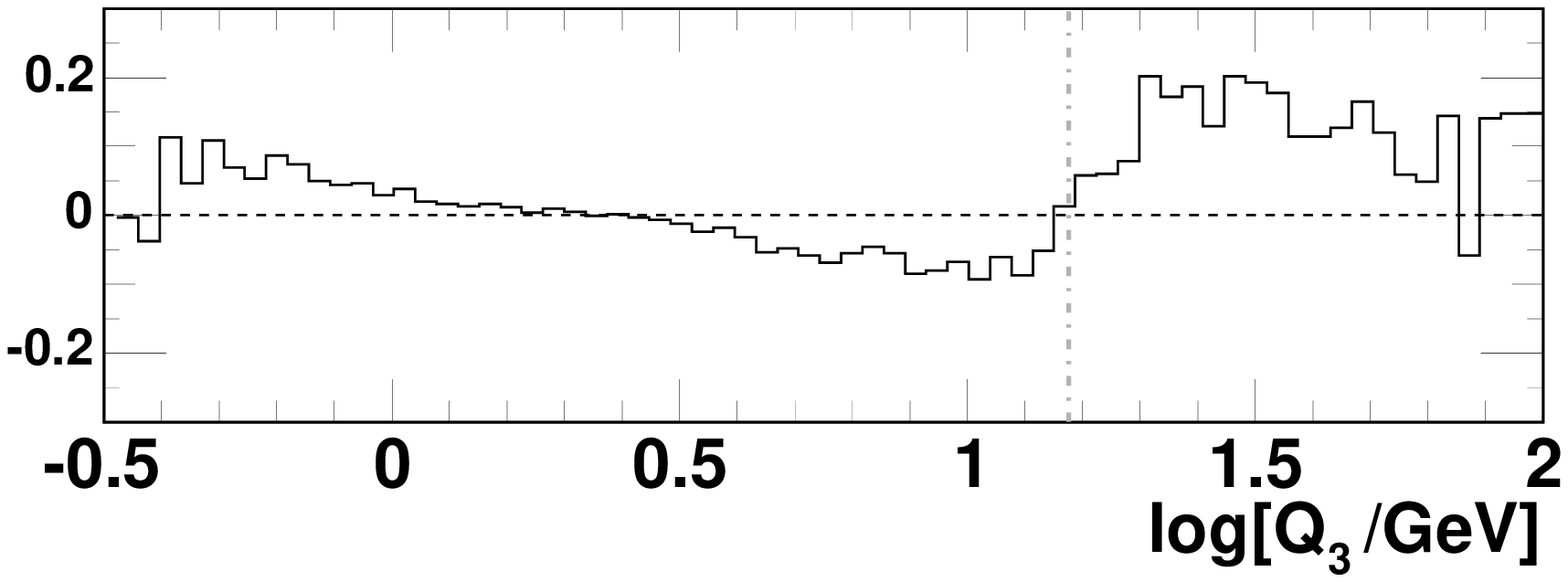}}
    \put(150,320){
      \includegraphics[width=54mm]{%
        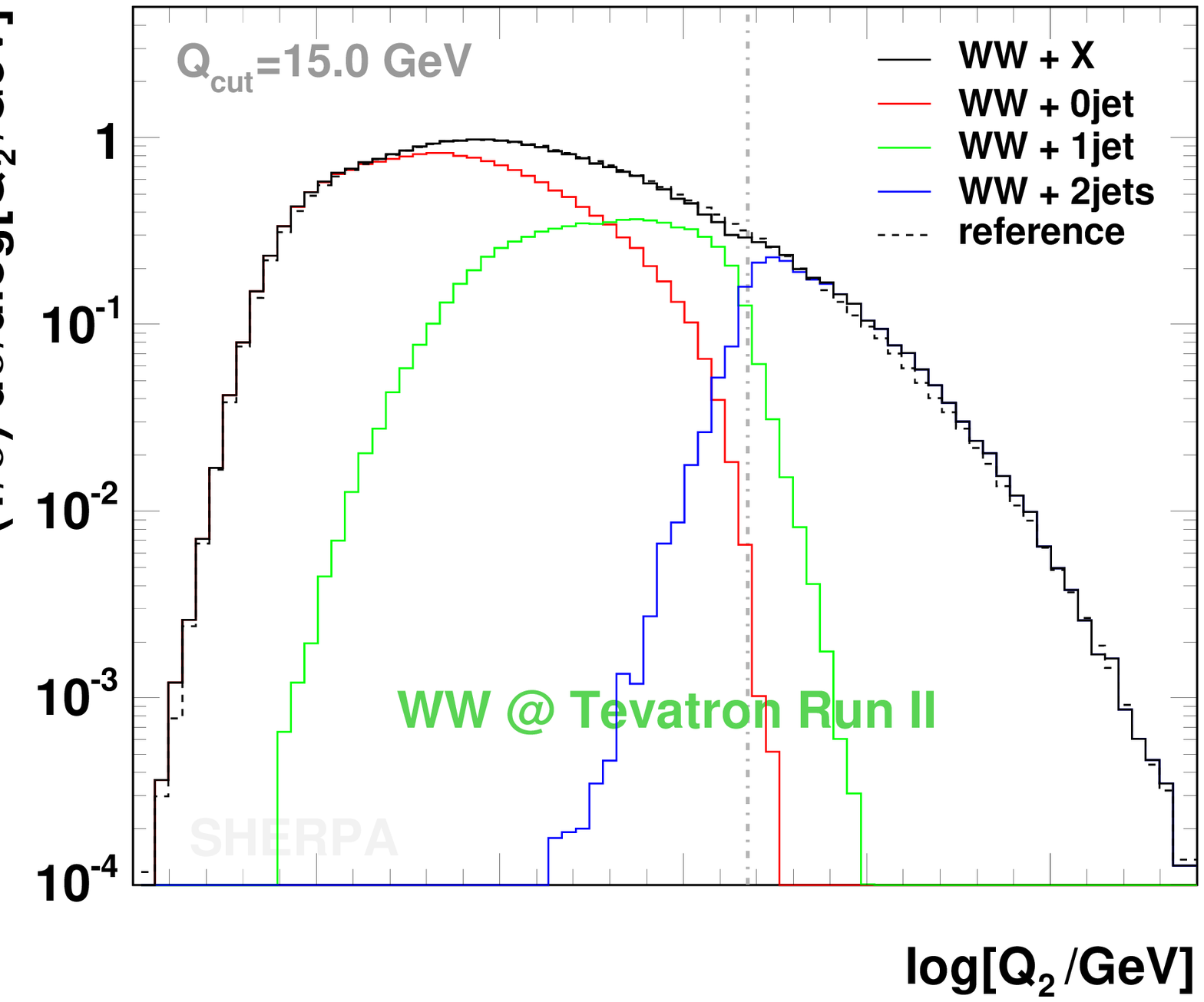}}
    \put(150,320){
      \includegraphics[width=54mm]{%
        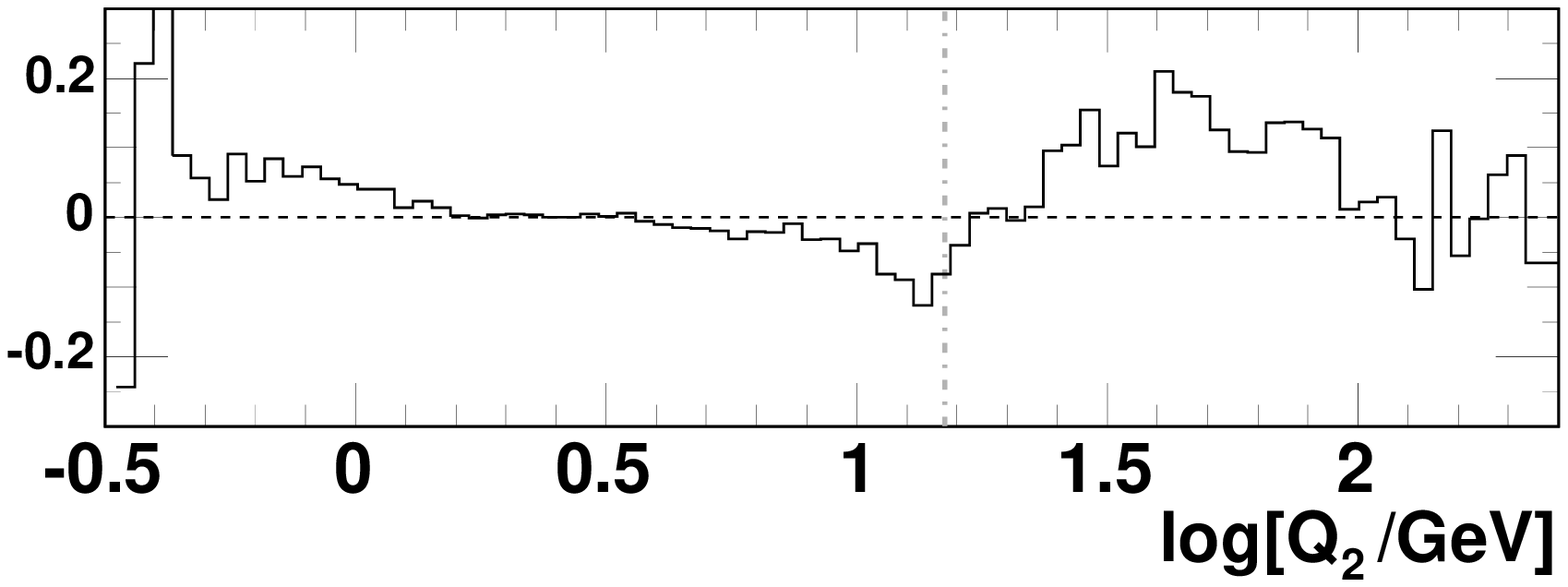}}
    \put(0,320){
      \includegraphics[width=54mm]{%
        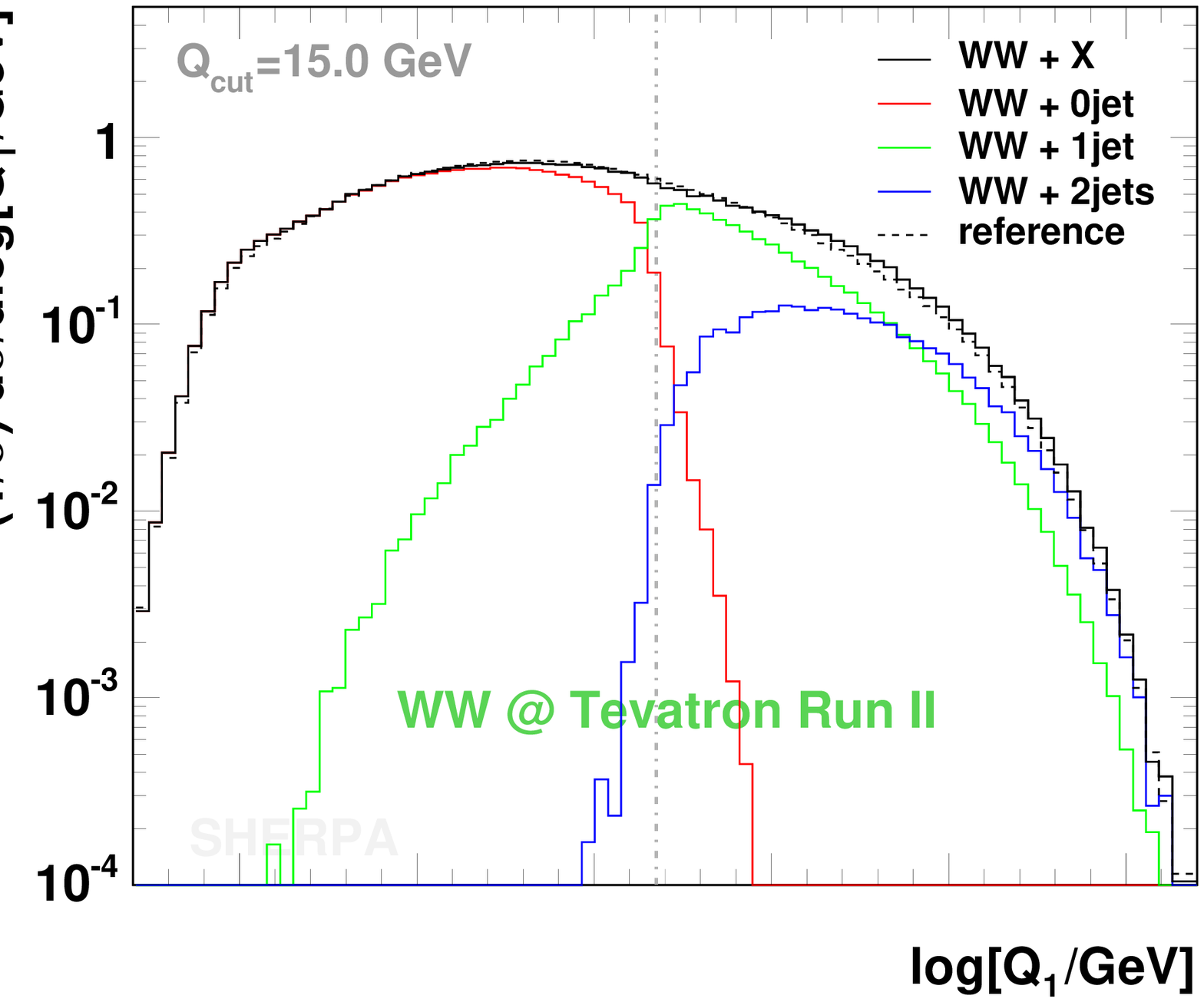}}
    \put(0,320){
      \includegraphics[width=54mm]{%
        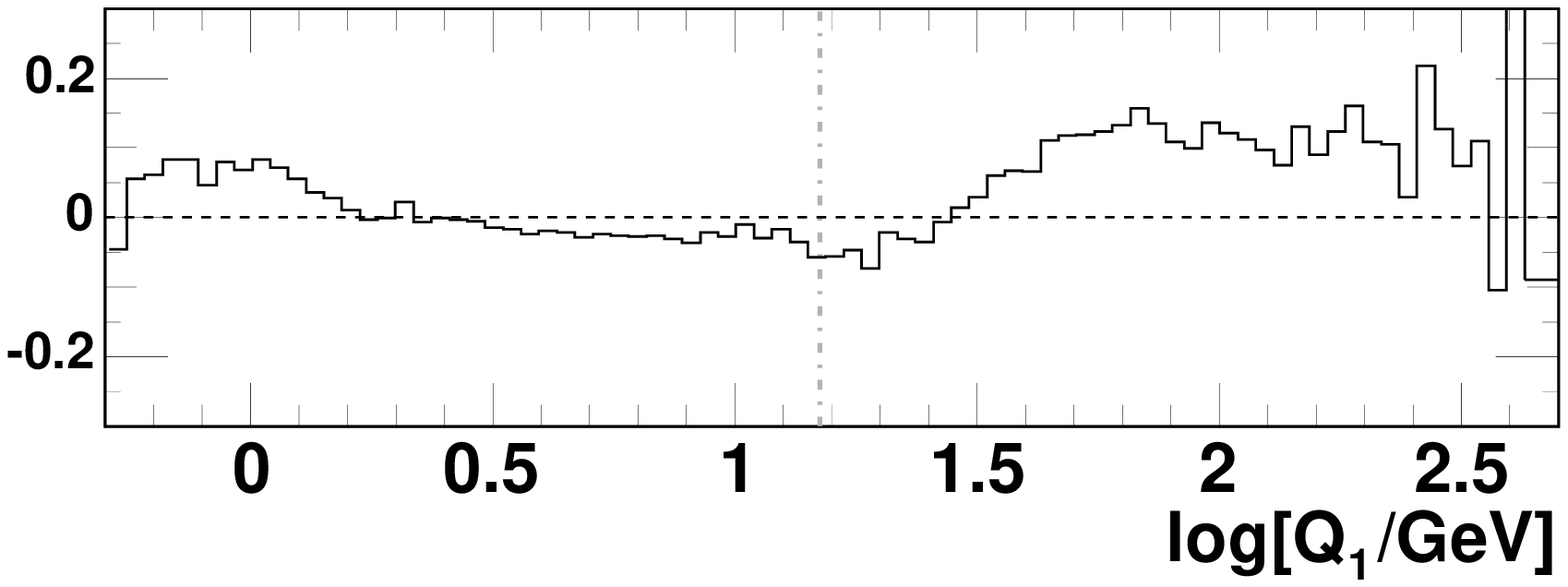}}
    \put(300,160){
      \includegraphics[width=54mm]{%
        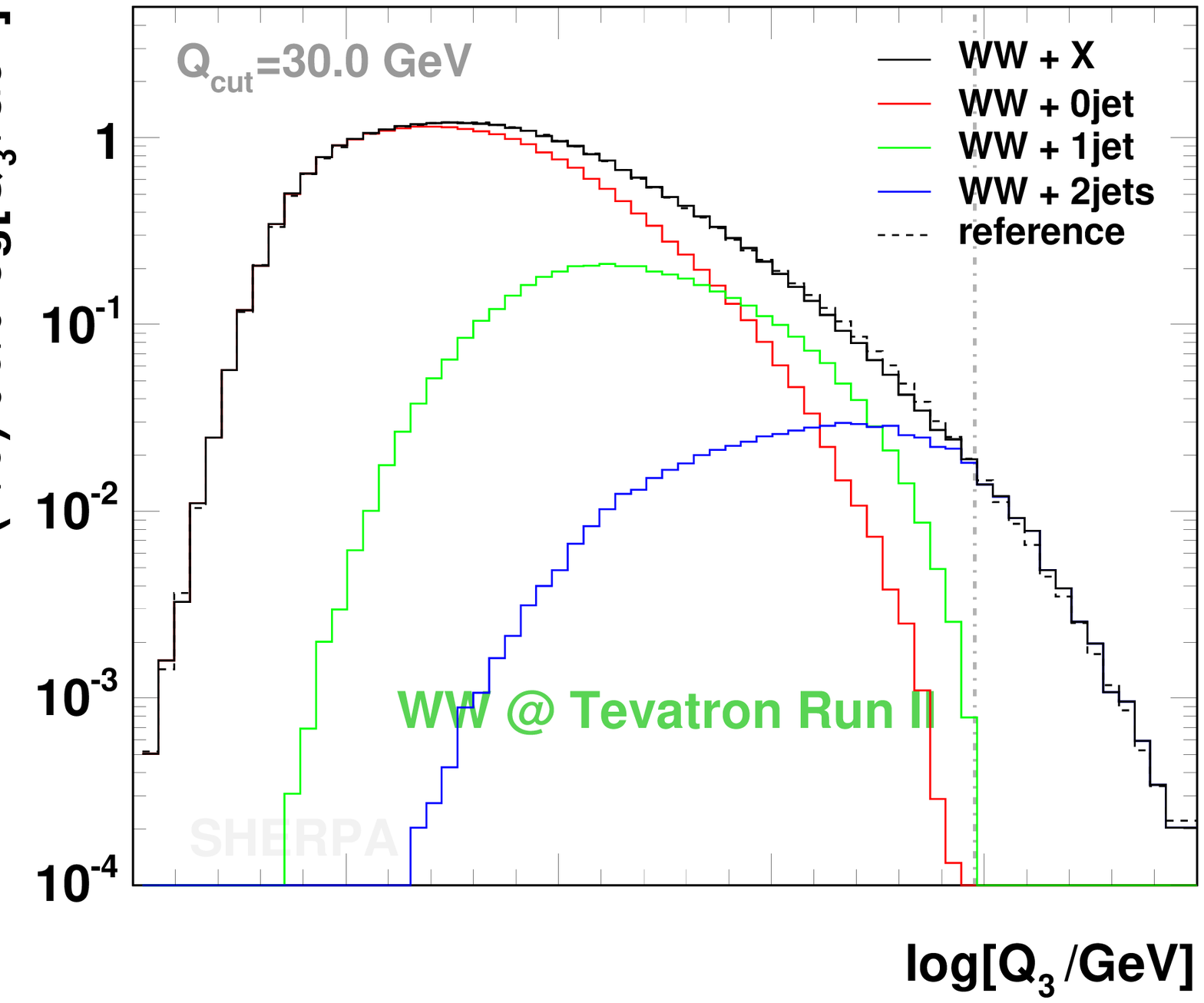}}
    \put(300,160){
      \includegraphics[width=54mm]{%
        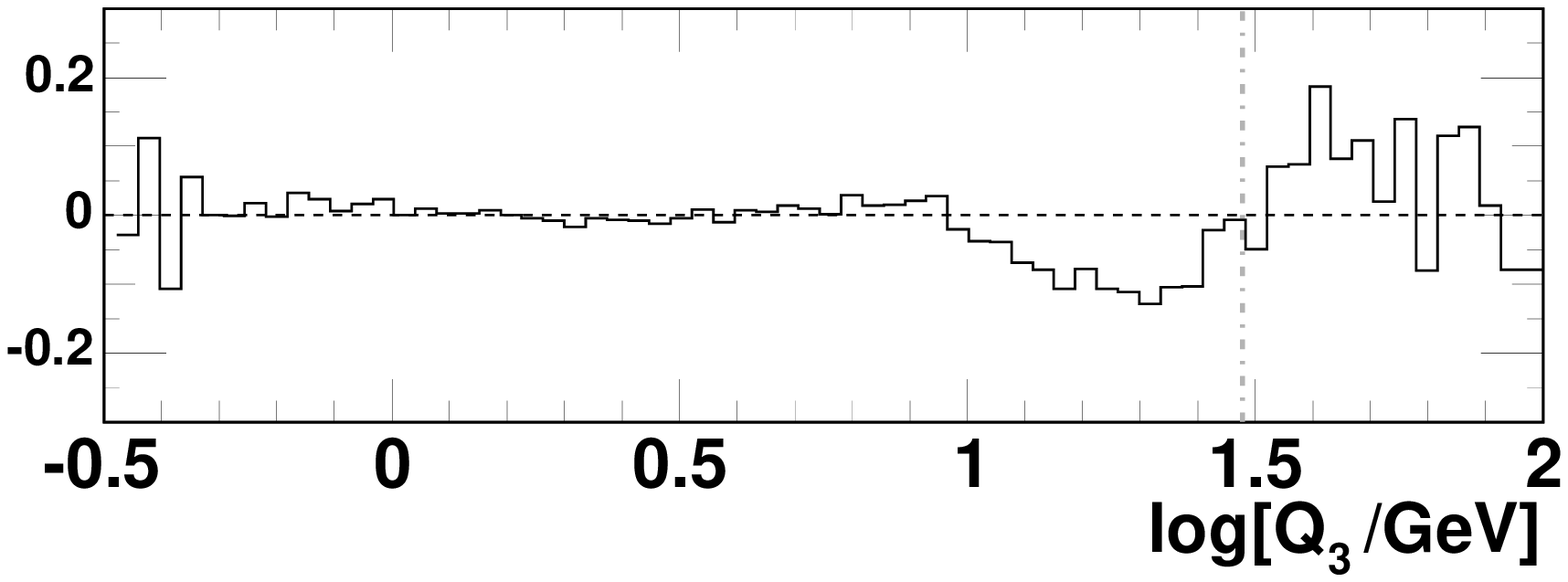}}
    \put(150,160){
      \includegraphics[width=54mm]{%
        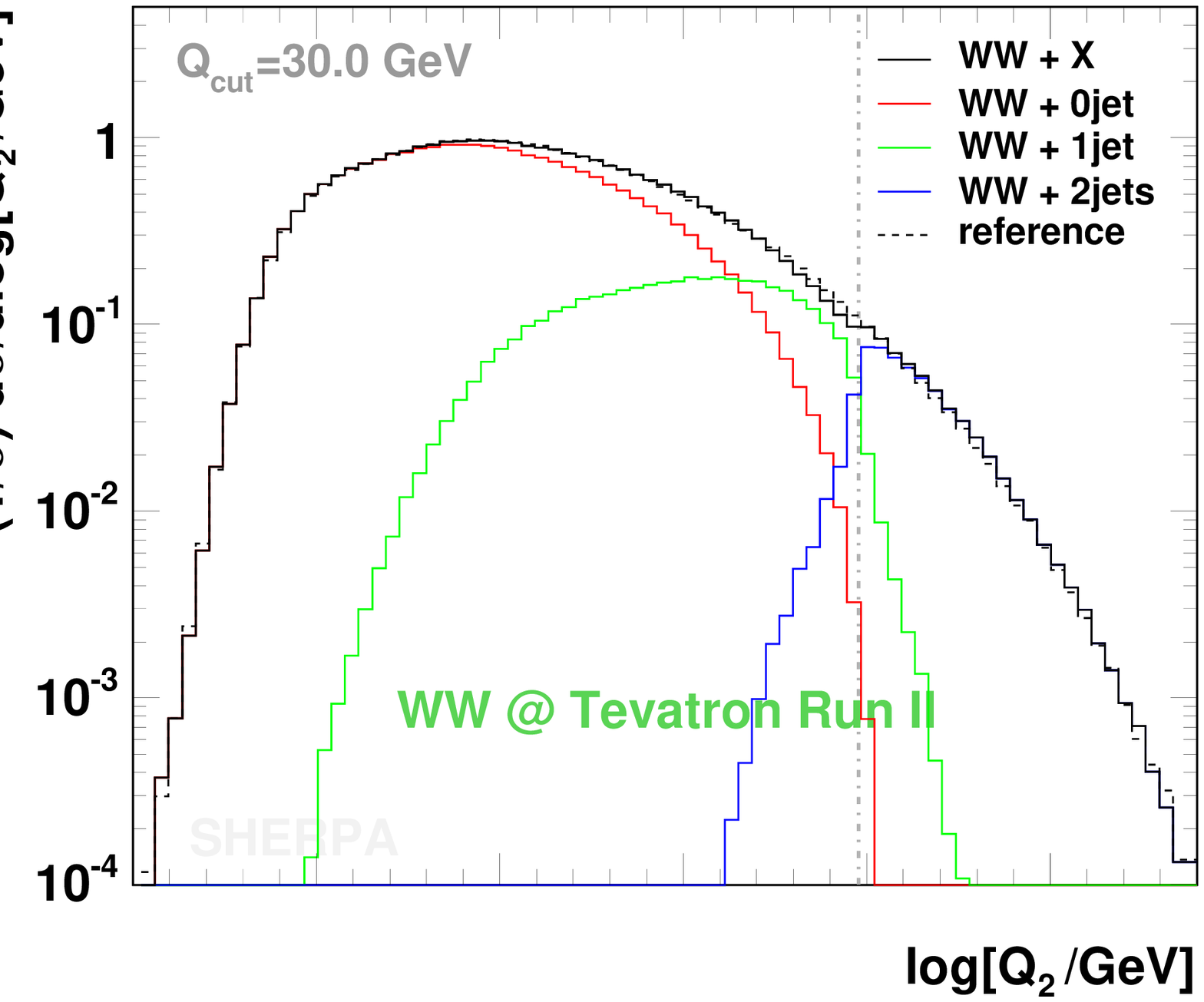}}
    \put(150,160){
      \includegraphics[width=54mm]{%
        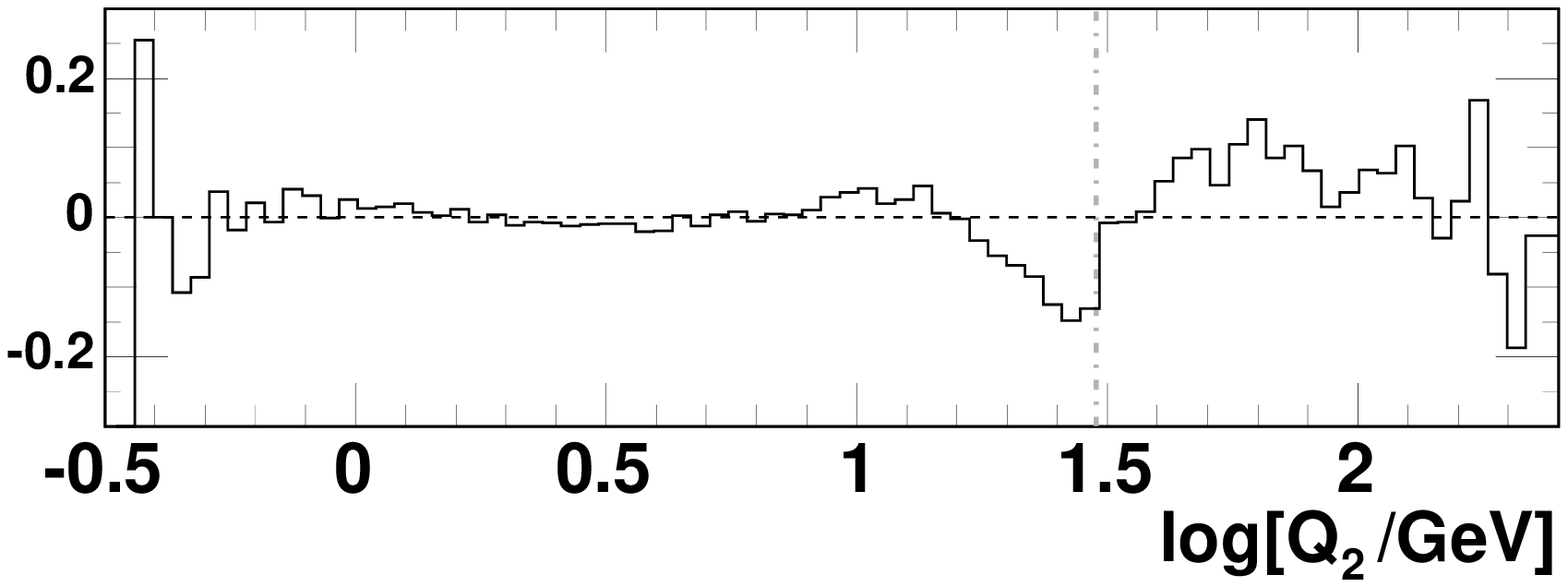}}
    \put(0,160){
      \includegraphics[width=54mm]{%
        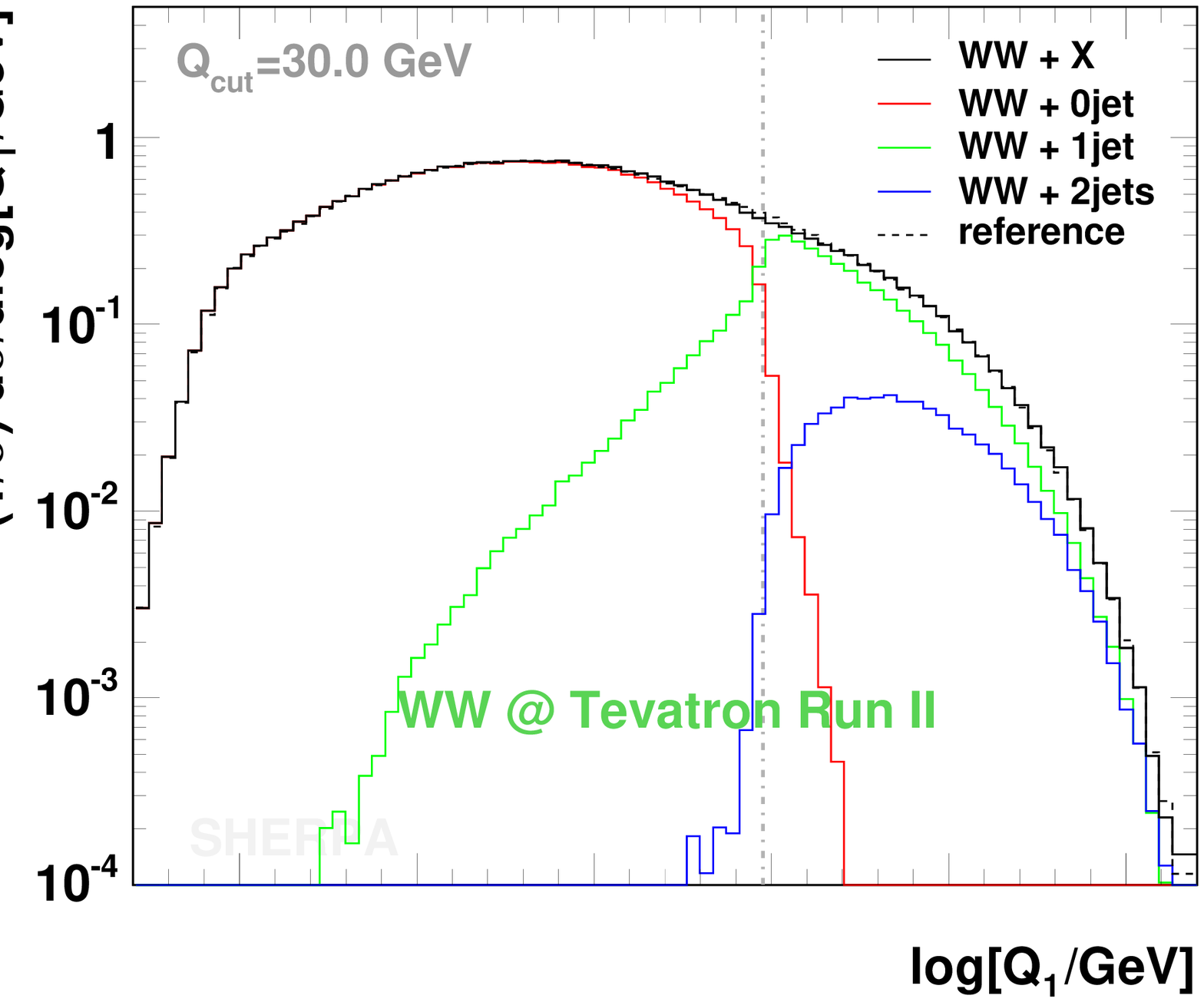}}
    \put(0,160){
      \includegraphics[width=54mm]{%
        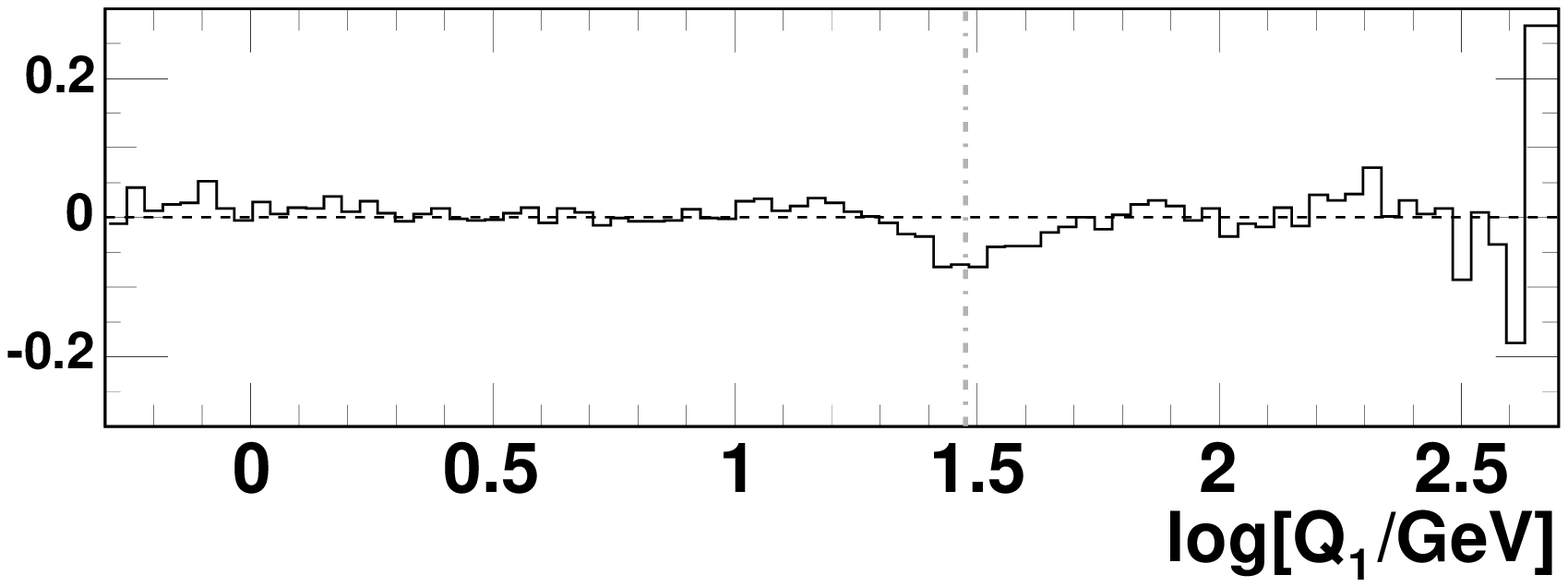}}
    \put(300,0){
      \includegraphics[width=54mm]{%
        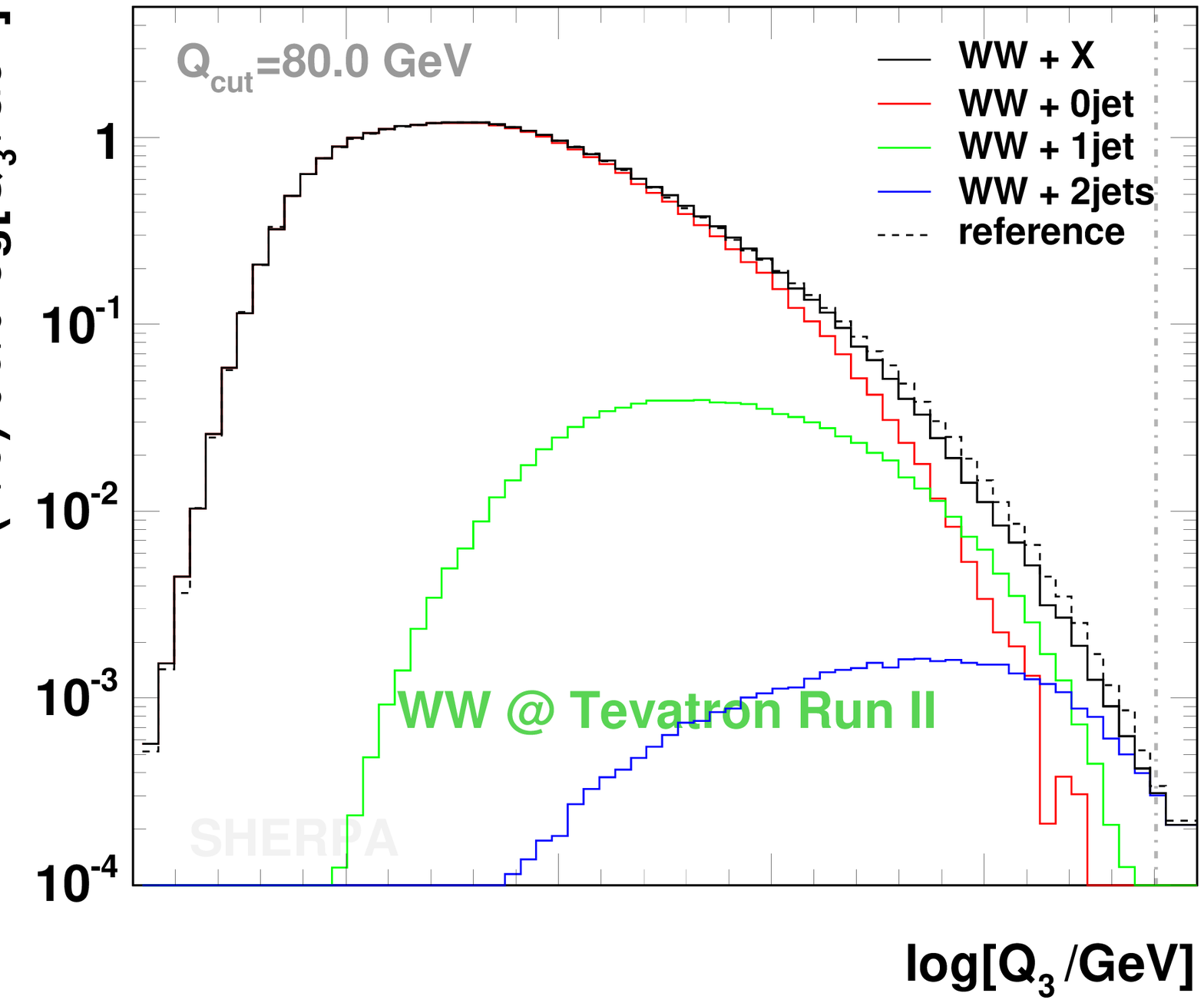}}
    \put(300,0){
      \includegraphics[width=54mm]{%
        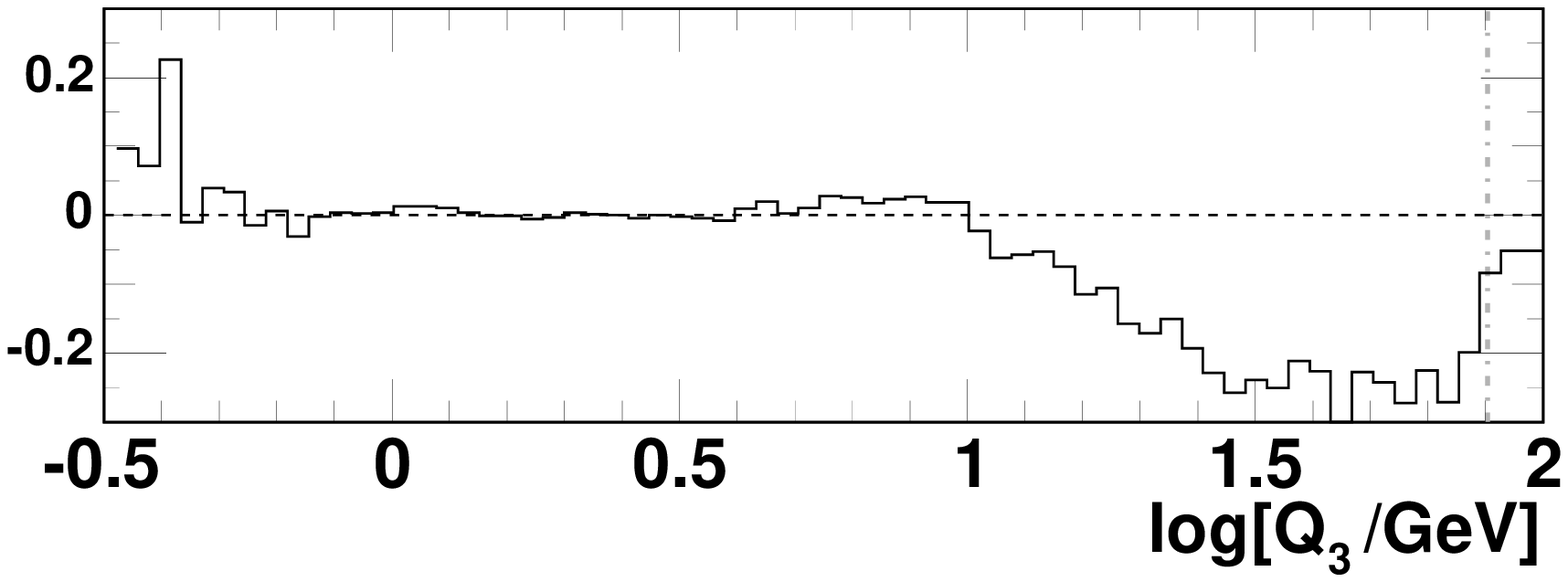}}
    \put(150,0){
      \includegraphics[width=54mm]{%
        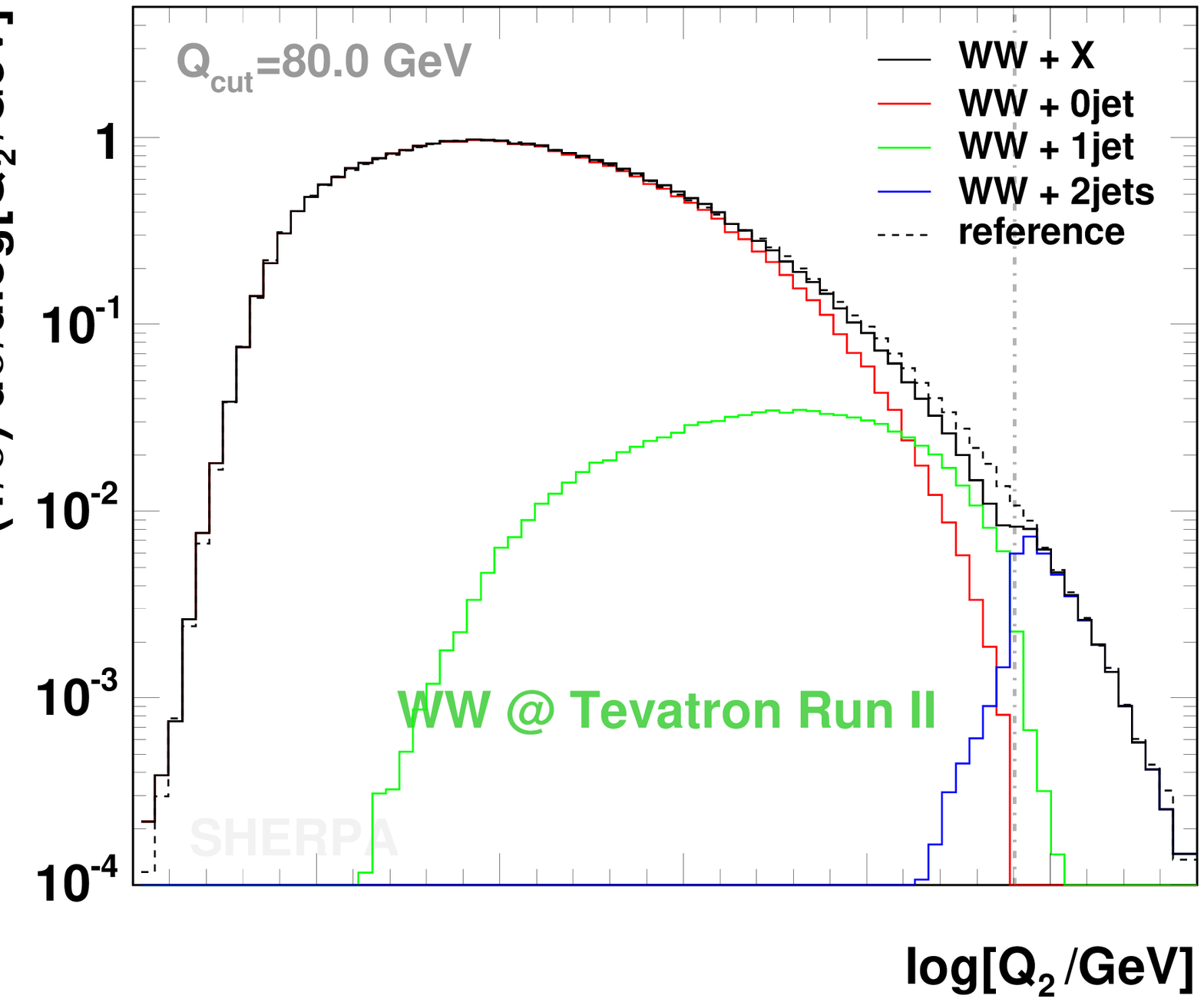}}
    \put(150,0){
      \includegraphics[width=54mm]{%
        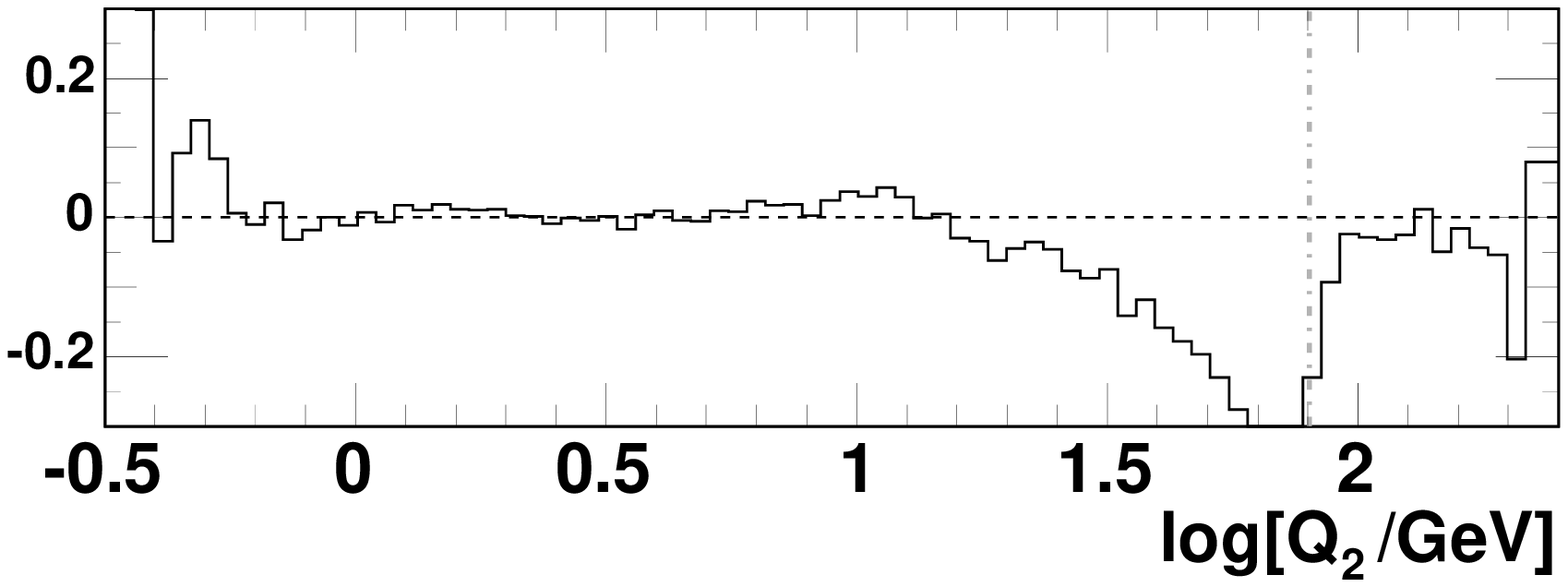}}
    \put(0,0){
      \includegraphics[width=54mm]{%
        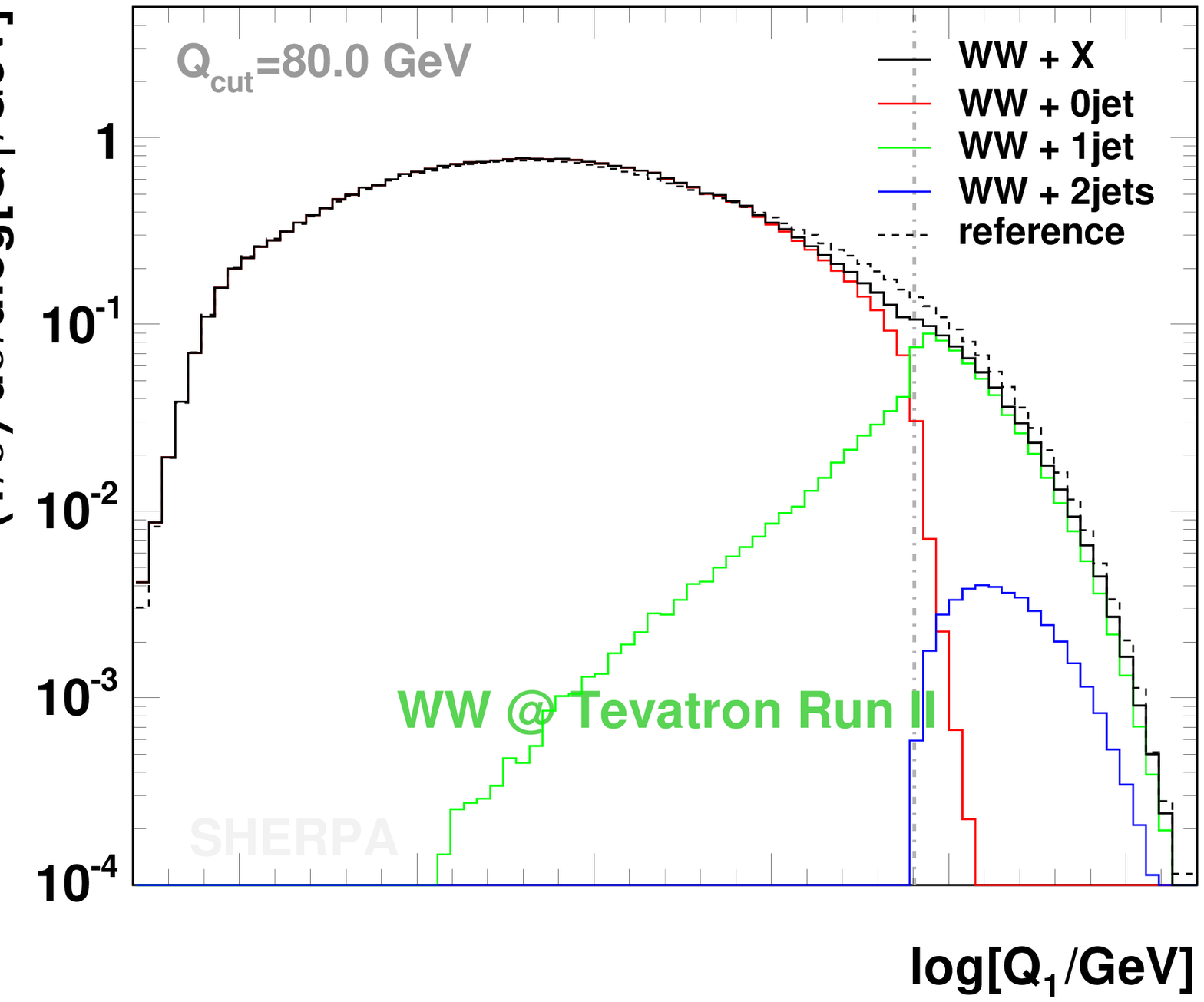}}
    \put(0,0){
      \includegraphics[width=54mm]{%
        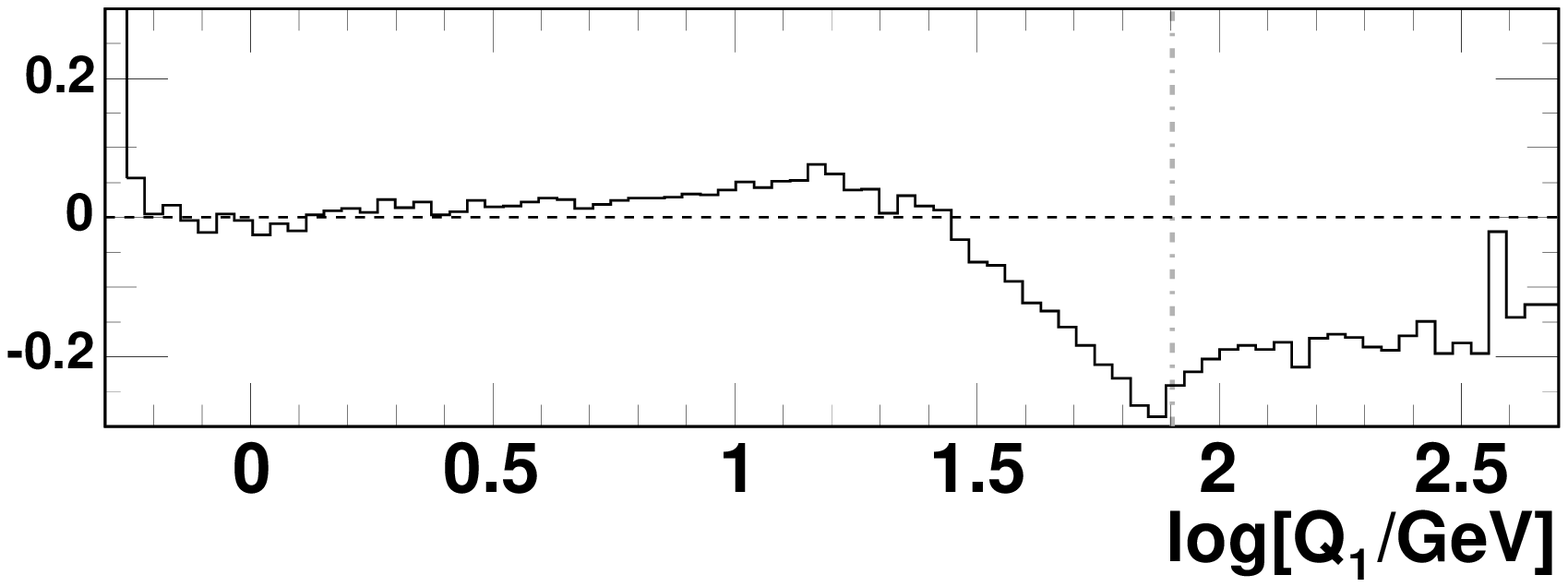}}
  \end{picture}
  \vspace{0mm}
  \caption{Differential $1\to0$ jet rate $Q_1$, $2\to1$ jet rate $Q_2$
    and $3\to2$ jet rate $Q_3$ (left to right) for the \she\ $n_{\rm
    max}=2$ configuration. The cut has been chosen to be $15$, $30$
    and $80$ GeV (from top to bottom). The black solid line shows the
    total result, the black dashed one is the reference obtained as
    the mean of different $Q_{\rm cut}$ runs and the coloured lines
    indicate the different multiplicity contributions.
    The vertical dashed dotted line indicates the
    separation cut position. The lower part in all plots pictures the
    normalized difference of the corresponding prediction with respect
    to the reference. For input parameters and cuts, see Apps.\
    \ref{app_input} and \ref{app_cuts}.}
  \label{djr2_coychk}
  \vspace{0mm}
\end{figure*}
%
%
\begin{figure*}[t!]
  \vspace{0mm}
  \begin{picture}(424,204)
    \put(240,0){
      \includegraphics[width=70mm]{%
        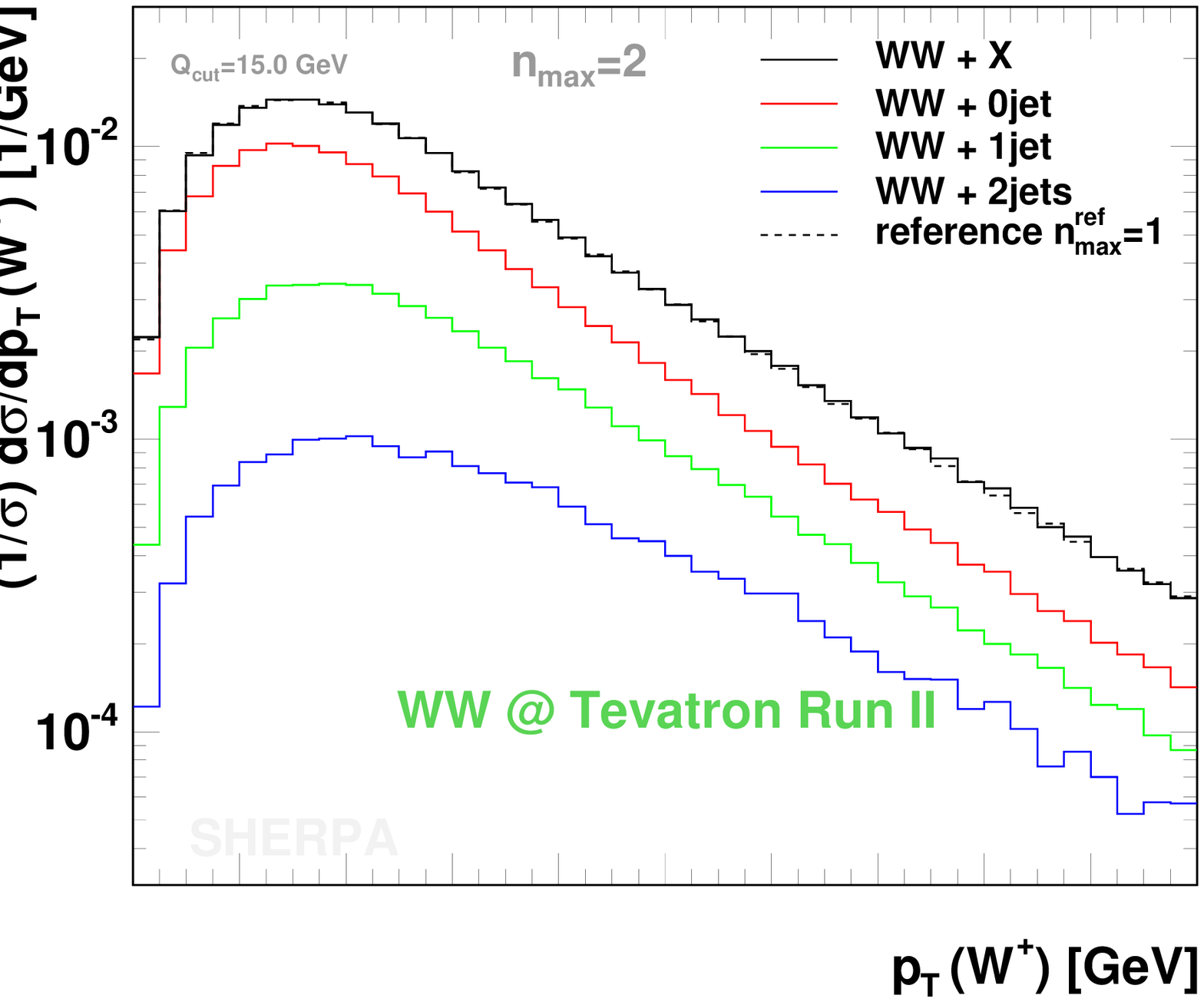}}
    \put(240,0){
      \includegraphics[width=70mm]{%
        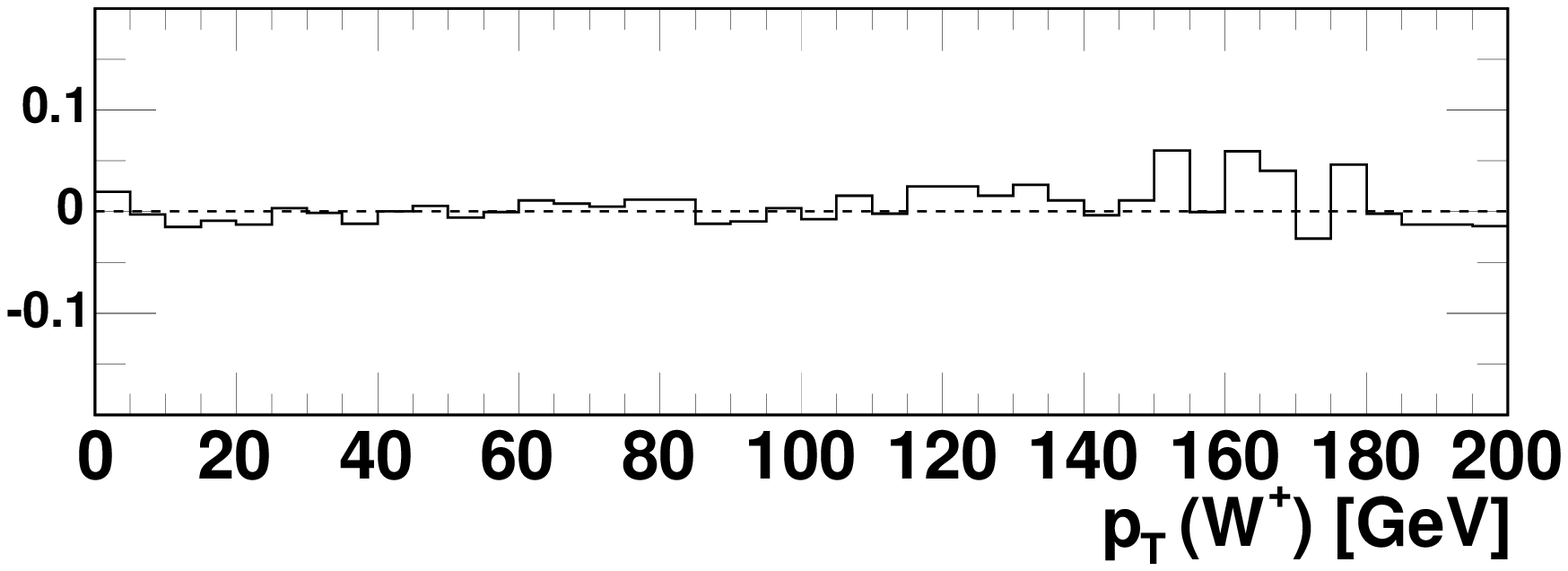}}
    \put(0,0){
      \includegraphics[width=70mm]{%
        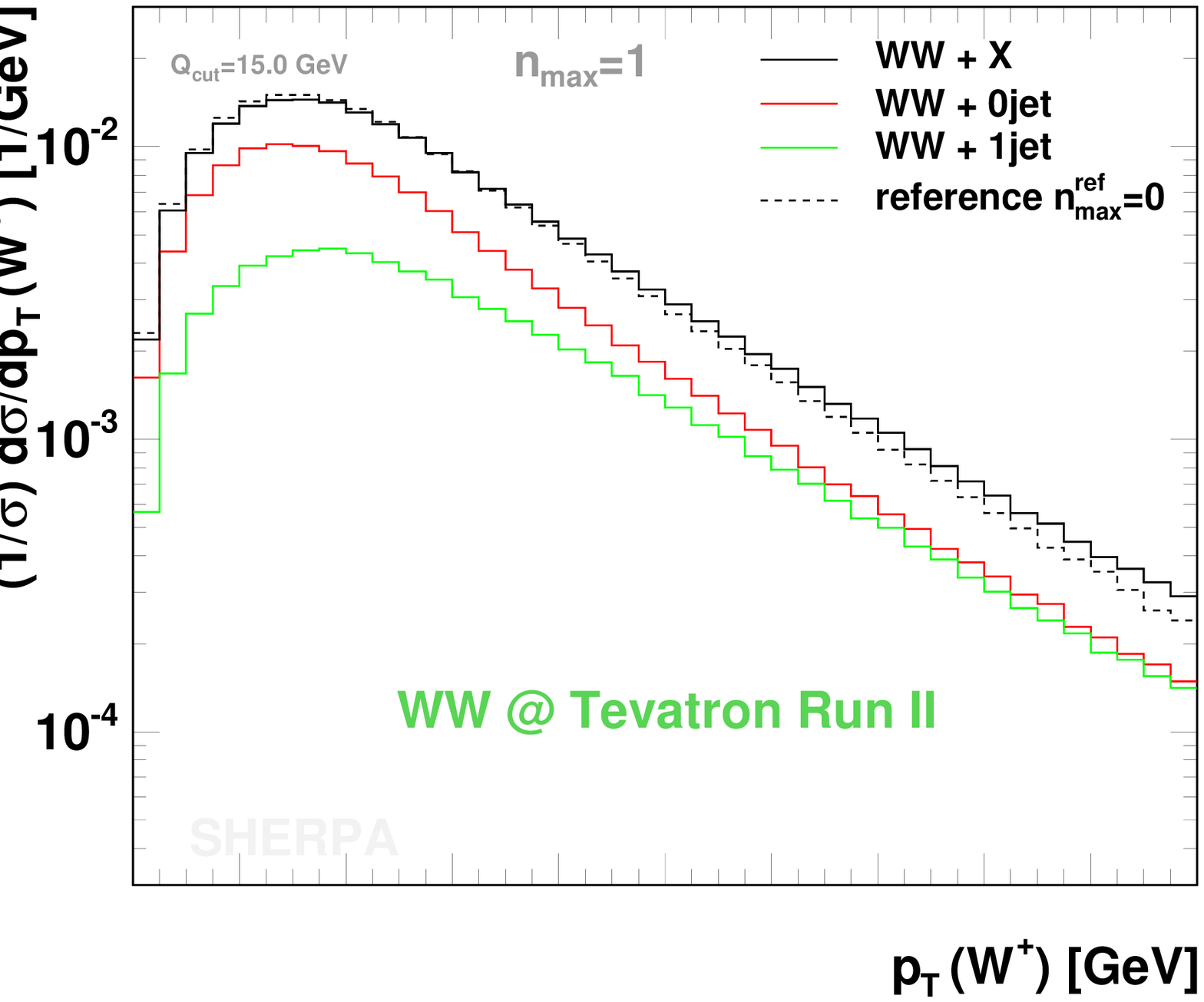}}
    \put(0,0){
      \includegraphics[width=70mm]{%
        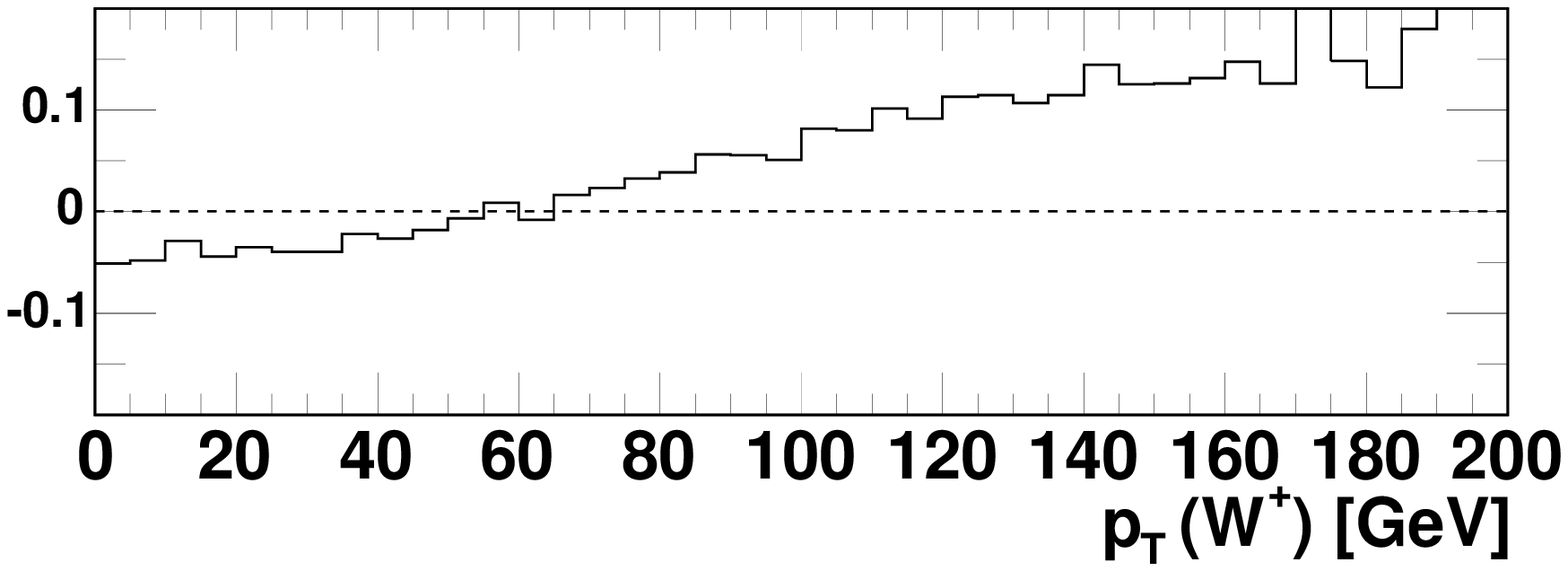}}
  \end{picture}
  \vspace{0mm}
  \caption{The $p_T$ distribution of the $W^+$ boson in dependence on
    the variation of the maximal jet number. The comparison is to a
    (black dashed) reference curve obtained with $n^{\rm ref}_{\rm
    max}=n_{\rm max}-1$. The cut has been chosen to be $15$ GeV. In
    both plots the black solid line shows the total result obtained
    with \she. The coloured lines indicate the different multiplicity
    contributions. The lower part in both plots visualizes the
    normalized difference of the corresponding prediction with respect
    to the reference. For input parameters and cuts, see Apps.\
    \ref{app_input} and \ref{app_cuts}.}
  \label{ptWplusN_coychk}
  \vspace{0mm}
\end{figure*}
\\
As a first result, consider the $p_T$ distribution of the $W^+$ boson,
presented in Fig.\ \ref{ptWplus2_coychk}. The distributions become
slightly softer for increasing cuts. However, this observable is very
stable under variation of $Q_{\rm cut}$ with maximal deviations on the
$\pm5\%$ level only. The shape of the $W^+$ boson's $p_T$ is already
described at LO (using a parton shower only). As it can be seen from
the figure, this LO dominance is nicely kept by the \she\ approach
under $Q_{\rm cut}$ variation. There the $1$jet (green line) and
$2$jet (blue line) contributions are reasonably -- for the $80$ GeV
run, even strongly -- suppressed with respect to the leading
contribution.
\\
In Fig.\ \ref{ptWW2_coychk} the transverse momentum spectrum of the
$W^+W^-$ system is depicted. Here, deviations show up, but they do not
amount to more than $\pm20\%$. Thus, the QCD radiation pattern depends
only mildly on $Q_{\rm cut}$ (indicated by a vertical dashed-dotted
line), which at the same time has been varied by nearly one order of
magnitude. For $Q_{\rm cut}=15$ GeV the matrix element domain is
enhanced with respect to the reference resulting in a harder $p_T$
tail. In contrast by using $Q_{\rm cut}=80$ GeV the hard tail of the
diboson transverse momentum is underestimated with respect to the
reference, since the parton shower attached only to the lowest order
matrix element starts to fail in the description of high-$p_T$ QCD
radiation at $p_T\approx30$ GeV. At $Q_{\rm cut}=80$ GeV a smooth
transition is required. The higher order matrix elements then stop the
decrease in the $p_T$ prediction.
\\
In previous publications it turned out that differential jet rates
most accurately probe the merging algorithm, since they most suitably
reflect the interplay of the matrix elements and the parton shower in
describing QCD radiation below and above the jet resolution cut.
Results obtained with the Run II $k_{\perp}$-algorithm using $R=1$ are
shown for the $1\to0$, $2\to1$ and $3\to2$ transition in the left,
middle and right panels of Fig.\ \ref{djr2_coychk}, respectively. The
value for the internal cut increases from $Q_{\rm cut}=15$ GeV (top)
to $Q_{\rm cut}=80$ GeV (bottom). Compared with the $p^{WW}_T$
spectra, similar characteristics of deviations from the reference
curve appear. However, here, they are moderately larger reaching up to
$\pm30\%$. The dashed dotted vertical line again marks the position of
$Q_{\rm cut}$, which also pictures the separation of the $n$jet from
the $n+1$jet contribution. Small holes visible around the respective
separation cuts are due to a mismatch of matrix element and parton
shower kinematics. For $Q_{\rm cut}=80$ GeV these holes are much more
pronounced, reflecting the failure of the parton shower in filling the
hard $p_T$ emission phase space appropriately.
\\
Taken together, the deviations found are very moderate; however, in
certain phase space regions they may reach up to $30\%$. This is
satisfactory, since the merging algorithm guarantees $Q_{\rm cut}$
independence on the leading logarithmic accuracy only. The residual
dependence of the results on $Q_{\rm cut}$ may be exploited to tune
the perturbative part of the Monte Carlo event generator.

\subsection*{Impact of the maximal number of included matrix elements}

\begin{figure}[t!]
  \vspace{0mm}
  \begin{picture}(188,204)
    \put(0,0){
      \includegraphics[width=70mm]{%
        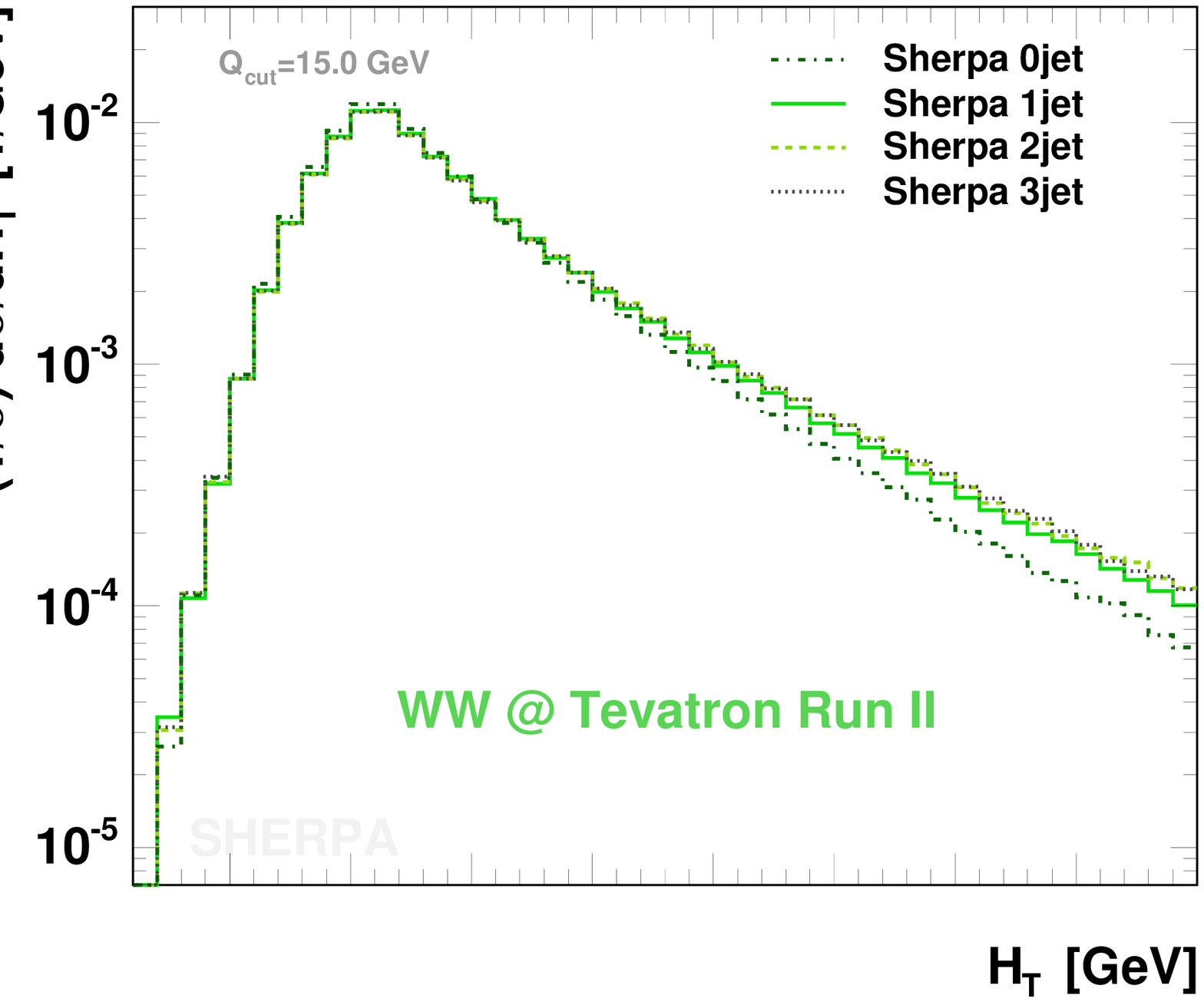}}
    \put(0,0){
      \includegraphics[width=70mm]{%
        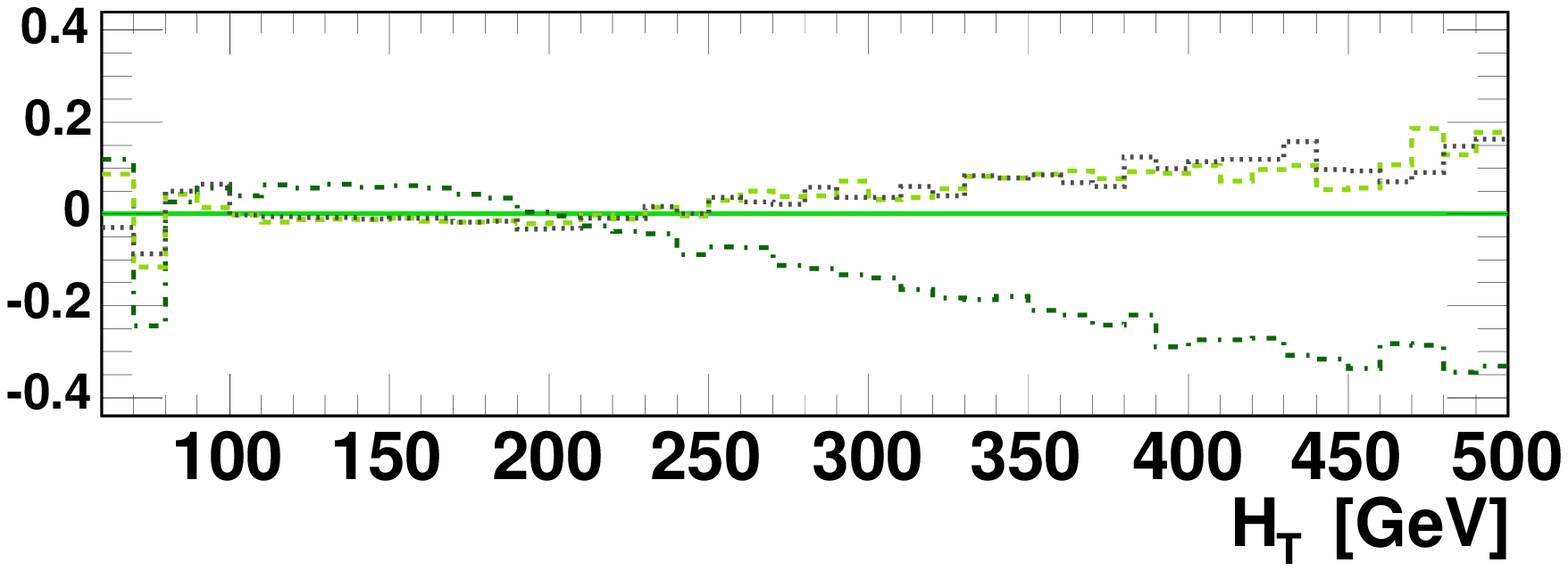}}
  \end{picture}
  \vspace{0mm}
  \caption{The $H_T$ distribution and its dependence on the variation
      of the maximal jet number. The separation cut has been chosen to
      be $15$ GeV. The green solid line shows the \she\ prediction
      obtained with $n_{\rm max}=1$, the lighter dashed and the grey
      dotted one stand for the $n_{\rm max}=2$ and $n_{\rm max}=3$
      prediction, respectively; the darkgreen dashed-dotted curve
      pictures the pure shower performance of \she\ starting off with
      the lowest order matrix element. The lower part of the plot
      shows the normalized differences with respect to the $n_{\rm
      max}=1$ case. For input parameters and cuts, see Apps.\
      \ref{app_input} and \ref{app_cuts}.}
  \label{htN_coychk}
  \vspace{0mm}
\end{figure}
\begin{figure}[t!]
  \vspace{0mm}
  \begin{picture}(188,410)
    \put(0,205){
      \includegraphics[width=70mm]{%
        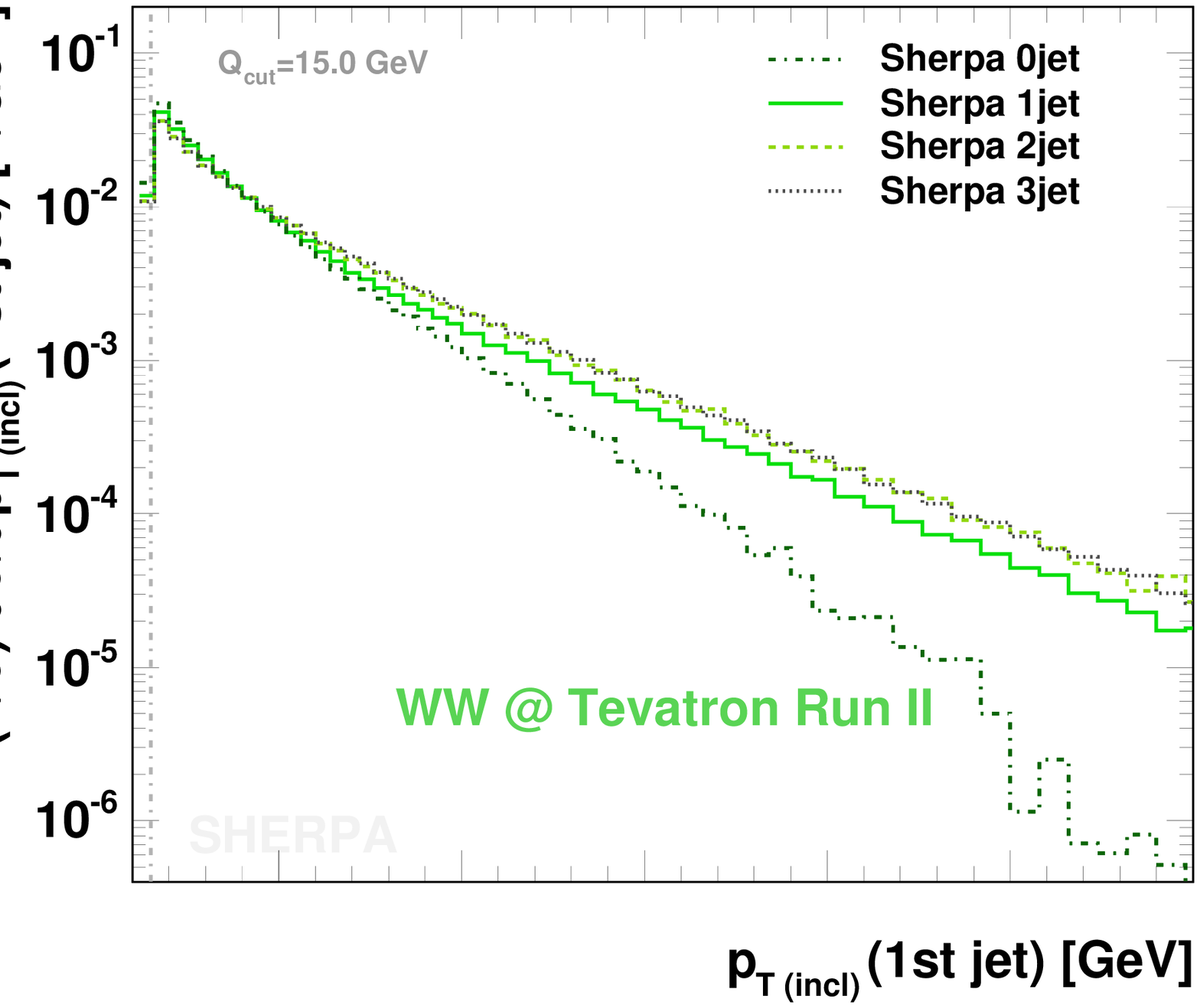}}
    \put(0,205){
      \includegraphics[width=70mm]{%
        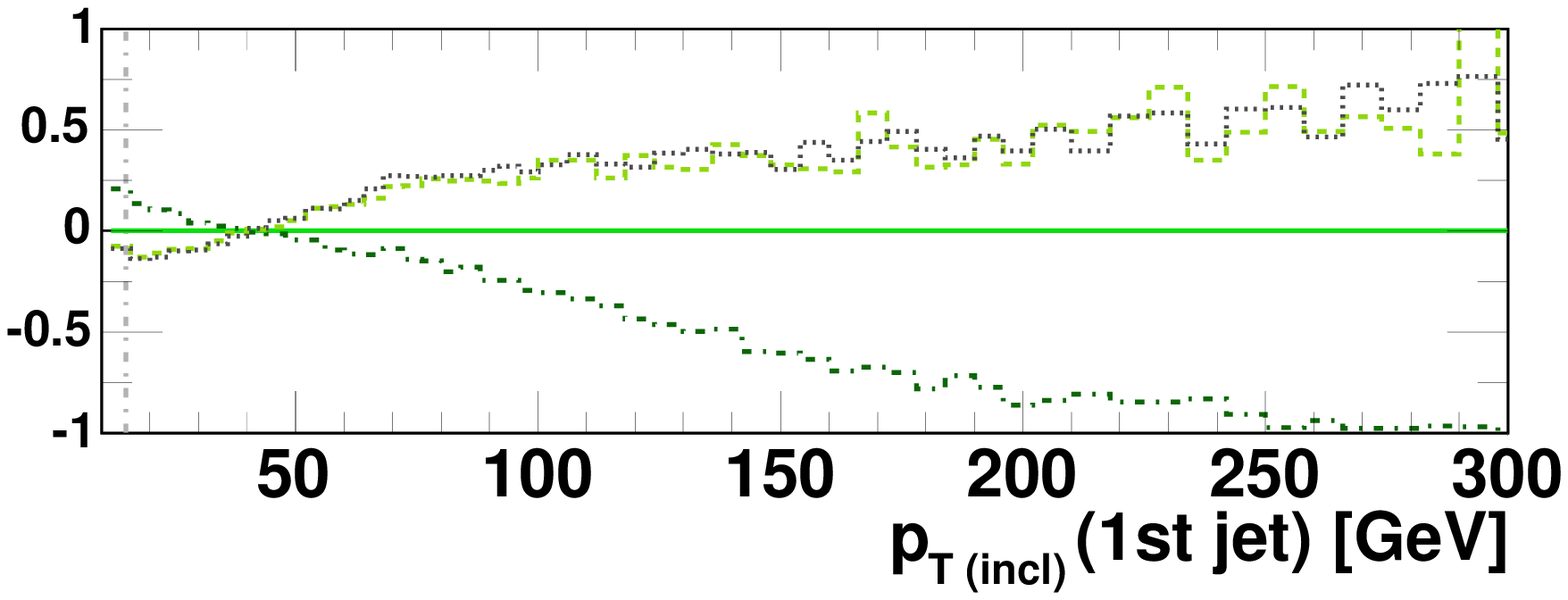}}
    \put(0,0){
      \includegraphics[width=70mm]{%
        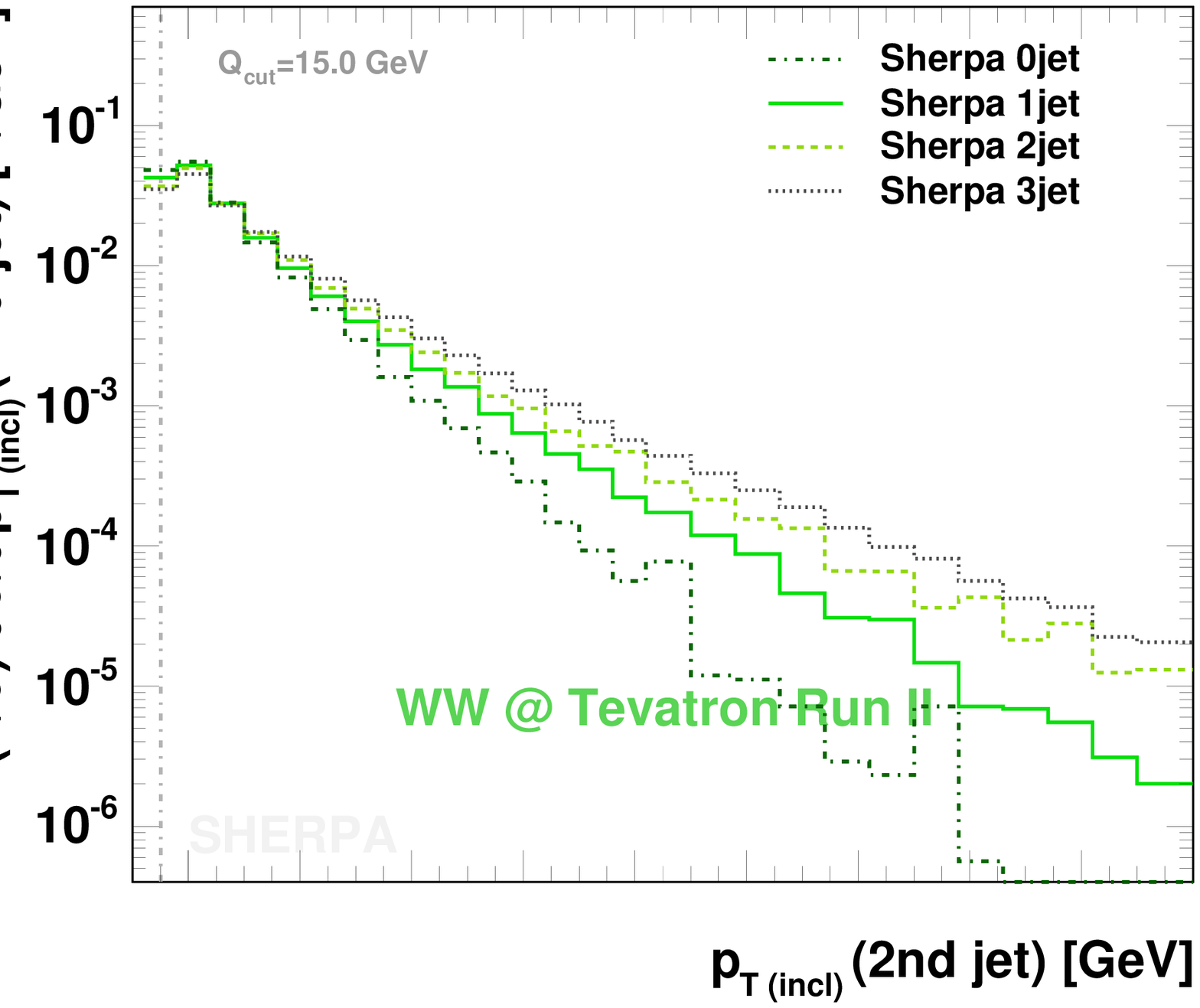}}
    \put(0,0){
      \includegraphics[width=70mm]{%
        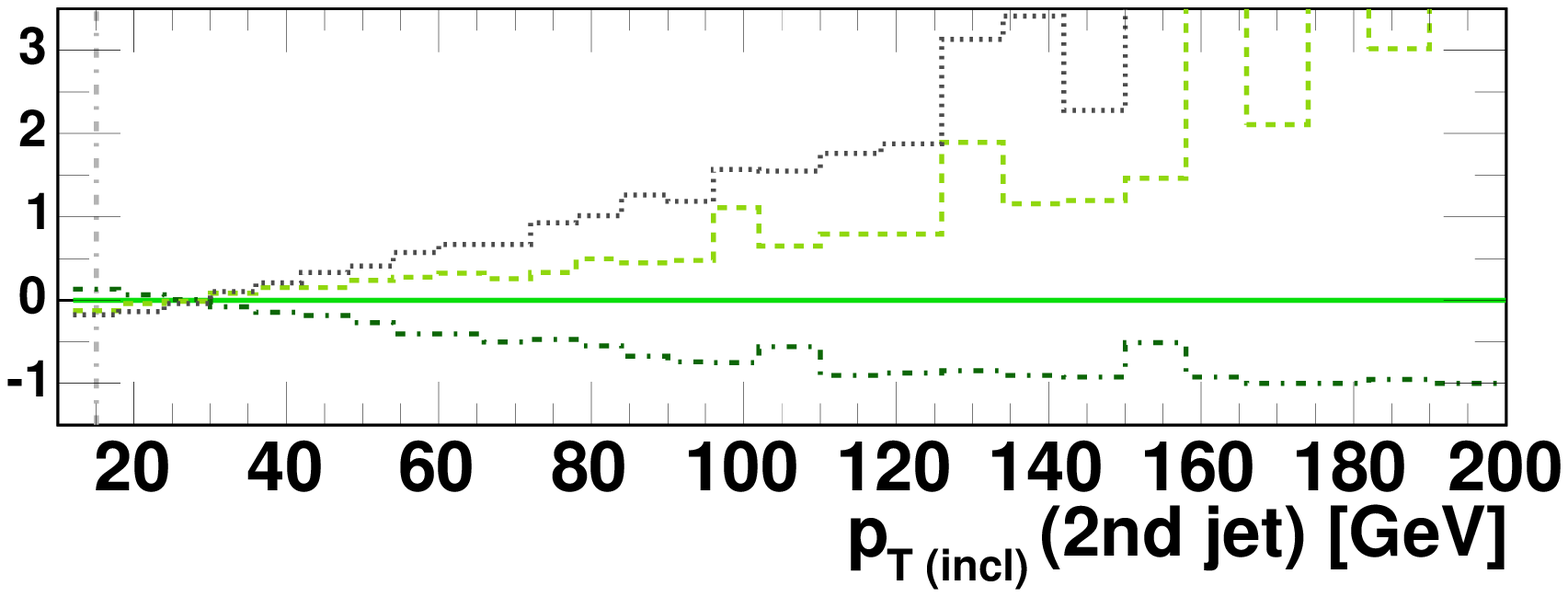}}
  \end{picture}
  \vspace{0mm}
  \caption{\she\ predictions of the inclusive $p_T$ of the
    associated jets considered in dependence on the variation of the
    maximal jet number. The spectra of the hardest and the second
    hardest jet are depicted in the upper and the lower panel,
    respectively. The jet resolution cut has been taken to be
    $15$ GeV. The green solid line shows the result of the $n_{\rm
    max}=1$ sample, the brighter dashed and the grey dotted one stand
    for the $n_{\rm max}=2$ and $n_{\rm max}=3$ sample, respectively;
    the darkgreen dashed-dotted curve depicts the pure shower
    performance. The lower part of the plot shows the normalized
    differences with respect to the $n_{\rm max}=1$ case. For the jet
    definition, the Run II $k_\perp$-algorithm with $R=0.7$ and
    $p^{\rm jet}_T>15$ GeV has been used. For more details, see Apps.\
    \ref{app_input} and \ref{app_cuts}.}
  \label{incjetptN_coychk}
  \vspace{0mm}
\end{figure}
\noindent
The approach of varying the maximal jet number $n_{\rm max}$ can be
exploited to further scrutinize the merging procedure. In all cases
considered here, $Q_{\rm cut}$ has been fixed to $Q_{\rm cut}=15$ GeV.
This maximizes the impact of higher order matrix elements. In spite of
this, for very inclusive observables, the rates differ very mildly,
the change is less than $2\%$.
\\
In Fig.\ \ref{ptWplusN_coychk}, once more the transverse momentum
distribution of the $W^+$ gauge boson is presented, illustrating that
the treatment of the highest multiplicity matrix elements (for more
details cf.\ \cite{Krauss:2004bs,Schaelicke:2005nv}) completely
compensates for the missing $2$jet matrix element in the $n_{\rm
max}=1$ case. The behaviour is almost unaltered when changing from the
$n_{\rm max}=1$ to the $n_{\rm max}=2$ prediction (cf.\ the right
panel). In contrast, $n_{\rm max}=0$ yields a considerably softer
distribution (cf.\ the left panel).
\\
Lepton $p_T$ spectra show similar characteristics like the $W^+$
distribution. However, there are a number of observables, which turned
out to be rather stable under the variation of $n_{\rm max}$, such as
the pseudo-rapidity spectra of the $W^+$ boson, the positron and muon
or correlations between the leptons, e.g.\ the $\Delta\phi$\/ or
$\Delta R$\/ distribution. In these cases, deviations turn out to be
smaller than $\pm5\%$ in total, \ie when considering the change
between the pure shower and the inclusive $3$jet production
performance of \she. Even the pseudo-rapidity spectra of the resolved
jets are rather unaffected.
\\
In contrast, three more observables are presented showing a sizeable
($<\pm30\%$) or even strong ($\approx\pm100\%$) dependence on the
variation of the maximal jet number, namely the $H_T$ distribution
depicted in Fig.\ \ref{htN_coychk} and the inclusive $p_{T}$ spectra
of associated jets exhibited in Fig.\ \ref{incjetptN_coychk}. The
upper and lower panel of Fig.\ \ref{incjetptN_coychk} shows the
spectra of the hardest and the second hardest jet, respectively. Owing
to the nature of these three observables to be sensitive on extra jet
emissions, predictions -- as expected -- become harder with the
increase of $n_{\rm max}$. However, a stabilization of the predictions
is clearly found with the inclusion of more higher order matrix
elements describing real QCD emissions.

\subsection*{Effects of renormalization and factorization scale variations}

\begin{figure}[t!]
  \vspace{0mm}
  \begin{picture}(188,204)
    \put(0,0){
      \includegraphics[width=70mm]{%
        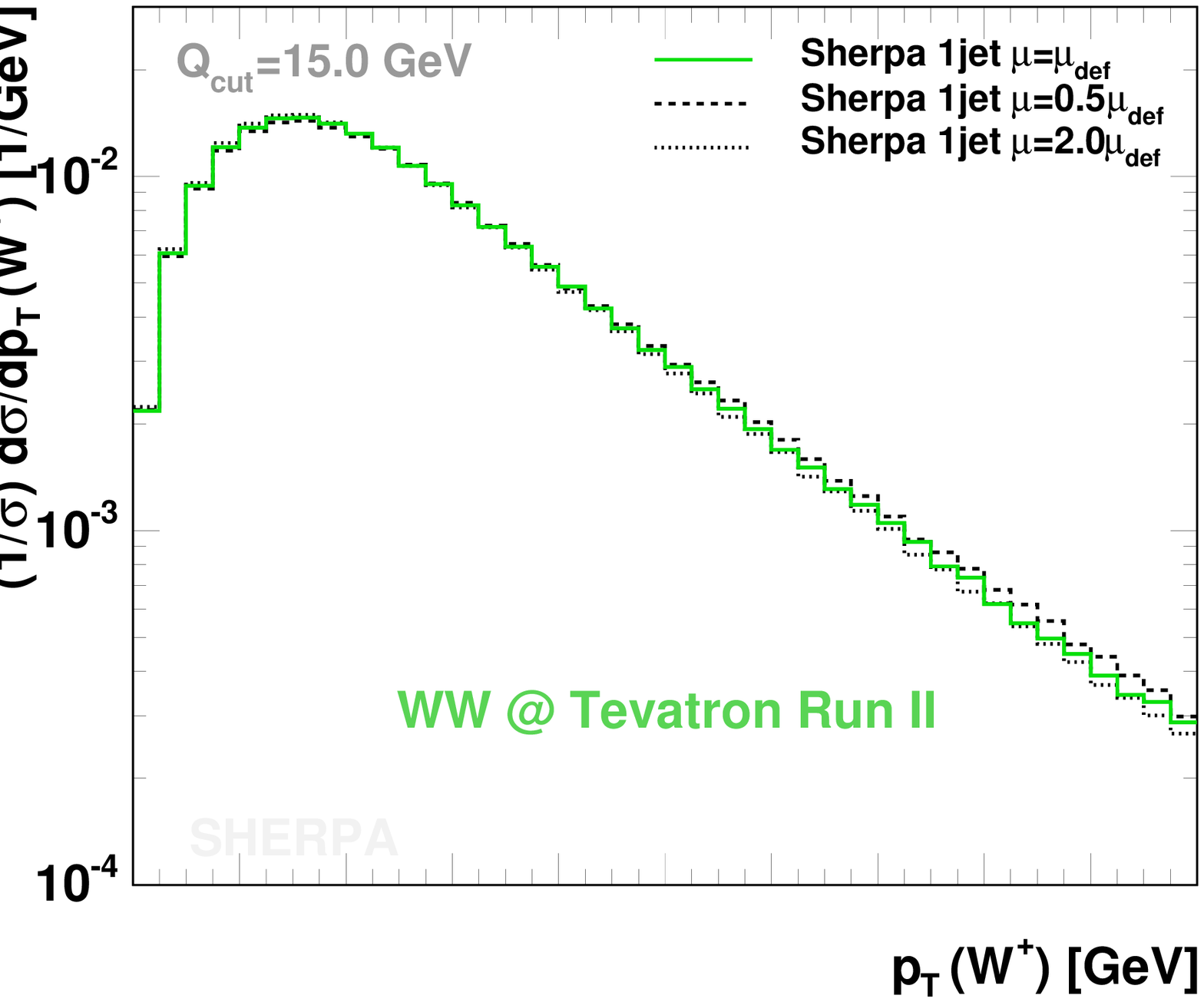}}
    \put(0,0){
      \includegraphics[width=70mm]{%
        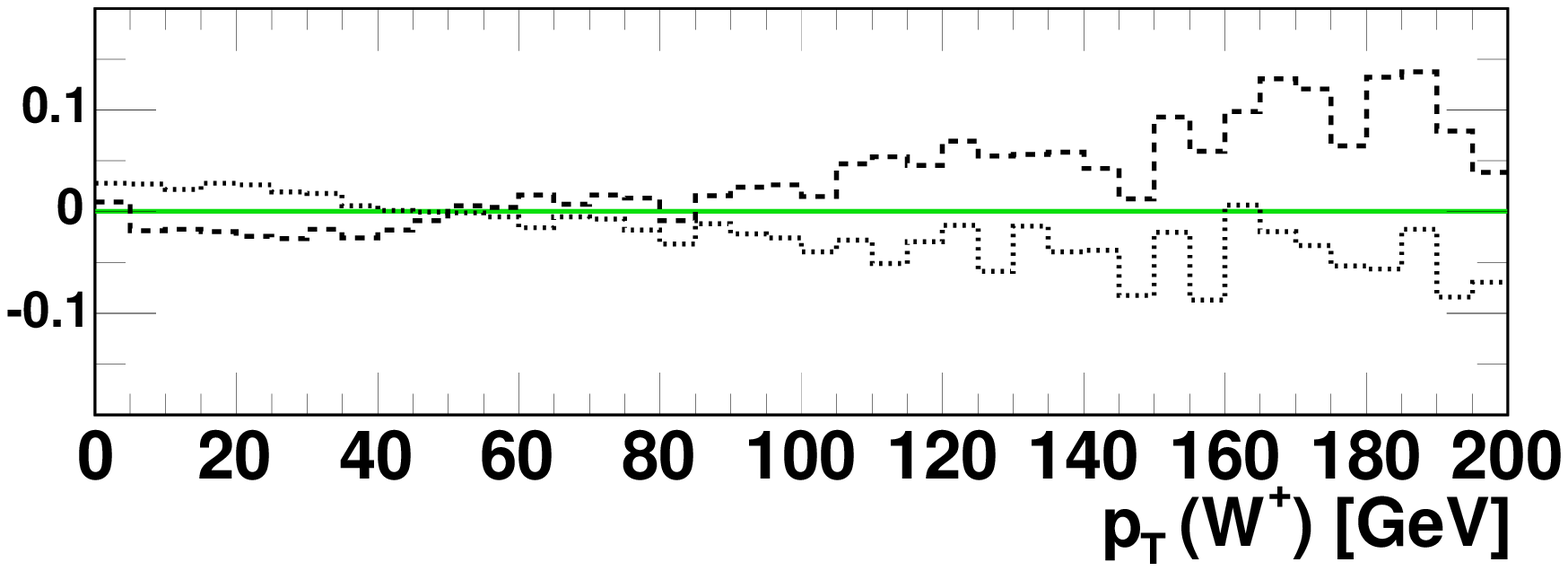}}
  \end{picture}
  \vspace{0mm}
  \caption{The $p_T$ distribution of the $W^+$ boson under scale
    variations. All predictions stem from \she\ with $n_{\rm max}=1$
    and $Q_{\rm cut}=15$ GeV. The green solid line shows the
    prediction under default scale choices for the merging procedure.
    For the black dashed and the black dotted curve, all scales for
    the coupling constants and PDFs have been multiplied by $0.5$ and
    $2.0$, respectively. The lower part of the plot presents the
    normalized differences with respect to the default choice. Input
    parameters and cuts are given in Apps.\ \ref{app_input} and
    \ref{app_cuts}.}
  \label{ptWplusSV_coychk}
  \vspace{0mm}
\end{figure}
\begin{figure}[t!]
  \vspace{4mm}
  \begin{picture}(188,204)
    \put(0,0){
      \includegraphics[width=70mm]{%
        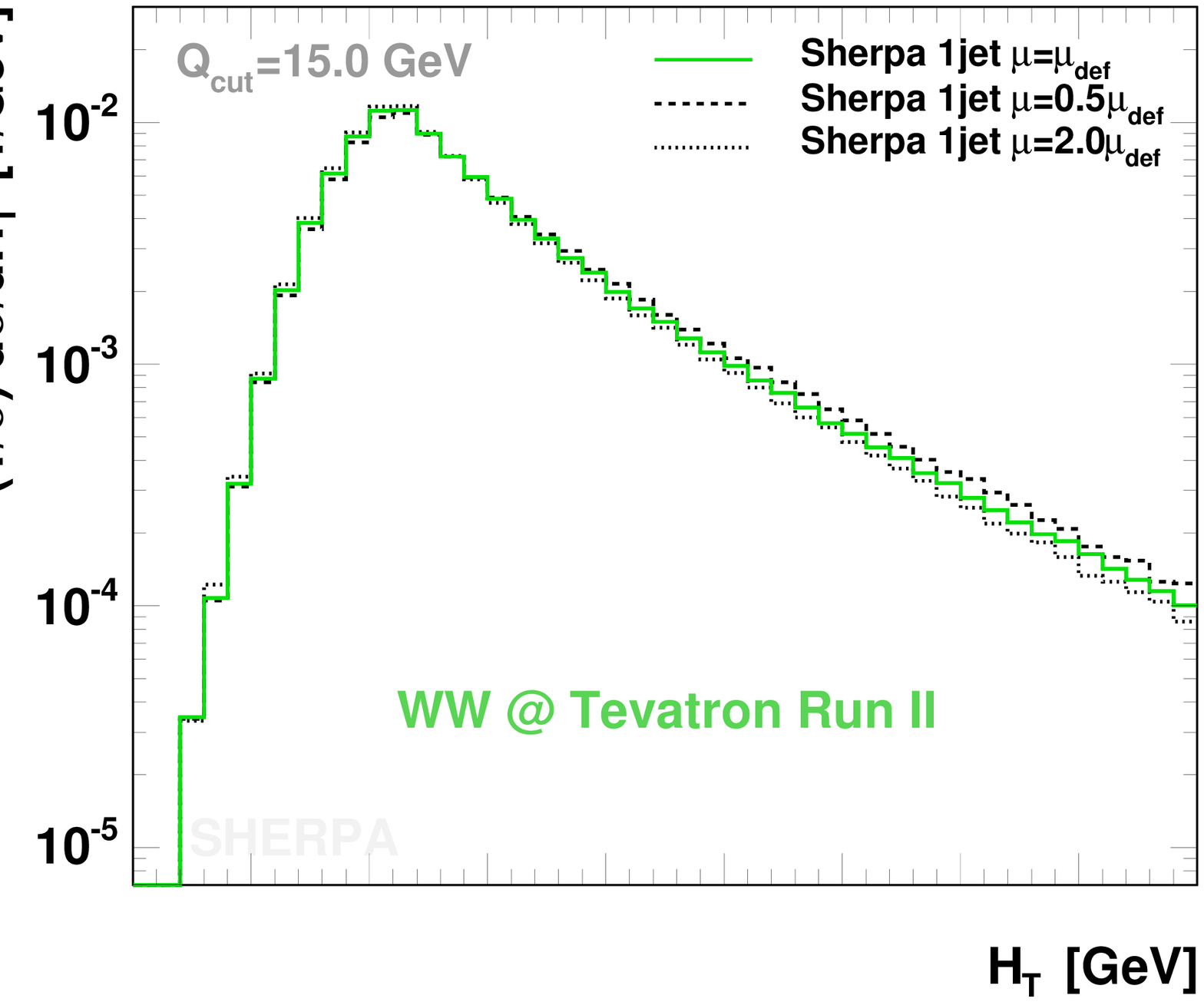}}
    \put(0,0){
      \includegraphics[width=70mm]{%
        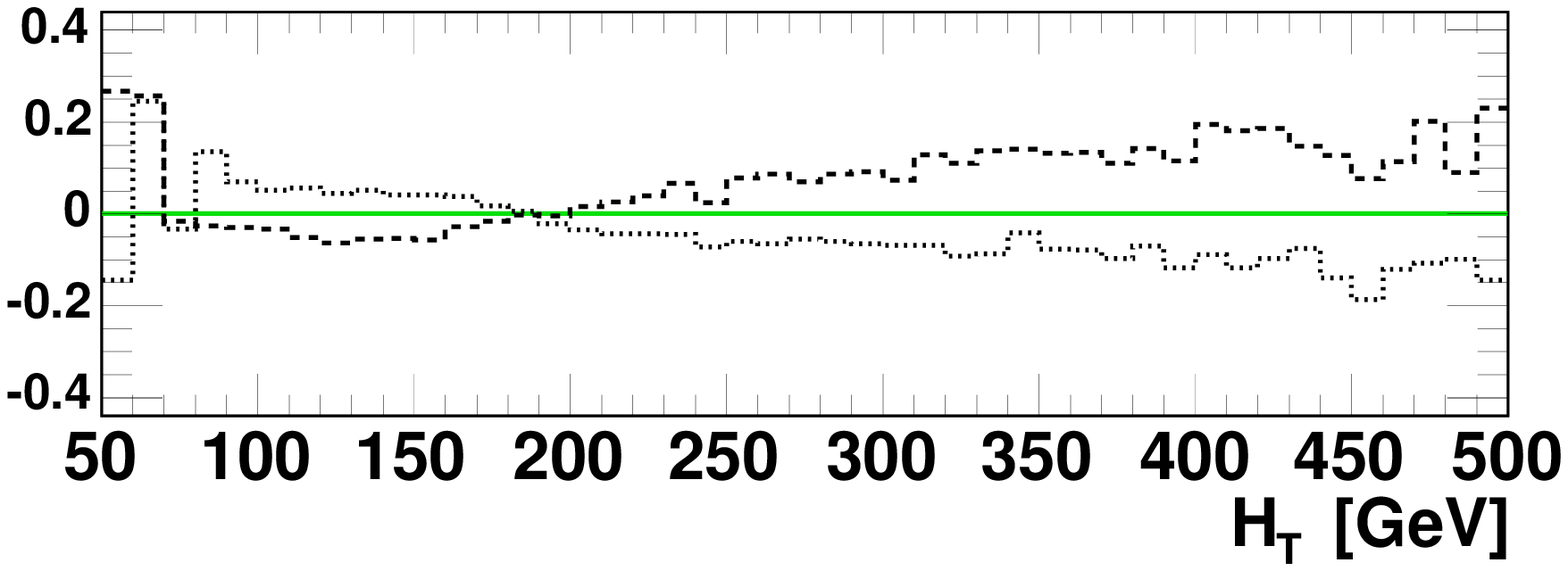}}
  \end{picture}
  \vspace{0mm}
  \caption{The $H_T$ distribution and its dependence on the variation
    of $\mu_R$ and $\mu_F$ in the merging prescription. Fixing $n_{\rm
    max}=1$ and $Q_{\rm cut}=15$ GeV, the green solid line shows the
    prediction under default scale choices. For the black dashed and
    the black dotted curve, all scales for the coupling constants and
    PDFs have been multiplied by $0.5$ and $2.0$, respectively. The
    lower part of the plot presents the normalized differences with
    respect to the default choice. Input parameters and analysis cuts
    are given in Apps.\ \ref{app_input} and \ref{app_cuts}.}
  \label{htSV_coychk}
  \vspace{0mm}
\end{figure}
\begin{figure}[t!]
  \vspace{0mm}
  \begin{picture}(188,204)
    \put(0,0){
      \includegraphics[width=70mm]{%
        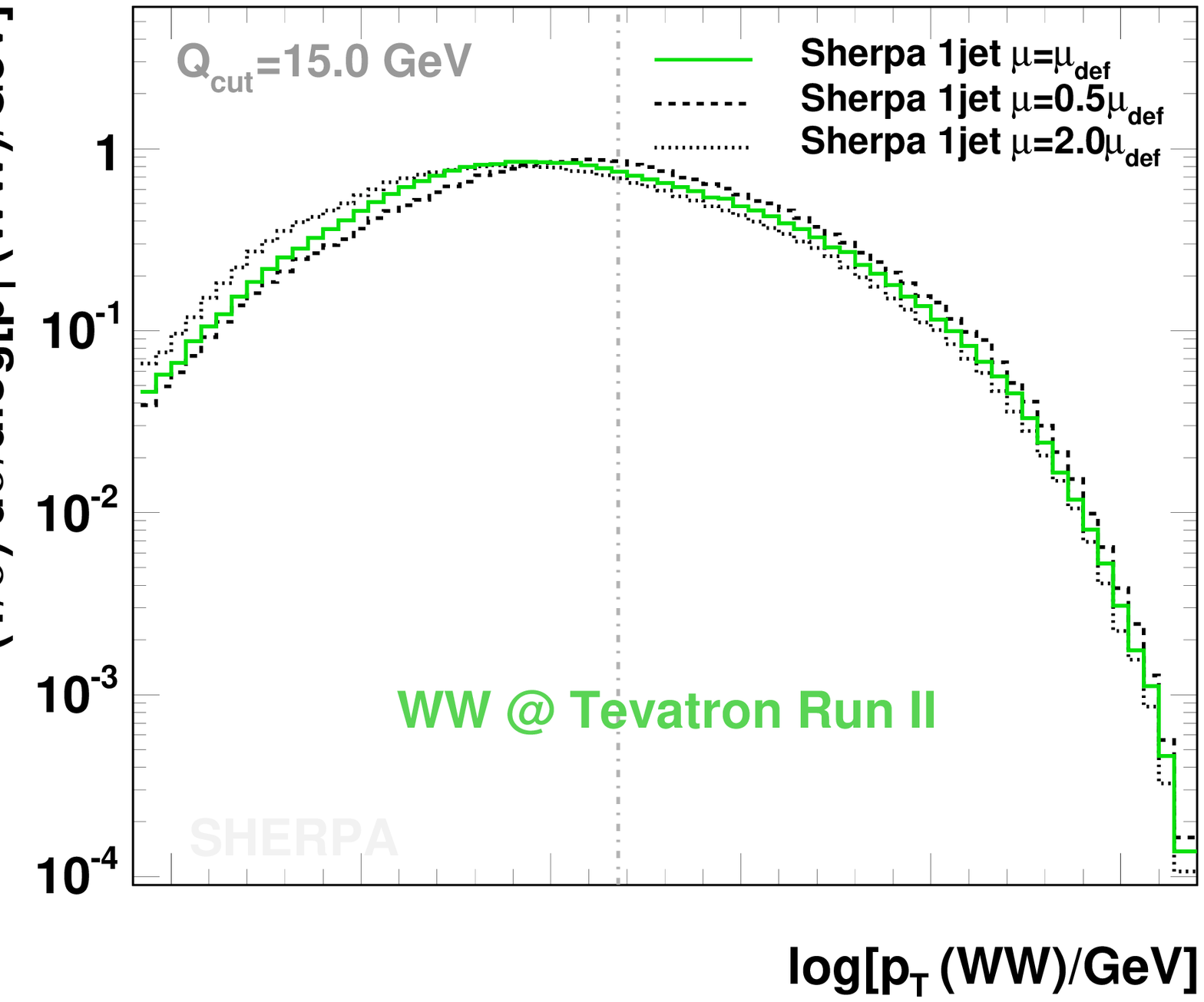}}
    \put(0,0){
      \includegraphics[width=70mm]{%
        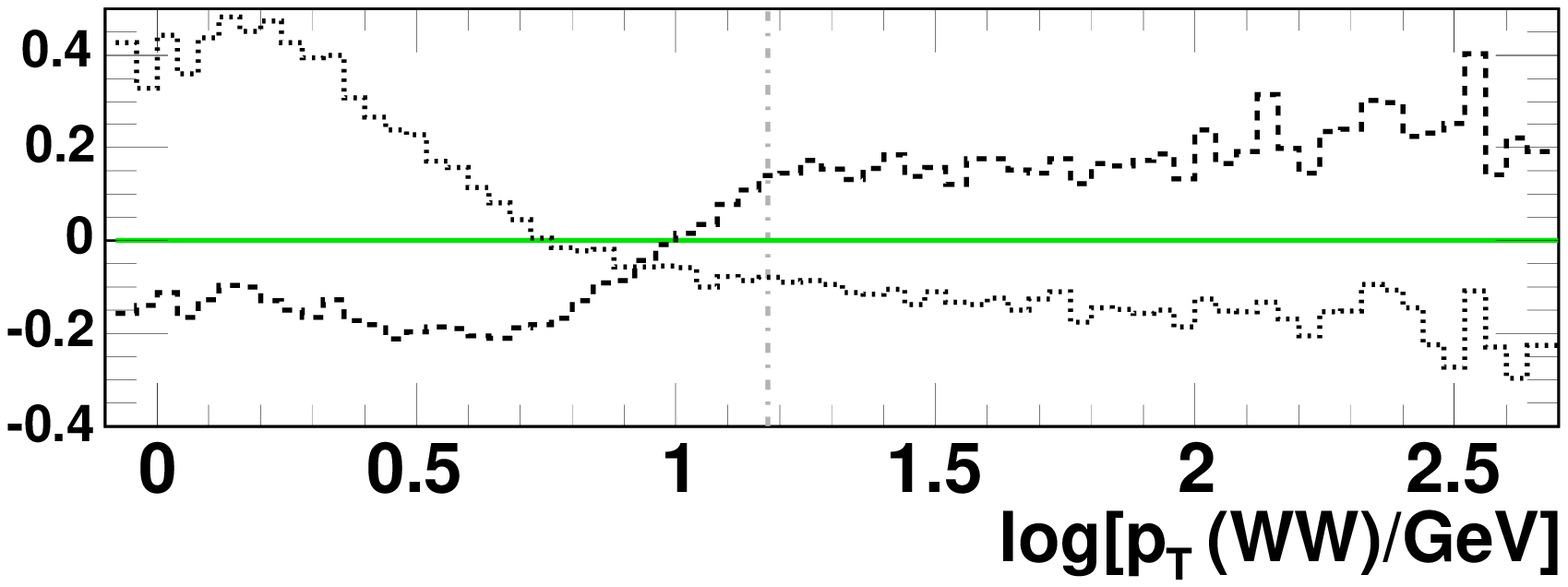}}
  \end{picture}
  \vspace{0mm}
  \caption{The $p_T$ distribution of the $W$\/ pair under variation of
    $\mu_R$ and $\mu_F$. Fixing $n_{\rm max}=1$ and $Q_{\rm cut}=15$
    GeV, the green solid line shows the prediction under default scale
    choices. The black dashed and the black dotted curve is generated
    when all scales used for the coupling constants and PDFs have been
    multiplied by $0.5$ and $2.0$, respectively. The lower part of the
    plot presents the normalized differences with respect to the
    default choice. Input parameters (including a primordial $k_\perp$
    smearing) and cuts are given in Apps.\ \ref{app_input} and
    \ref{app_cuts}.}
  \label{ptWWSV_coychk}
  \vspace{0mm}
\end{figure}
\noindent
In the following the impact of renormalization and factorization scale
variations is discussed. For the \she\ merging approach, this
variation (also cf.\ \cite{wz@LHC}) is performed by multiplying
all scales with a constant factor in all coupling constants and PDFs,
which are relevant for the matrix element evaluation, the Sudakov
weights and for the parton shower evolution.
\\
For this study, the \she\ samples are produced with $n_{\rm max}=1$
and $Q_{\rm cut}=15$ GeV. In all figures the green solid line
represents \she's default scale choices, whereas the black dashed and
the black dotted curve show the outcome for scale multiplications by
$0.5$ and $2.0$, respectively. The total rate as provided by the
merging algorithm is again remarkably stable, varying with respect to
the default only by $\pm4.2\%$, thereby increasing for smaller scales.
\\
The transverse momentum distribution of the $W^+$ boson is
investigated in Fig.\ \ref{ptWplusSV_coychk}. Scale variations
slightly distort the shape, shifting it towards harder $p_T$ for
smaller scales and vice versa. The effect is more pronounced in
the $H_T$ distribution, shown in Fig.\ \ref{htSV_coychk}, and in the
transverse momentum distribution of the diboson system, depicted in
Fig.\ \ref{ptWWSV_coychk}. However, the deviations maximally found
reach up to $\pm30\%$. In contrast to the findings stated so far, jet
transverse momentum spectra do not feature shape distortions under
scale variations.
\\
The pattern found from these investigations can be explained as
follows. The single matrix element contributions -- here the $0$jet
and $1$jet contribution -- have their own rate and shape dependencies
under scale variations. In their interplay these differences transfer
to changing the admixture of the single contributions. Hence, shape
modifications can appear as soon as different phase space regions are
dominated by a single contribution. This also explains the behaviour
found for jet $p_T$s. In the case studied here, they are solely
described by the $1$jet matrix element with the parton shower
attached, thus, their different rates cancel out due to normalization
and their shapes are not affected.
\\
Taken together, the dependencies found here, together with the ones on
$Q_{\rm cut}$ and $n_{\rm max}$, yield an estimate for the uncertainty
related to the \she\ predictions.

\section{SHERPA comparison with MCFM\label{sec_MCFM}}

\begin{figure}[t!]
  \vspace{-8.5mm}
  \bc\includegraphics[width=70mm,angle=-90.0]{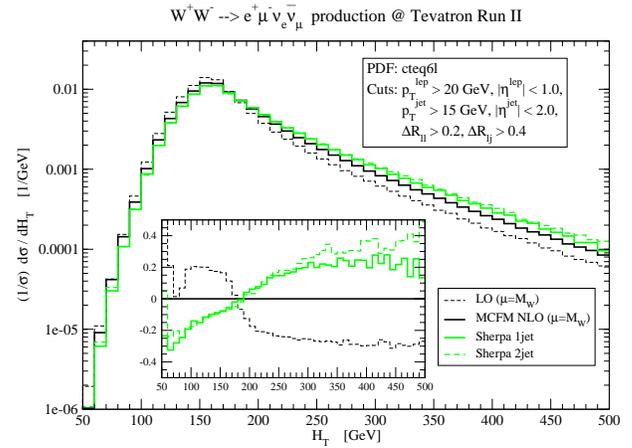}\ec
  \vspace{-7mm}
  \caption{Normalized $H_T$ distribution. \she\ results are
    shown for $n_{\rm max}=1$ (green solid line) and $n_{\rm max}=2$
    (green dashed line) and compared to the QCD NLO result of {\tt
    MCFM} (black solid line). The LO result with the same scale choice
    is depicted as a thin black dashed line. A difference plot with
    the {\tt MCFM} NLO prediction as reference is given within the
    figure.}
  \label{ht_mcfmi}
  \vspace{0mm}
\end{figure}

\noindent
In this section, the focus shifts from internal sanity checks to
comparisons with a full NLO calculation. For this, the {\tt MCFM}
program \cite{Campbell:1999ah} has been used. In both, {\tt MCFM} and
\she\, the CKM matrix has been taken diagonal, and no $b$\/ quarks are
considered in the partonic initial state of the hard process. If not
stated otherwise, in {\tt MCFM} the renormalization and factorization
scale have been chosen as $\mu_R=\mu_F=M_W$, according to the choice
made in \cite{Campbell:1999ah}. For more details on the input
parameters and setups, see Apps.\ \ref{app_input} and \ref{app_sets}.
In the following the results of {\tt MCFM} are confronted with those
of \she\ (using $Q_{\rm cut}=15$ GeV) obtained at the parton shower
level. Furthermore, for this analysis, realistic experimental cuts
(cf.\ App.\ \ref{app_cuts}) have been applied and all distributions
have been normalized to one.
\\
First the $H_T$ distribution, depicted in Fig.\ \ref{ht_mcfmi}, is
considered. Clearly, higher order corrections affect the $H_T$ shape.
This is due to two reasons. First of all, the additional QCD radiation
may manifest itself as jet(s), which thus contribute to $H_T$.
Otherwise the additional partons still form a system against which the
$W$\/ pair may recoil. Quantitatively, the inclusion of NLO results in
a shift of the $H_T$ distribution at harder values by up to $20\%$; in
\she\ this trend is amplified by roughly the same amount. The
differences between {\tt MCFM} and \she, however, are due to the
different scale choices in both codes. In {\tt MCFM} all scales have
been fixed to $\mu=M_W$, whereas, forced by the merging procedure, in
\she\ the scales are set dynamically. In view of the scale variation
results discussed in the previous section for $H_T$ (cf.\ Fig.\
\ref{htSV_coychk}) deviations of this magnitude owing to different
scale choices are possible.
\begin{figure}[b!]
  \vspace{-2mm}
  \bc\includegraphics[width=70mm,angle=-90.0]{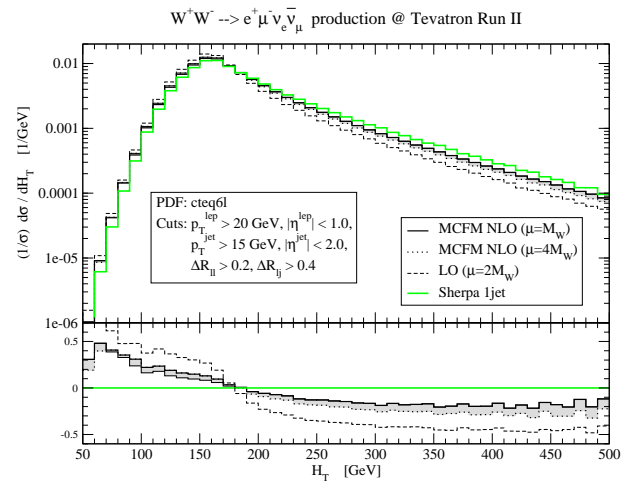}\ec
  \vspace{-6mm}
  \caption{Normalized $H_T$ distribution. Here both, the
    renormalization and factorization scale of the NLO calculation
    have been varied in the range $\mu_R=\mu_F=M_W\ldots4\,M_W$,
    indicated by the shaded area. These {\tt MCFM} results are
    compared with the leading order result at $\mu_R=\mu_F=2\,M_W$
    (thin black dashed line) and with the result of \she\ where
    $n_{\rm max}=1$ (green solid line). The lower part of the plot
    shows the normalized differences with respect to the \she\
    result.}
  \label{ht_mcfms}
  \vspace{0mm}
\end{figure}
\begin{figure}[t!]
  \vspace{-8.5mm}
  \bc\includegraphics[width=70mm,angle=-90.0]{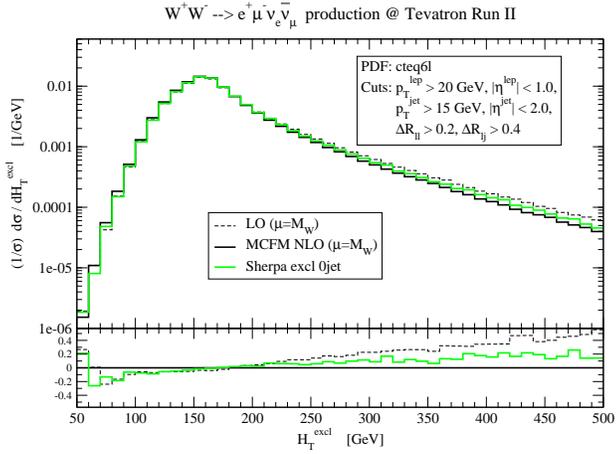}\ec
  \vspace{-7mm}
  \caption{Normalized $H_T$ distribution in exclusive $p\bar p\to
    e^+\mu^-\nu_e\bar\nu_\mu$ production. The \she\ result (green
    solid line) is obtained with $n_{\rm max}=0$ and a parton shower
    constrained not to produce any extra jets. This result is compared
    with the {\tt MCFM} result at NLO in $\alpha_s$ (black solid line)
    and with the LO result (thin black dashed line). The latter two
    are taken for the default scale choices. The lower part of the
    plot shows the normalized differences with respect to the NLO
    result obtained from {\tt MCFM}.}
  \label{ht_mcfmx}
  \vspace{0mm}
\end{figure}
\begin{figure}[t!]
  \vspace{-2mm}
  \bc\includegraphics[width=70mm,angle=-90.0]{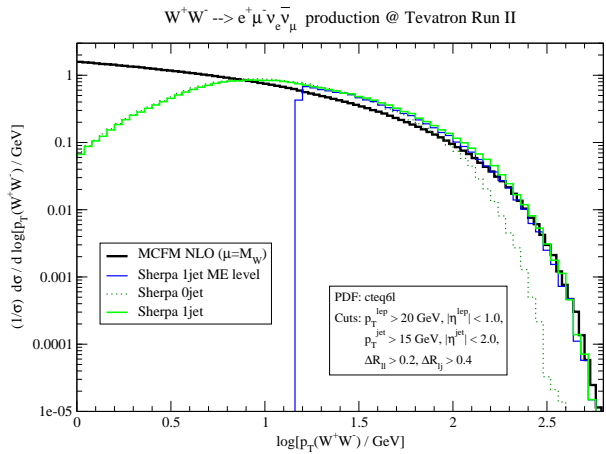}\ec
  \vspace{-7mm}
  \caption{Normalized $p_T$ distribution of the $W$\/ pair. The {\tt
      MCFM} result at $\mu_R=\mu_F=M_W$ (black line) is contrasted
      with the predictions made by \she\, both at the matrix element
      level (blue line) and at the parton shower level with $n_{\rm
      max}=0$ (darkgreen dotted line) and $n_{\rm max}=1$ (green solid
      line). A primordial $k_\perp$ smearing has been used to obtain
      the \she\ shower results.}
  \label{ptWW_mcfm}
  \vspace{0mm}
\end{figure}
\begin{figure}[t!]
  \vspace{-8.5mm}
  \bc\includegraphics[width=70mm,angle=-90.0]{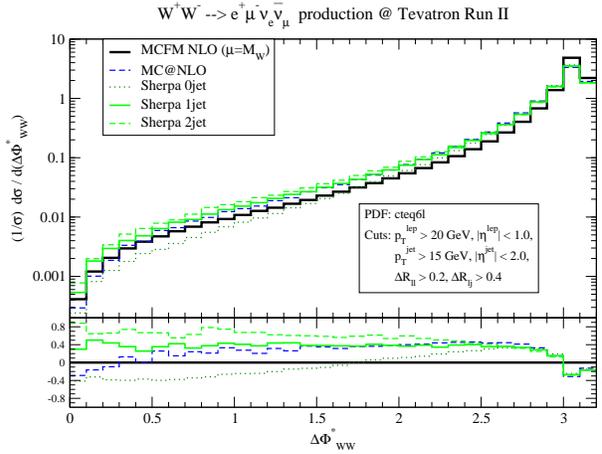}\ec
  \vspace{-7mm}
  \caption{Normalized $\Delta\Phi^\star_{WW}$ distribution of the
    $W$\/ boson system. The {\tt MCFM} result (black line) is
    contrasted with results from \she\ at the parton shower level with
    $n_{\rm max}=0$ (darkgreen dotted line), $n_{\rm max}=1$ (green
    solid line) and $n_{\rm max}=2$ (green dashed line). Again a
    primordial $k_\perp$ smearing has been used. Additionally, the
    blue dashed curve represents a prediction obtained with {\tt
    MC@NLO}. The lower part of the plot shows the normalized
    differences with respect to the result of {\tt MCFM}.}
  \label{DPhiWW_mcfm}
  \vspace{0mm}
\end{figure}
\begin{figure}[t!]
  \vspace{-7mm}
  \bc\includegraphics[width=70mm,angle=-90.0]{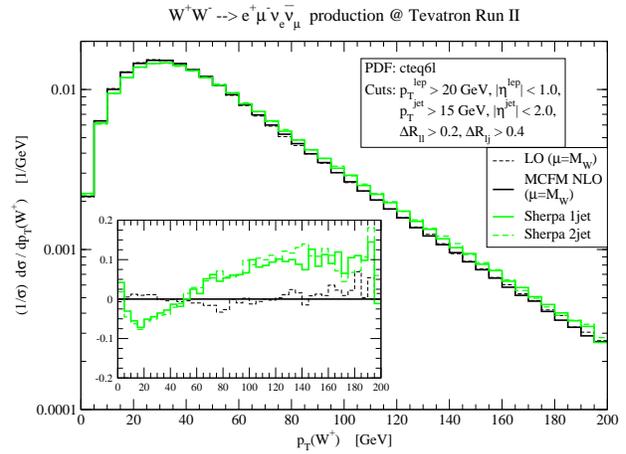}\ec
  \vspace{-7mm}
  \caption{Normalized transverse momentum distribution of the $W^+$
    boson. The results of \she\ for $n_{\rm max}=1$ (green solid line)
    and for $n_{\rm max}=2$ (green dashed line) are compared with the
    QCD NLO result obtained by {\tt MCFM} (black solid line) and with
    the LO result (thin black dashed line) for the default scale
    choices, \ie $\mu_R=\mu_F=M_W$. Within the plot the normalized
    differences with respect to the NLO result of {\tt MCFM} are
    given.}
  \label{ptW_mcfmi}
  \vspace{0mm}
\end{figure}
\begin{figure}[t!]
  \vspace{-8.5mm}
  \bc\includegraphics[width=70mm,angle=-90.0]{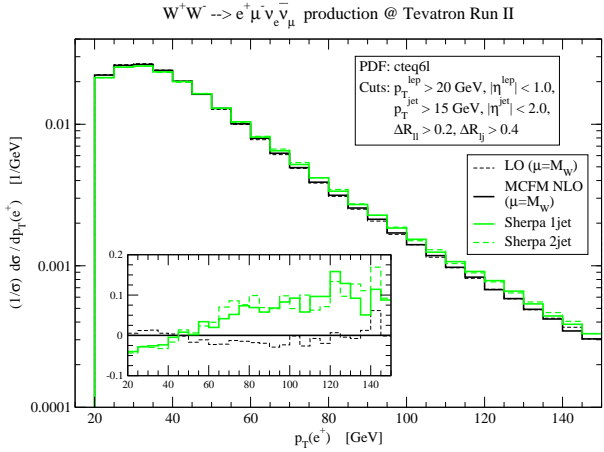}\ec
  \vspace{-7mm}
  \caption{Normalized transverse momentum distribution of the $e^+$
    produced in the decay of the $W^+$. The results of \she\ for
    $n_{\rm max}=1$ (green solid line) and for $n_{\rm max}=2$ (green
    dashed line) are confronted with the QCD NLO result obtained by
    {\tt MCFM} (black solid line) and with the LO result (thin black
    dashed line). For the latter two, the scales are again fixed
    according to the default choices, \ie $\mu_R=\mu_F=M_W$. Within
    the plot the normalized differences with respect to the NLO result
    of {\tt MCFM} are shown.}
  \label{pte_mcfmi}
  \vspace{0mm}
\end{figure}
\begin{figure}[t!]
  \vspace{-6mm}
  \bc\includegraphics[width=70mm,angle=-90.0]{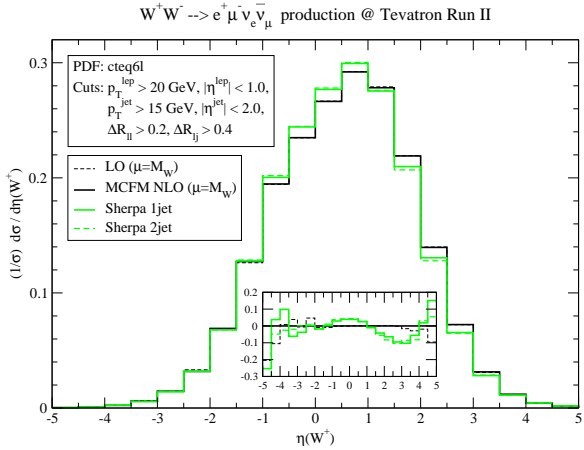}\ec
  \vspace{-7mm}
  \caption{Normalized $\eta$\/ distribution of the $W^+$ boson. The
    \she\ results for $n_{\rm max}=1$ (green solid line) and $n_{\rm
    max}=2$ (green dashed line) are confronted with those of {\tt
    MCFM} (black solid line) and with the LO result (thin black dashed
    line). Again, in the latter two the scales are chosen as
    $\mu_R=\mu_F=M_W$. The normalized differences with respect to the
    NLO result of {\tt MCFM} are also shown.}
  \label{etaW_mcfmi}
  \vspace{0mm}
\end{figure}
\begin{figure}[t!]
  \vspace{-8.5mm}
  \bc\includegraphics[width=70mm,angle=-90.0]{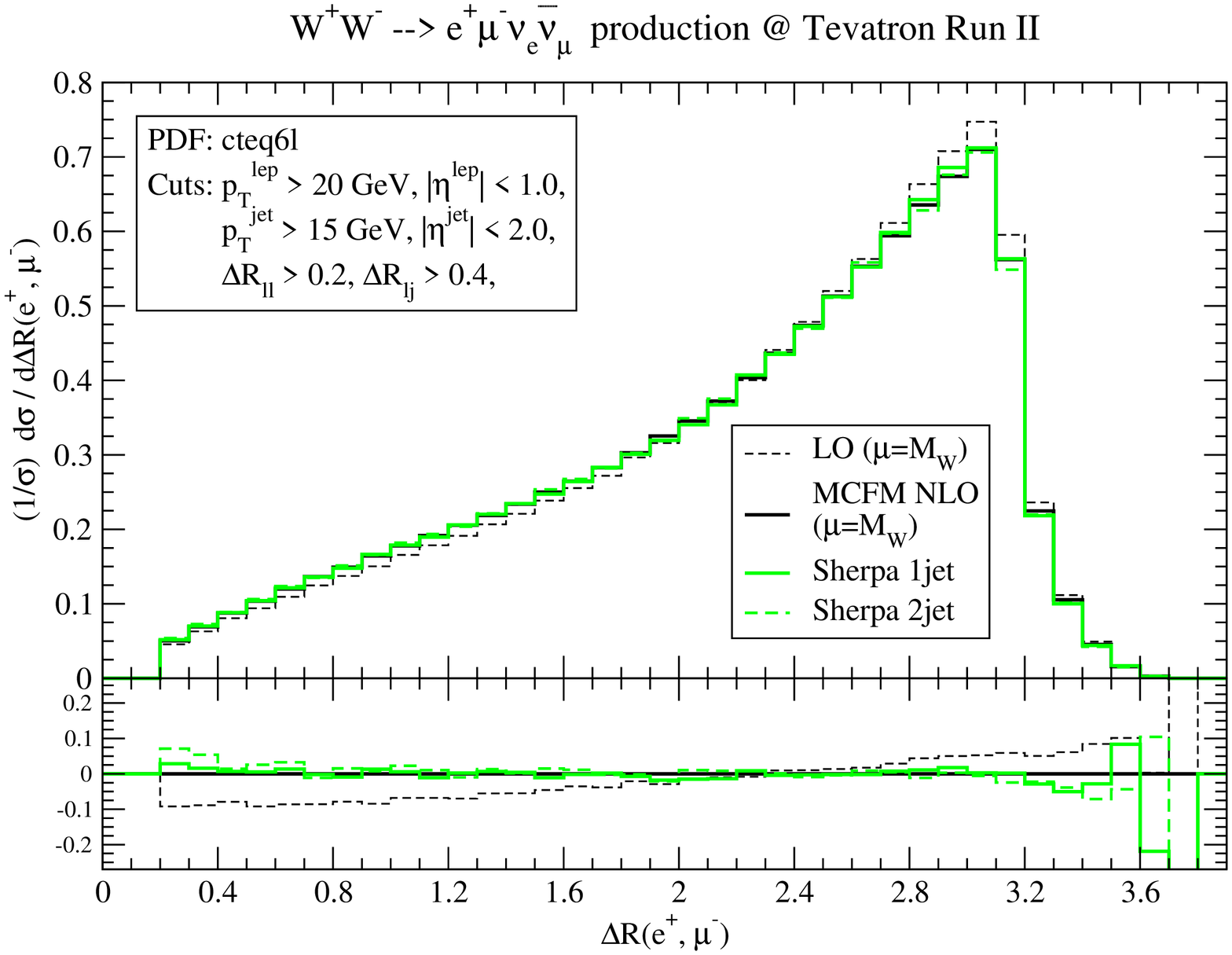}\ec
  \vspace{-7mm}
  \caption{Normalized $\Delta R$\/ distribution between the two
    charged leptons, the positron and the muon, emerging from the
    $W$\/ decays. \she\ results for $n_{\rm max}=1$ (green solid line)
    and $n_{\rm max}=2$ (green dashed line) are compared to those
    predicted by {\tt MCFM} (black solid line). The LO result with the
    same scale choice, is shown as a black dashed line. The lower part
    of the plot shows the normalized differences with respect to the
    NLO result of {\tt MCFM}.}
  \label{remu_mcfmi}
  \vspace{0mm}
\end{figure}
\\
The impact of scale variations on the shape of the same observable is
quantified in Fig.\ \ref{ht_mcfms}. This time, however, the \she\
result with $n_{\rm max}=1$ is compared to NLO results obtained from
{\tt MCFM} with scale choices in the range
$\mu_R=\mu_F=M_W\ldots4\,M_W$ and with a LO result taken at
$\mu_R=\mu_F=2M_W$. Obviously, the smaller choice of scale results in
the {\tt MCFM} outcome to be closer to the one of \she. As expected,
in comparison to the scale variation results found for \she, the shape
uncertainties of the full NLO prediction due to varying the scales are
smaller.
\\
In Fig.\ \ref{ht_mcfmx}, $H_T$ is depicted again, this time for the
case of exclusive $p\bar p\to e^+\mu^-\nu_e\bar\nu_\mu$ production.
There, the real part of the NLO correction in {\tt MCFM} is
constrained such that it does not produce an extra jet (for jet
definition, see App.\ \ref{app_cuts}). In \she\ the $0$jet matrix
element with the parton shower attached is considered exclusively, \ie
the parton shower is now forced not to produce any jet at all. In this
case, the higher order corrections lead to a softer $H_T$ distribution
compared to the leading order prediction, and the results of {\tt
MCFM} and \she\ show the same deviations as before (cf.\ Fig.\
\ref{ht_mcfmi}).
\\
The effect of QCD radiation is best observed in the $p_T$ distribution
of the $W$\/ pair, depicted in Fig.\ \ref{ptWW_mcfm}. Clearly, without
any radiation, the $p_T$ of the $W$\/ pair is exactly zero, and only
the emission of partons leads to a recoil of the diboson system. In
the NLO calculation of {\tt MCFM}, however, the spectrum is therefore
described at lowest order, in this particular case taken at
$\mu_R=\mu_F=M_W$. In contrast, in the \she\ matrix element result,
subjected to the explicit jet cut, Sudakov form factors and $\alpha_s$
reweighting are applied with a variable scale choice, explaining
the differences between the two matrix-element type results in this
figure. Contrasting this with the parton shower approach, it is clear
that parton emission through the shower alone is not sufficient to
generate sizeable $p_T$ of the $W$\/ pair in the hard region. For
this, the corresponding matrix element has to be employed, leading to
a very good agreement with the {\tt MCFM} outcome in the high-$p_T$
tail of the distribution. In the soft regime the result of the bare
{\tt MCFM} matrix element is unphysical. Due to the cascade emission
of soft and collinear partons, \she\ accounts for resummation effects,
which clearly yield the depopulation of the softest-$p_T$ region.
\\
Another way to look at the effects of QCD radiation is to consider the
relative angle between the two $W$\/ bosons\footnote{%
  The angle is measured in the frame, where the $W^+W^-$ system rests
  at the beam axis, \ie the diboson system is corrected on its initial
  $\hat{\rm z}$\/ boost.},
see Fig.\ \ref{DPhiWW_mcfm}. Of course, when they decay into leptons
plus neutrinos this is not an experimental observable, on the
generator level, however, it is very nice to visualize the effect of
QCD radiation in this way. Without any QCD radiation, the two $W$s
would be oriented back-to-back, at $\Delta\Phi^\star_{WW}=\pi$.
Including QCD radiation, this washes out, as depicted in the figure.
Again, resummation effects alter the result of the matrix element
alone by decreasing the amount of softest radiation, this time
corresponding to the back-to-back region around
$\Delta\Phi^\star\approx\pi$. The effect of high-$p_T$ radiation can
be clearly seen for small $\Delta\Phi^\star$ by comparing the
different $n_{\rm max}$ predictions of \she. The larger $n_{\rm max}$
is chosen, the harder the prediction for small $\Delta\Phi^\star$. On
the other hand to better value the influence of the parton shower a
prediction made by {\tt MC@NLO} (see App.\ \ref{app_sets}) has been
included. For a wide region of $\Delta\Phi^\star$, it well agrees with
the \she\ result for $n_{\rm max}=1$.
\\
Figs.\ \ref{ptW_mcfmi} and \ref{pte_mcfmi} exhibit the transverse
momentum distributions of the $W^+$ and of the $e^+$ produced in its
decay, respectively. Only mild deviations less than $10\%$ between
{\tt MCFM} and \she\ are found, which again can be traced back to
different scale choices in both approaches. These differences recur as
and, therefore, explain part of the deviations found in the $H_T$
spectrum, cf.\ Fig.\ \ref{ht_mcfmi}. As expected, the inclusion of the
$2$jet contribution in \she\ gives no further alterations of the
$n_{\rm max}=1$ result. Of course, the different radiation patterns
also have some minor effects on the $\eta$\/ distribution of the $W^+$
depicted in Fig.\ \ref{etaW_mcfmi}. In the $\Delta R_{e\mu}$
distribution presented in Fig.\ \ref{remu_mcfmi}, the NLO result of
{\tt MCFM} and the parton shower level results of \she\ are in nearly
perfect agreement with each other. Higher order effects tend to change
the shape of the LO prediction with respect to the NLO one by roughly
$10\%$. The interesting observation here is that this change is
seemingly not related to the transverse hardness of a jet system
against which the $W$\/ pair recoils. This gives rise to the
assumption that the change with respect to the LO result is due to
some altered spin structure in the $2\to5$ matrix element.

\section{Comparison with other event generators\label{sec_MC}}

\noindent
In this section a comparison of \she\ with other hadron level event
generators, in particular {\tt PYTHIA} and {\tt MC@NLO} will be
discussed. Details on how their respective samples have been produced
can be found in the Apps.\ \ref{app_input} and \ref{app_sets}. The
\she\ samples have been generated with $n_{\rm max}=1$ and $Q_{\rm
cut}=15$ GeV. The comparison is again on inclusive distributions --
normalized to one -- under the influence of realistic experimental
cuts, for details see App.\ \ref{app_cuts}.

\subsection*{Comparison of the QCD activity}

\begin{figure}[b!]
  \vspace{0mm}
  \begin{picture}(188,204)
    \put(0,0){
      \includegraphics[width=70mm]{%
        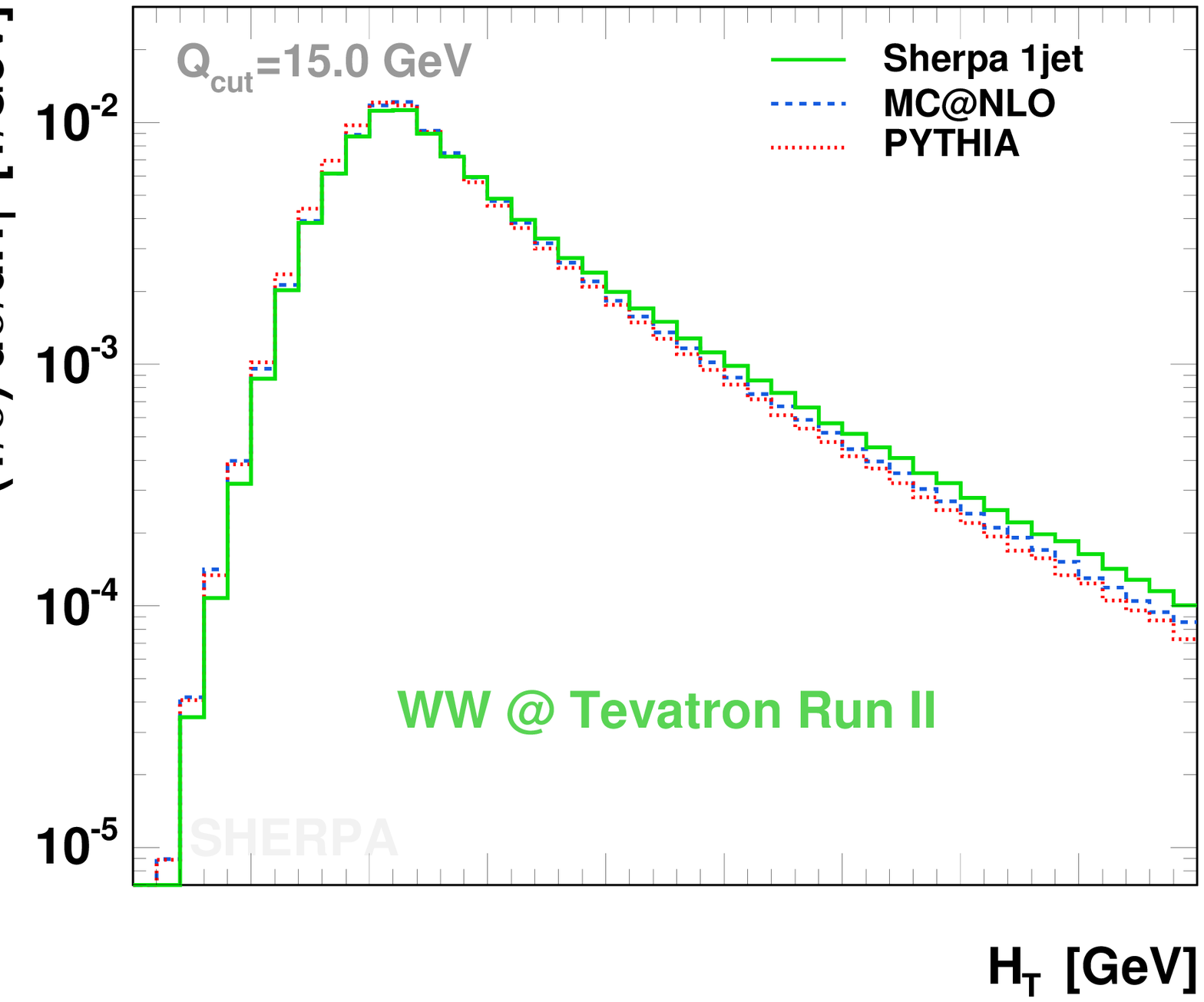}}
    \put(0,0){
      \includegraphics[width=70mm]{%
        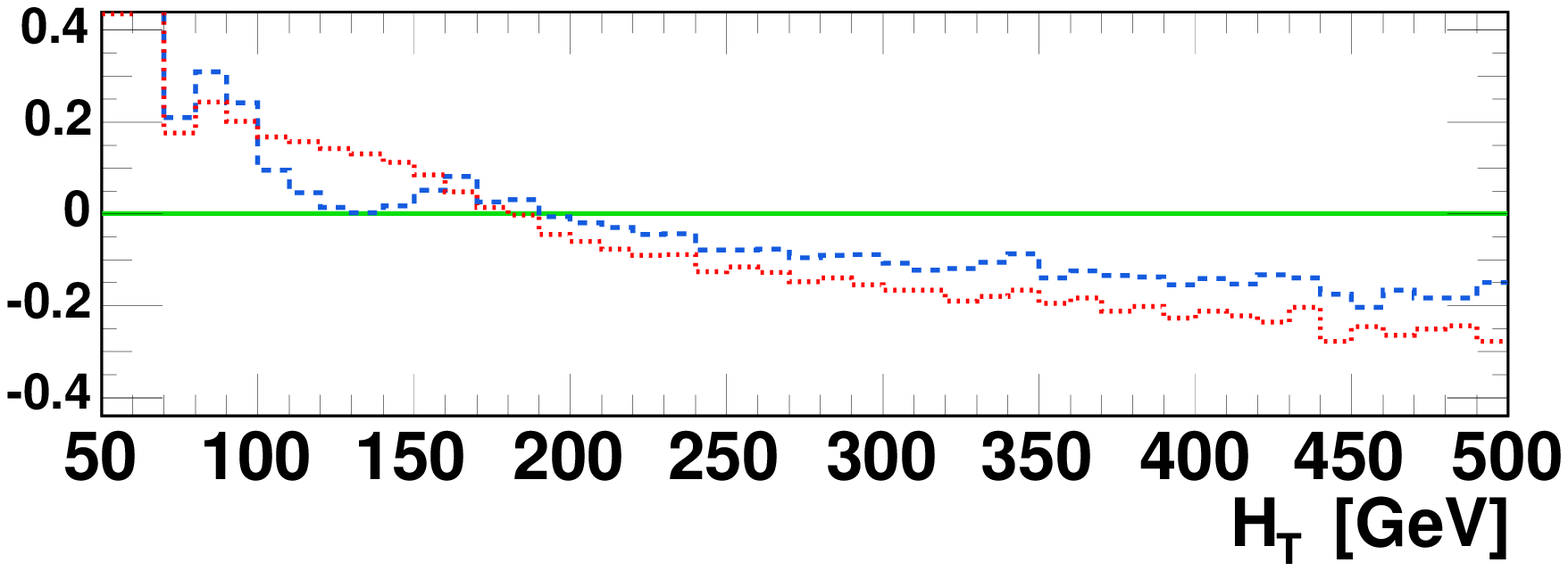}}
  \end{picture}
  \vspace{0mm}
  \caption{Normalized $H_T$ distribution obtained from {\tt PYTHIA}
    (red dotted line), {\tt MC@NLO} (blue dashed line) and \she\
    (green solid line). For the generation of the \she\ sample,
    $n_{\rm max}=1$ and $Q_{\rm cut}=15$ GeV have been chosen. The
    lower part of the plot exhibits the normalized differences with
    respect to the \she\ prediction. Input parameters and the employed
    cuts are specified in the Apps.\ \ref{app_input} and
    \ref{app_cuts}.}
  \label{ht_mocas}
  \vspace{0mm}
\end{figure}
\begin{figure}[b!]
  \vspace{0mm}
  \begin{picture}(188,204)
    \put(0,0){
      \includegraphics[width=70mm]{%
        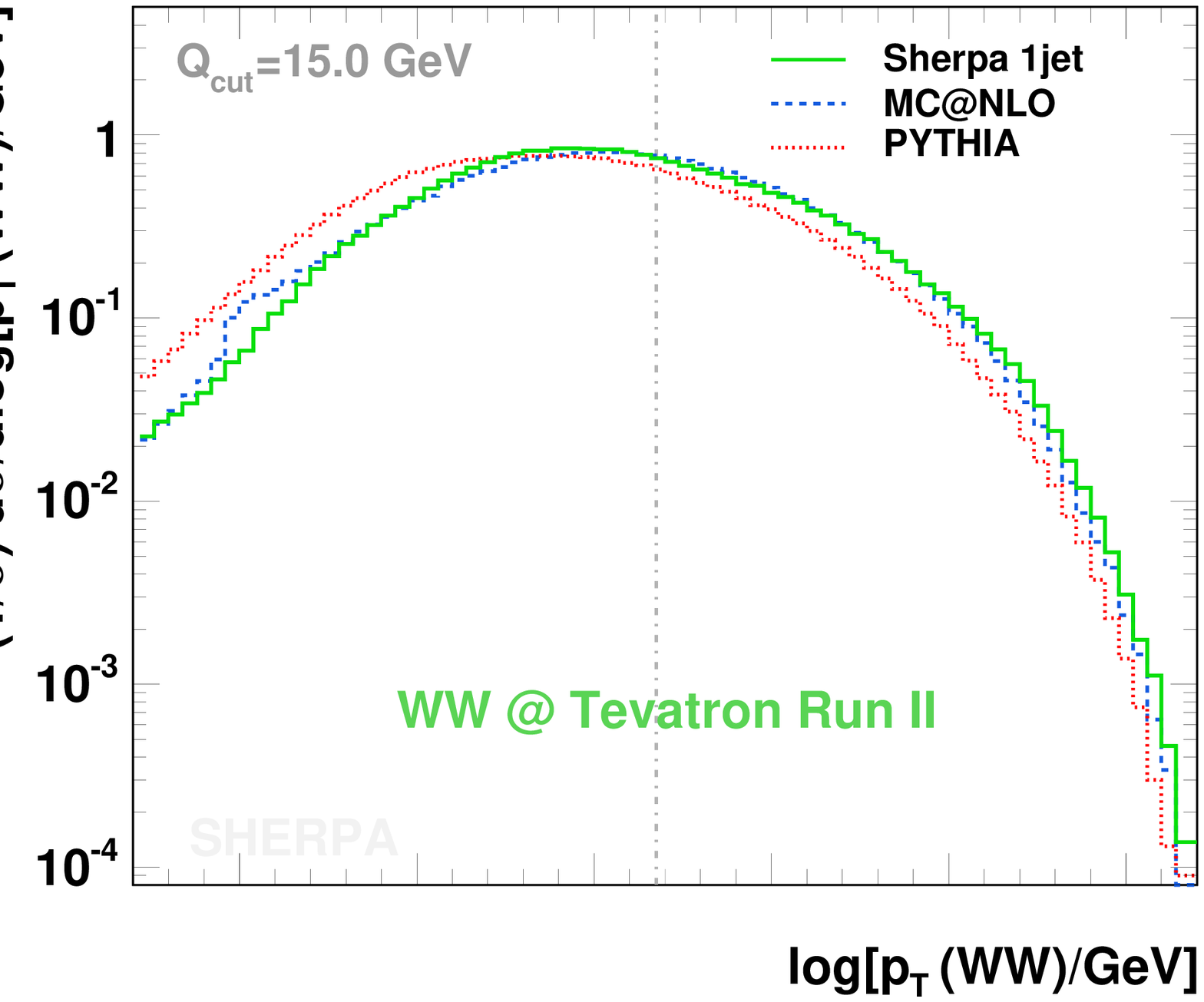}}
    \put(0,0){
      \includegraphics[width=70mm]{%
        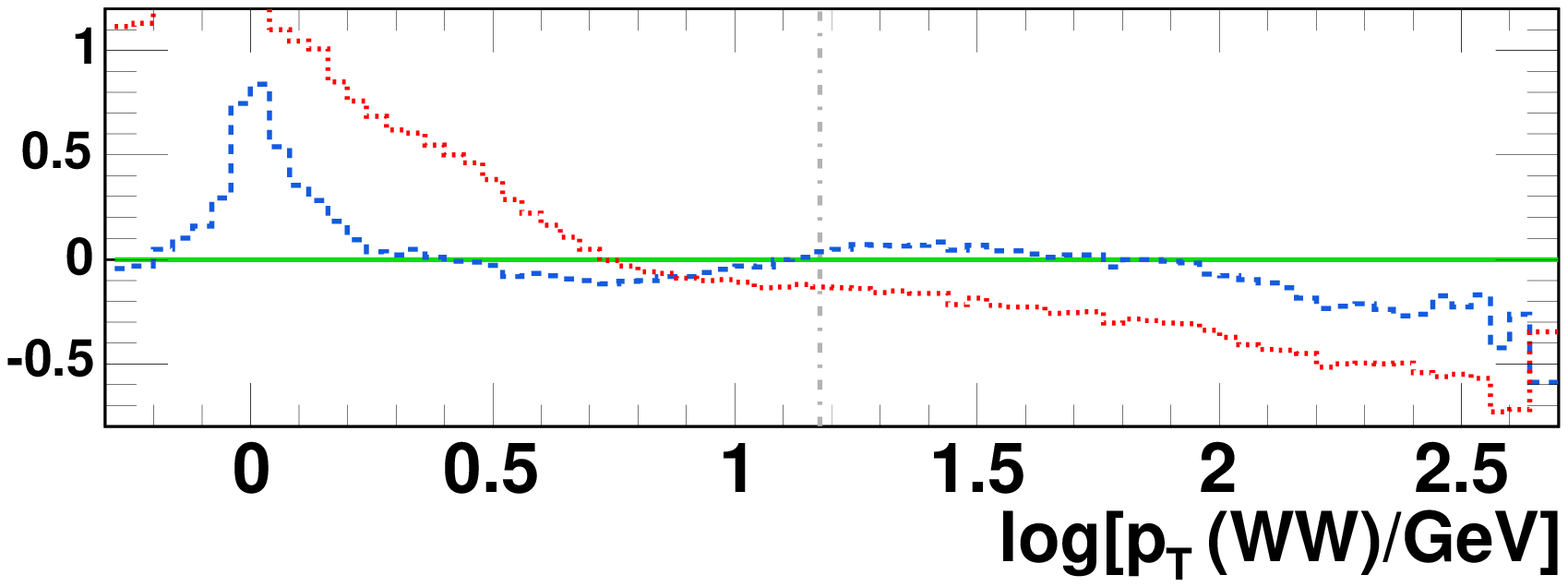}}
  \end{picture}
  \vspace{0mm}
  \caption{Normalized $p_T$ distribution of the $W^+W^-$ system.
    Results from {\tt PYTHIA} (red dotted line), {\tt MC@NLO} (blue
    dashed line) and \she\ (green solid line) are compared. For the
    generation of the latter, $n_{\rm max}=1$ and $Q_{\rm cut}=15$ GeV
    have been chosen. The lower part of the plot presents the
    normalized differences with respect to the \she\ prediction. Input
    parameters (including a primordial $k_\perp$ smearing) and the
    employed cuts are specified in the Apps.\ \ref{app_input} and
    \ref{app_cuts}.}
  \label{ptWW_mocas}
  \vspace{0mm}
\end{figure}
\begin{figure}[b!]
  \vspace{2mm}
  \begin{picture}(188,204)
    \put(0,0){
      \includegraphics[width=70mm]{%
        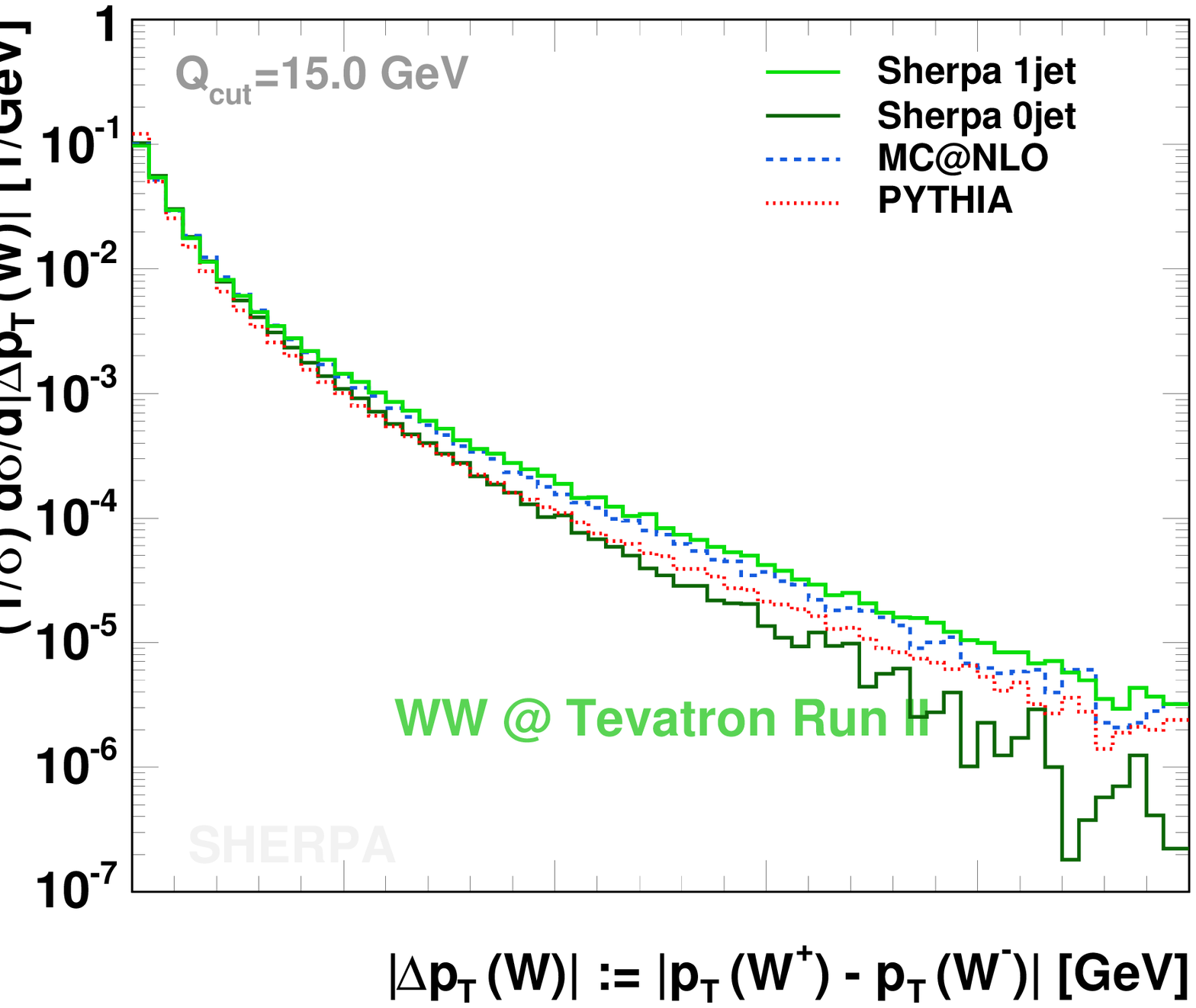}}
    \put(0,0){
      \includegraphics[width=70mm]{%
        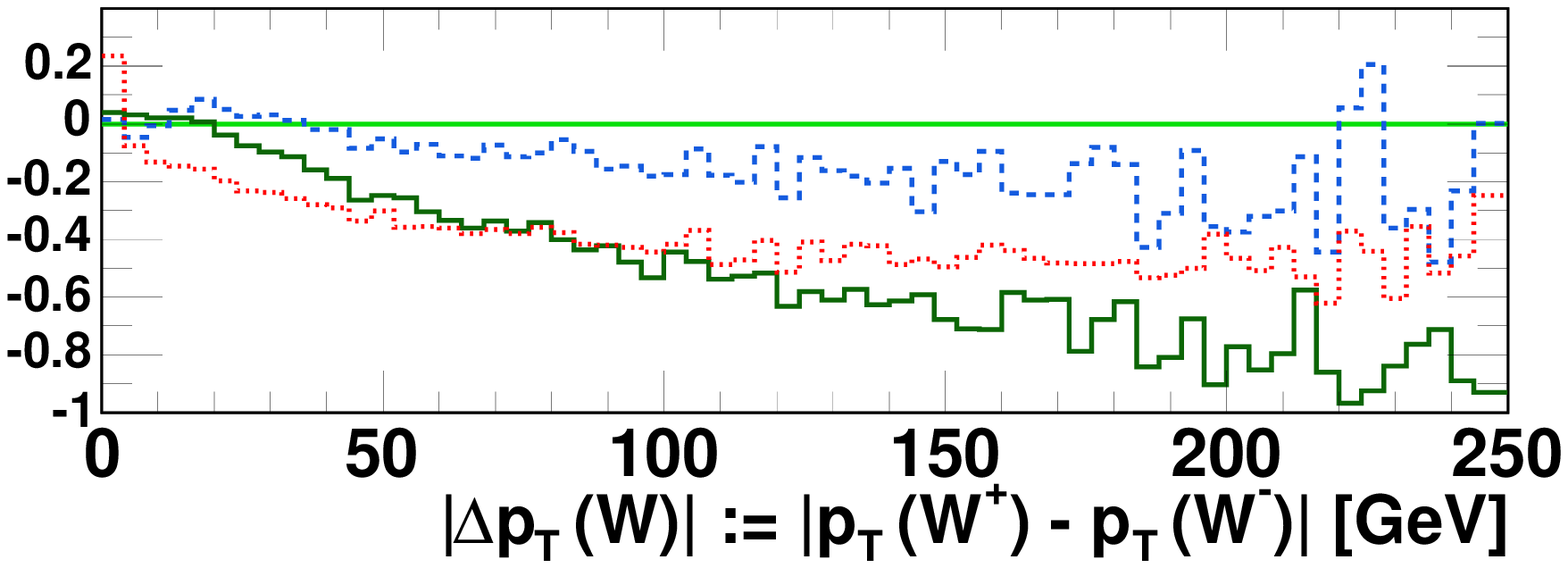}}
  \end{picture}
  \vspace{0mm}
  \caption{Difference of the scalar transverse momenta of the two
    $W$\/ bosons, $|p^{W^+}_T-p^{W^-}_T|$. The predictions compared
    are: {\tt PYTHIA} given as a red dotted curve, {\tt MC@NLO}
    depicted by the blue dashed line and \she\ in inclusive $1$jet
    production at $Q_{\rm cut}=15$ GeV drawn as a green solid line as
    well as \she\ in pure shower performance shown as a darkgreen
    solid line. The lower part of the plot shows the normalized
    differences with respect to the \she\ prediction with $n_{\rm
    max}=1$. Input parameters and the employed cuts are summarized in
    the Apps.\ \ref{app_input} and \ref{app_cuts}.}
  \label{ptdiff_mocas}
  \vspace{0mm}
\end{figure}
\begin{figure}[t!]
  \vspace{0mm}
  \begin{picture}(188,410)
    \put(0,205){
      \includegraphics[width=70mm]{%
        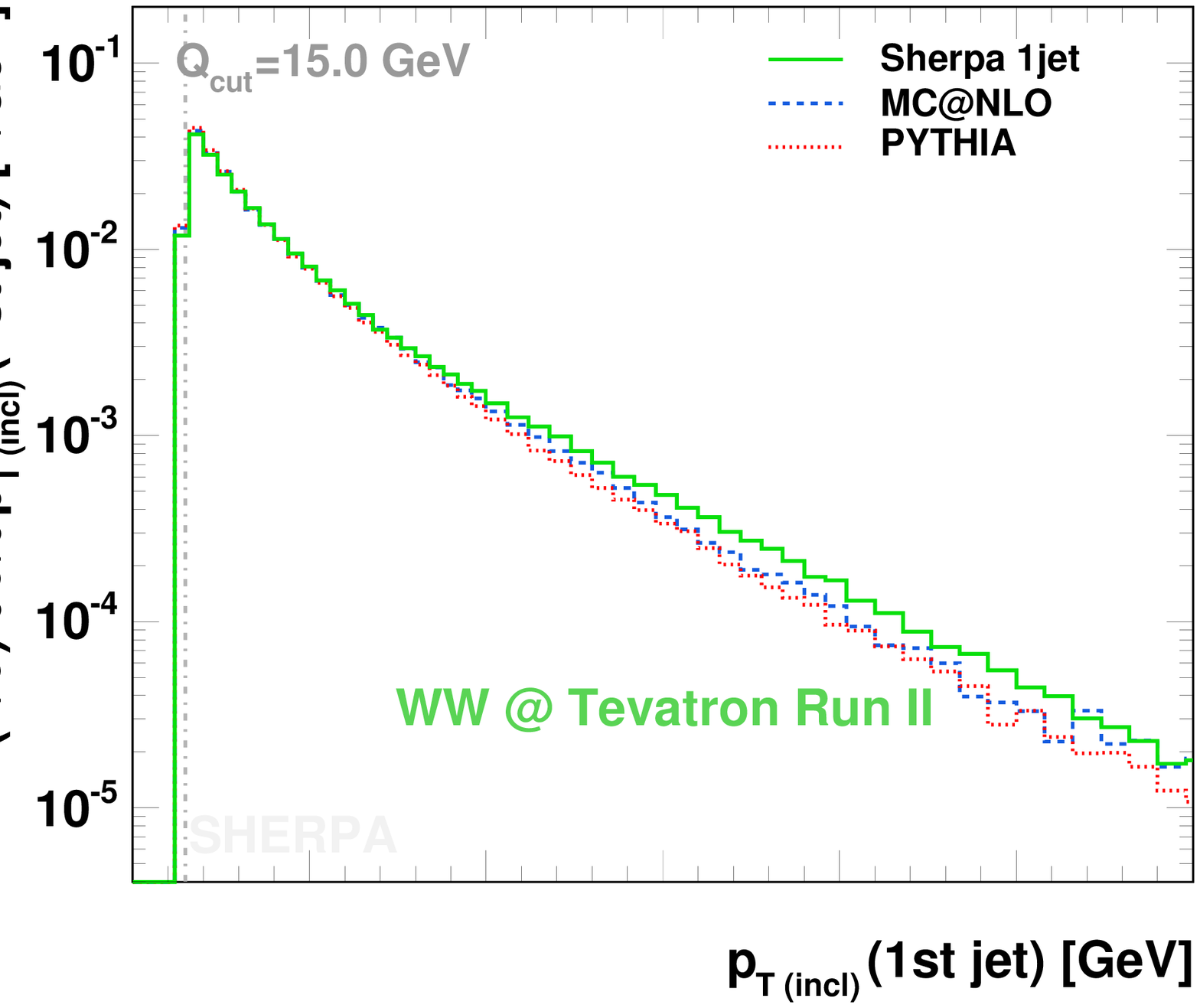}}
    \put(0,205){
      \includegraphics[width=70mm]{%
        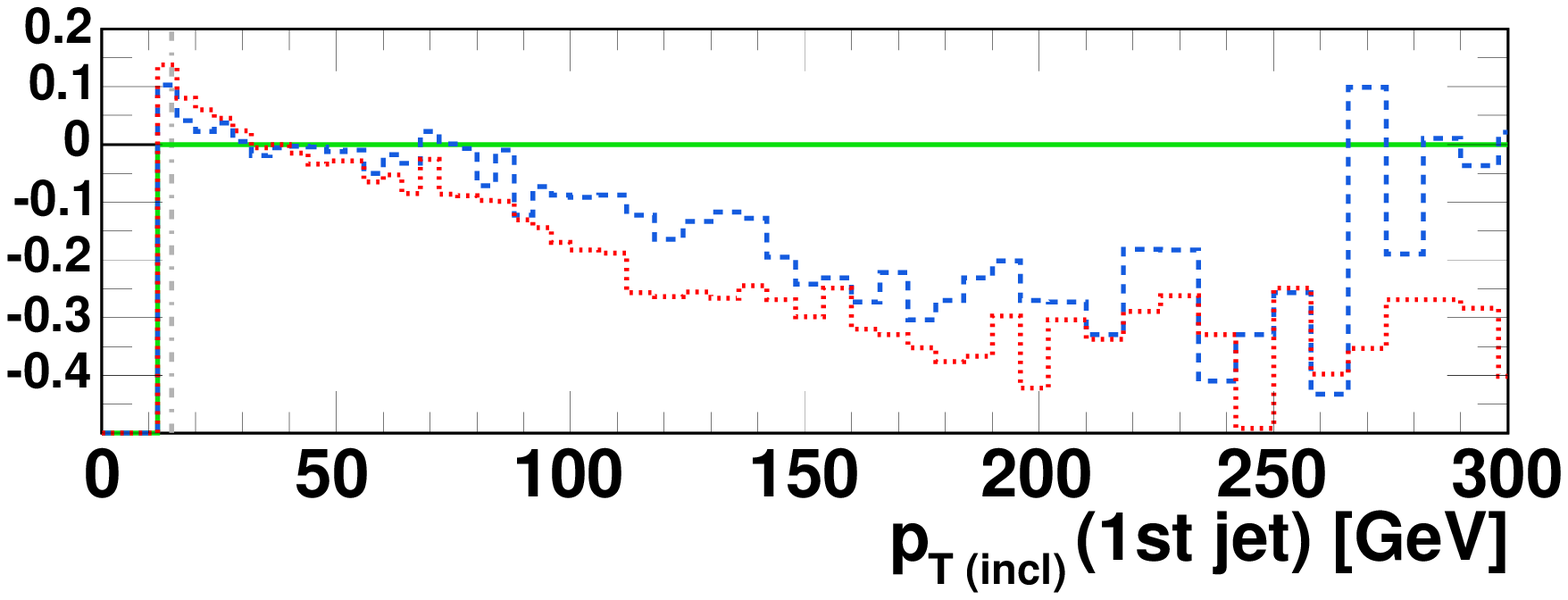}}
    \put(0,0){
      \includegraphics[width=70mm]{%
        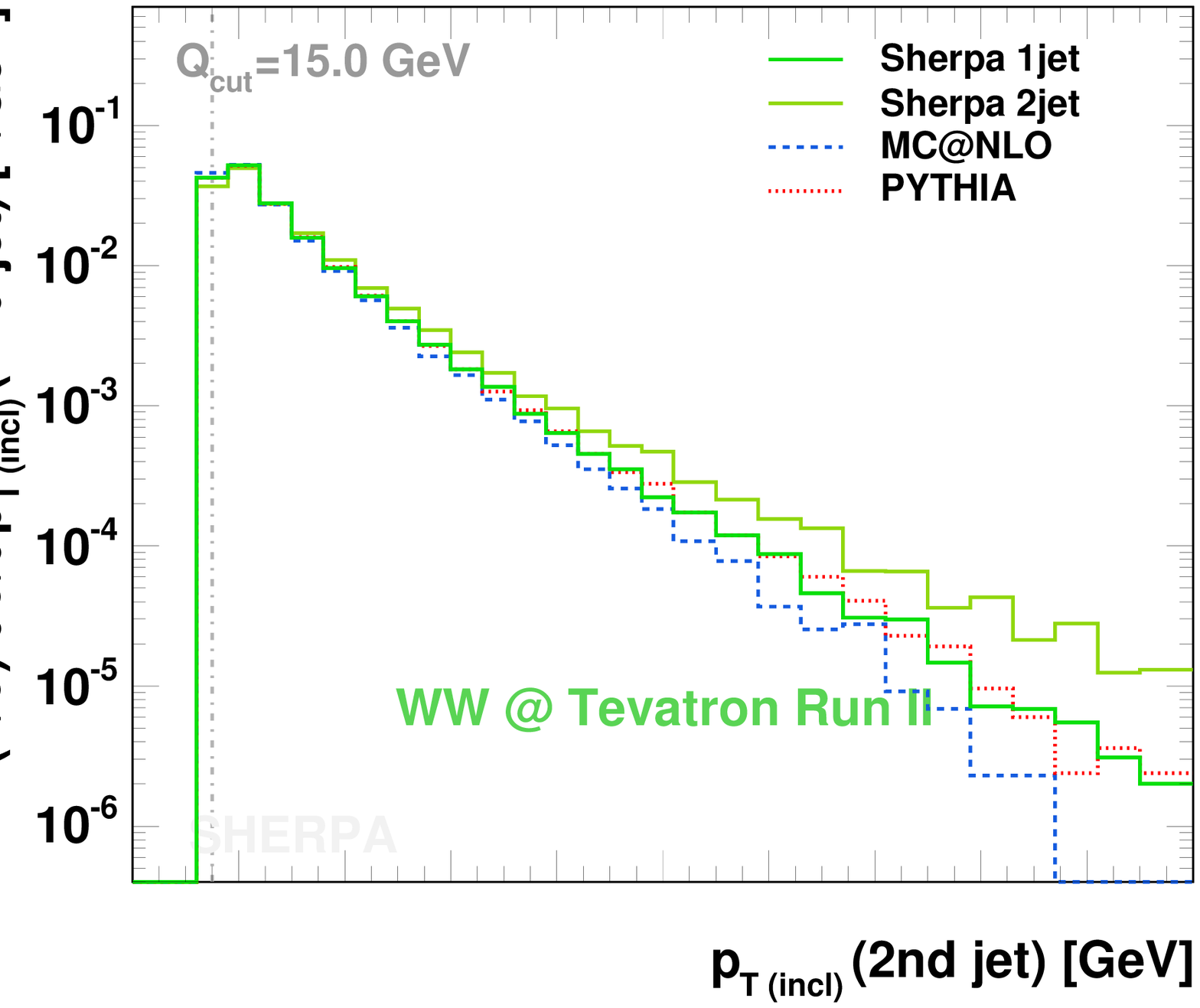}}
    \put(0,0){
      \includegraphics[width=70mm]{%
        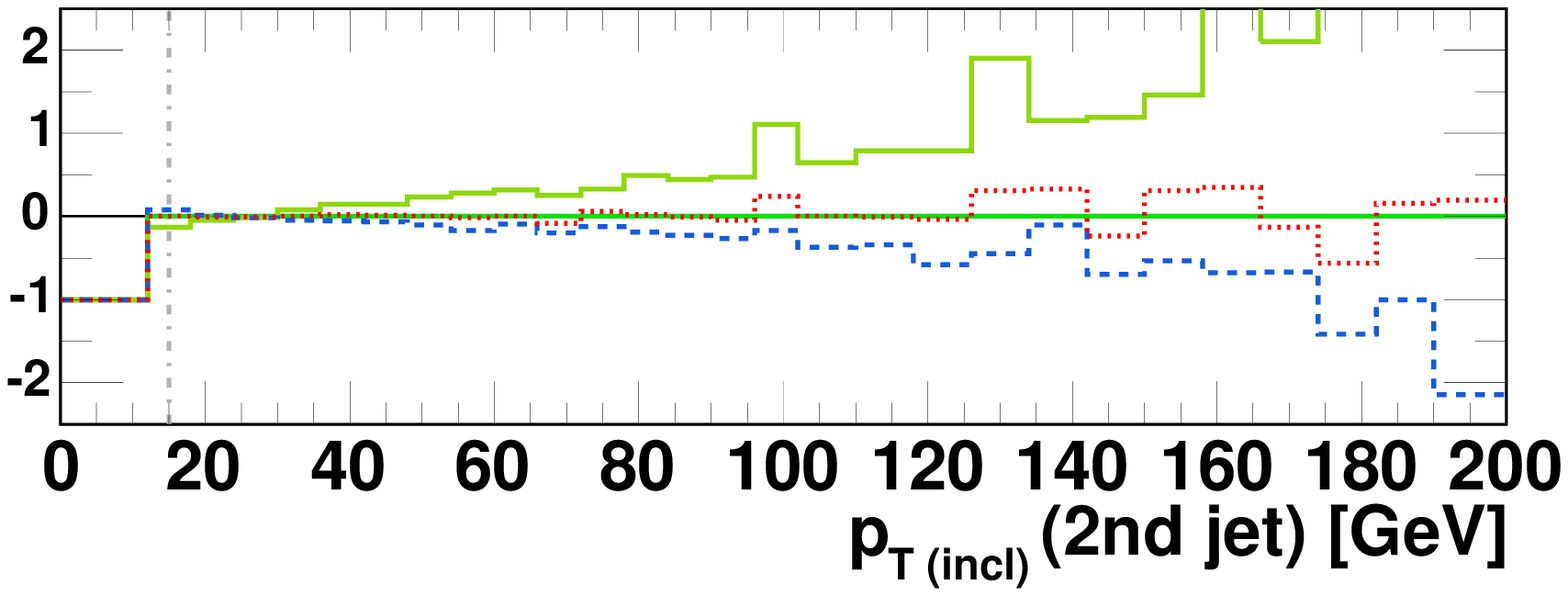}}
  \end{picture}
  \vspace{0mm}
  \caption{Transverse momentum distributions of the associated jets,
    in the upper panel, the inclusive $p_T$ of the hardest jet is
    depicted, whereas in the lower panel that one of the second
    hardest jet is displayed. Again, results from {\tt PYTHIA} are
    given by the red dotted lines, {\tt MC@NLO} results are
    represented as blue dashed lines and \she\ results are the green
    solid lines. For the generation of the latter, $n_{\rm max}=1$ and
    $Q_{\rm cut}=15$ GeV have been used. The lightgreen solid line in
    the lower panel corresponds to the \she\ result obtained with
    $n_{\rm max}=2$. The lower part of both plots shows the normalized
    differences with respect to the \she\ $n_{\rm max}=1$ performance.
    The input parameters and the employed cuts are summarized in the
    Apps.\ \ref{app_input} and \ref{app_cuts}.}
  \label{jet_mocas}
  \vspace{0mm}
\end{figure}
\noindent
As before, the starting point is the discussion of the radiation
activity predicted by the various codes. In Fig.\ \ref{ht_mocas},
results for the $H_T$ observable obtained from {\tt PYTHIA}, {\tt
MC@NLO} and \she\ are displayed. The predictions of the former two
codes nicely agree with each other. Similar to the \she\ {\tt MCFM}
comparison, \she\ again predicts a slightly harder spectrum, with
relative deviations of up to $20\%$.
\\
Closer inspection of the reason for the differences in the $H_T$
spectrum reveals that the agreement of {\tt PYTHIA} and {\tt MC@NLO}
is presumably a little bit accidental. A first hint into that
direction can be read off Fig.\ \ref{ptWW_mocas}, where the $p_T$
spectrum of the $W$\/ pair is displayed. In the region of low $p_T$
(up to $100$ GeV), the results of {\tt MC@NLO} and \she\ are in fairly
good agreement\footnote{%
  Apart from the very soft region, where the difference is due to
  parton shower cutoff effects in {\tt HERWIG}.},
and sizeable differences larger than $10\%$ appear only for $p_T>100$
GeV. In contrast, the {\tt PYTHIA} result for this observable shows a
significant enhancement of the low-$p_T$ region and stays well below
the other predictions for $p_T>10$ GeV. This comparison of the three
differential cross sections clearly underlines that the three codes
differ in their description of the QCD emissions.
\\
Fig.\ \ref{ptdiff_mocas} depicts the norm of the scalar difference of
the transverse momenta of the $W^+$ and $W^-$ gauge boson,
$|p^{W^+}_T-p^{W^-}_T|$. This observable is sensitive to higher order
effects, since at LO it merely has a delta peak at $p_T=0$ GeV. Again,
the hardest prediction is delivered by \she\ with $n_{\rm max}=1$,
results from {\tt MC@NLO}, {\tt PYTHIA}, and the pure shower
performance of \she\ are increasingly softer. For $|\Delta p_T|>60$
GeV, this observable seems to depend more and more on the quality of
modelling the hardest emission, which is intrinsically better
described by {\tt MC@NLO} and by \she\ with $n_{\rm max}=1$. The fact
that the {\tt PYTHIA} shower performs better than the pure \she\
shower for high $p_T$ differences can be traced back to the choice of
starting scale for the shower evolution, which is either $s_{p\bar p}$
({\tt PYTHIA}) or $s_{WW}$ (\she).
\\
In fact, differences appear in the $p_T$ distributions of the
hardest two jets, see Fig.\ \ref{jet_mocas}. The upper part of this
figure depicts the transverse momentum spectrum of the hardest jet.
Surprisingly, although {\tt MC@NLO} contains a matrix element for the
emission of an extra jet, its $p_T$ distribution is considerably
softer (by up to $40\%$) than the result of \she\ generated with
$n_{\rm max}=1$.
This trend is greatly amplified when going to the spectrum of the
second hardest jet. There, clear shape differences of the order of a
factor $2$ between the \she\ $1$jet sample and {\tt MC@NLO} show up
for $p_T\approx180$ GeV. The surprise according to this figure is that
{\tt PYTHIA} and \she\ using $n_{\rm max}=1$ almost agree on the $p_T$
distribution of the second jet, although they were different for the
hardest jet. At that point it should be noted that the second jet in
both cases, {\tt PYTHIA} and \she\ with $n_{\rm max}=1$, is produced
by the parton shower only. Given the drastically larger shower start
scale of {\tt PYTHIA}, it seems plausible to achieve to some extent a
compensation for the intrinsic parton shower deficiencies in filling
the hard emission phase space\footnote{%
  {\tt PYTHIA}'s ability to account for harder second jets with
  respect to {\tt MC@NLO} is a hint for the similarity of their $H_T$
  predictions.}.
However, in the very moment, \she\ events are generated with
appropriate matrix elements, \ie with $n_{\rm max}=2$, this
distribution is dramatically different for the three codes with
deviations larger than a factor $2$ for $p_T\approx120$ GeV.
\\
Taken together, these findings hint that the three codes differ in
their modelling of the QCD activity, especially in those of the
hardest QCD emission. For {\tt MC@NLO} and \she\ the latter can be
traced back to the different ansatz in including the matrix element
for this emission, where again different scale choices may trigger
effects on the $20\%$ level.

\subsection*{Comparison of lepton observables}

\noindent
Finally, the leptons in the final state as described by the three
event generators {\tt PYTHIA}, {\tt MC@NLO} and \she\ will be
investigated. There, some significant differences appear between \she\
and {\tt PYTHIA} on the one hand, and {\tt MC@NLO} on the other hand.
These differences are due to the fact that at the moment spin
correlations of the $W$\/ decay products are not implemented in {\tt
MC@NLO}\footnote{%
  This situation is currently being cured by the authors of {\tt
  MC@NLO} who prepare a new version of their code including spin
  correlations \cite{Frixione::private}.}.
To validate that effects are indeed due to the lack of spin
correlations, \she\ samples have been prepared, where these
correlations are artificially switched off. Furthermore, in order to
quantify these effects without any bias, results have been obtained
without the application of any lepton and jet cuts.
\begin{figure}[b!]
  \vspace{0mm}
  \begin{picture}(188,204)
    \put(0,0){
      \includegraphics[width=70mm]{%
        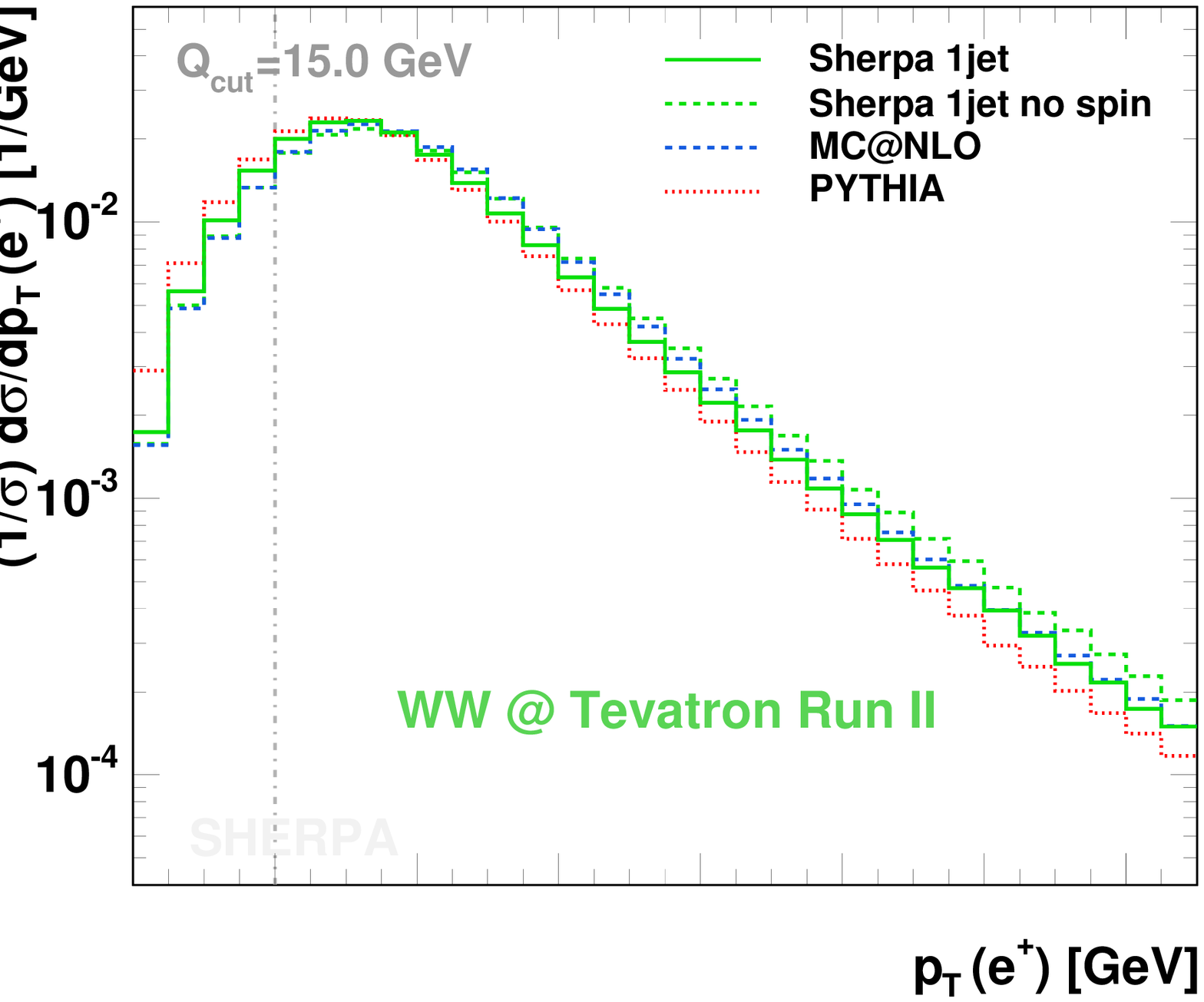}}
    \put(0,0){
      \includegraphics[width=70mm]{%
        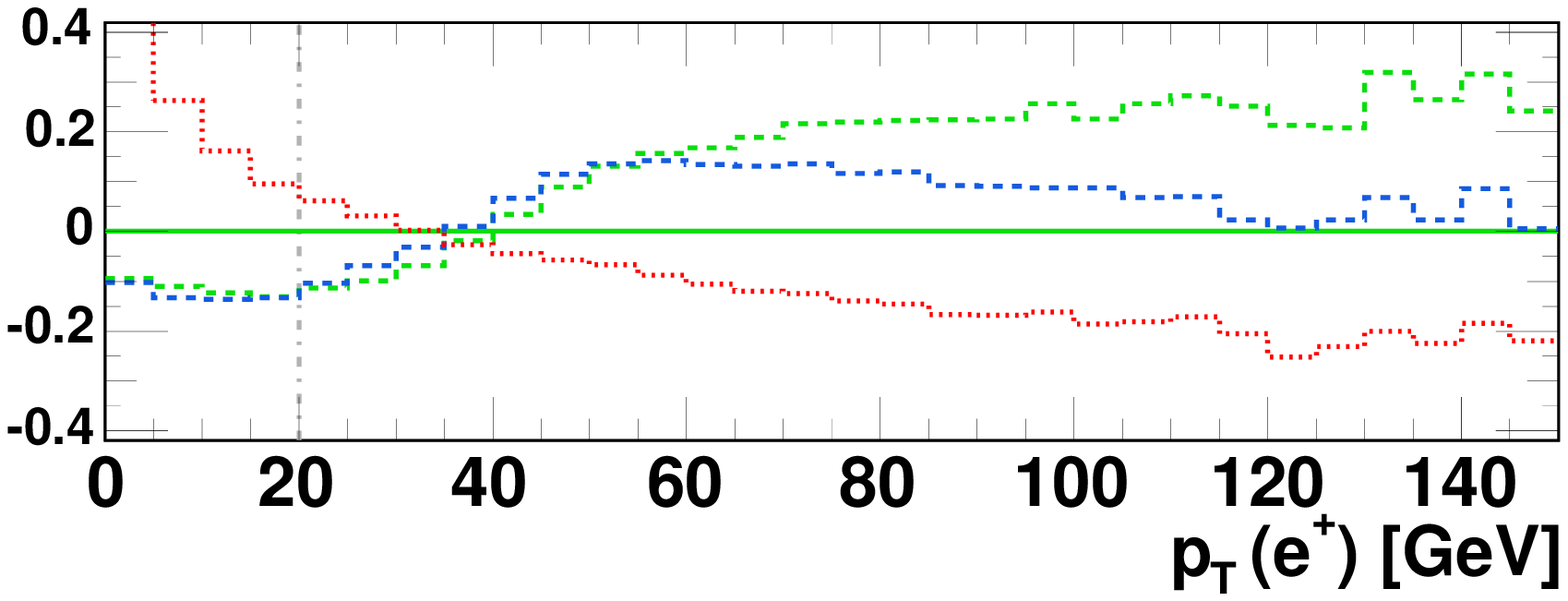}}
  \end{picture}
  \vspace{0mm}
  \caption{Normalized $p_T$ spectrum of the positron. Results of {\tt
    PYTHIA} (red dotted line) and \she\ (green solid line) including
    spin correlations are confronted with those obtained from {\tt
    MC@NLO} (blue dashed line) and with results from \she, where spin
    correlations have been switched off (green dashed line).
    All predictions are generated without the use of cuts. The
    vertical dashed-dotted line is added to indicate the position of
    the usually employed lepton $p_T$ cut. For input parameters, see
    App.\ \ref{app_input}. The lower part of the plot shows the
    normalized differences with respect to the \she\ prediction
    including spin correlations.}
  \label{pte_spini_mocas}
  \vspace{0mm}
\end{figure}
\begin{figure}[b!]
  \vspace{0mm}
  \begin{picture}(188,204)
    \put(0,0){
      \includegraphics[width=70mm]{%
        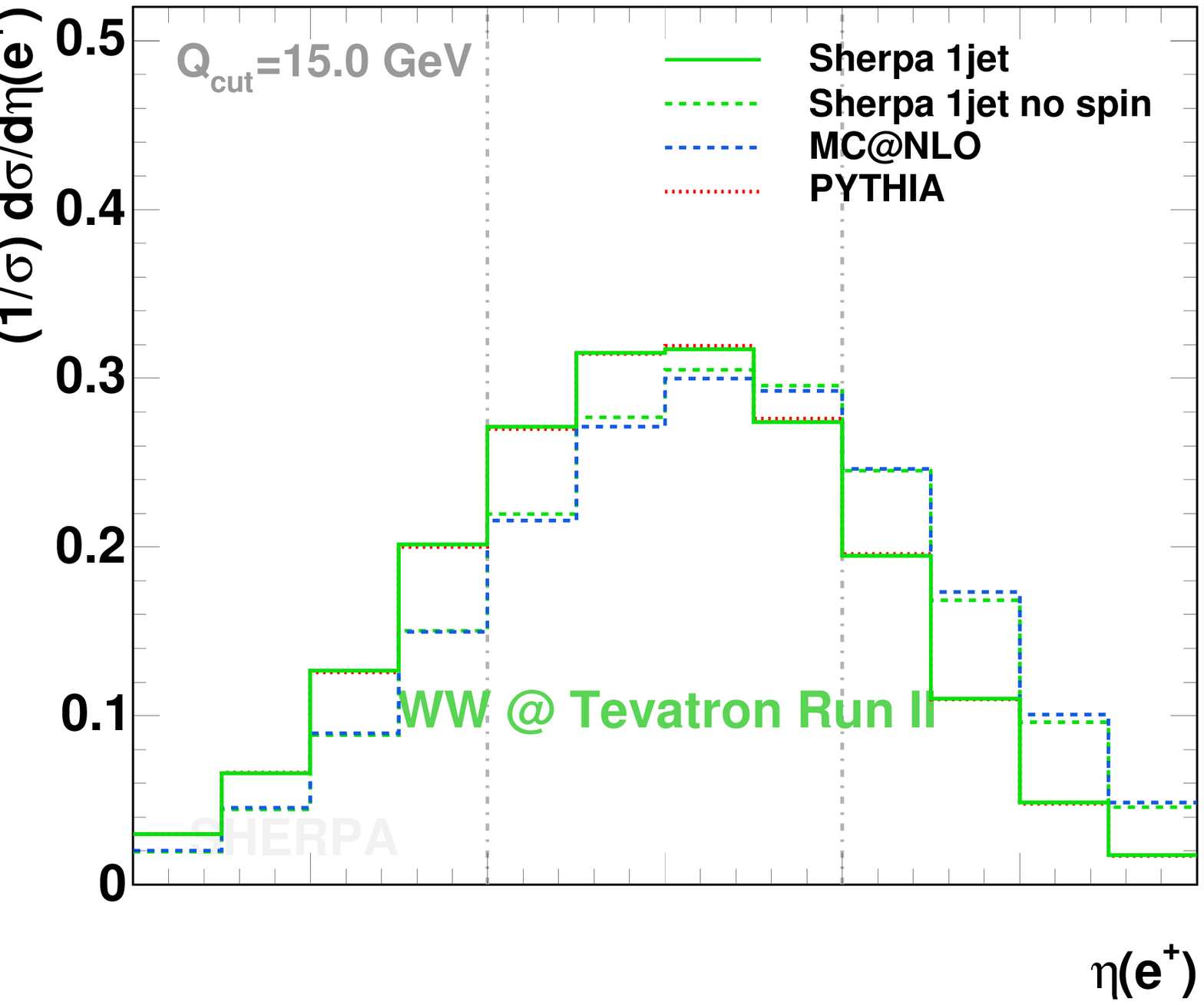}}
    \put(0,0){
      \includegraphics[width=70mm]{%
        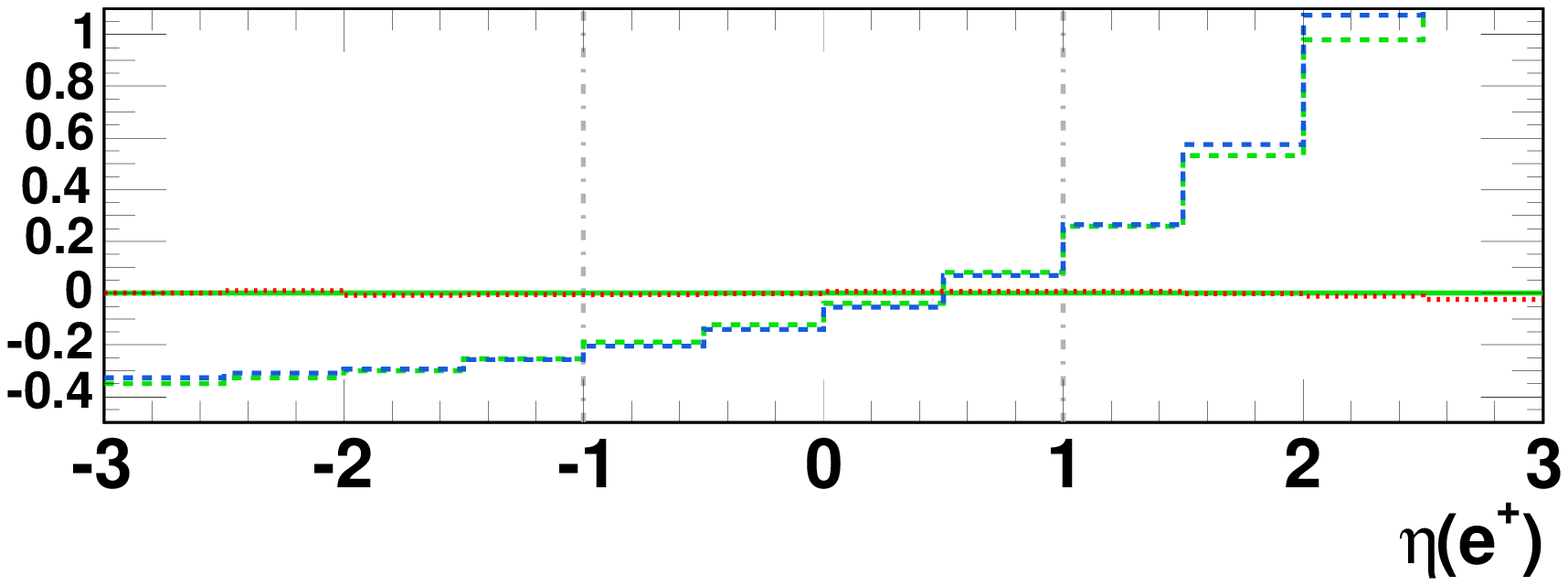}}
  \end{picture}
  \vspace{0mm}
  \caption{Normalized $\eta$\/ spectrum of the positron. Results of
    {\tt PYTHIA} (red dotted line) and \she\ (green solid line)
    including spin correlations are compared with those obtained from
    {\tt MC@NLO} (blue dashed line) and with results from \she, where
    spin correlations have been switched off (green dashed line).
    All predictions are generated without any restriction.
    The vertical dashed-dotted lines are added to indicate the
    position of the usually employed lepton $\eta$\/ cuts.
    For input parameters, see App.\ \ref{app_input}. The lower part of
    the plot shows the normalized differences with respect to the
    \she\ prediction including spin correlations.}
  \label{etae_spini_mocas}
  \vspace{0mm}
\end{figure}
\begin{figure}[b!]
  \vspace{0mm}
  \begin{picture}(188,204)
    \put(0,0){
      \includegraphics[width=70mm]{%
        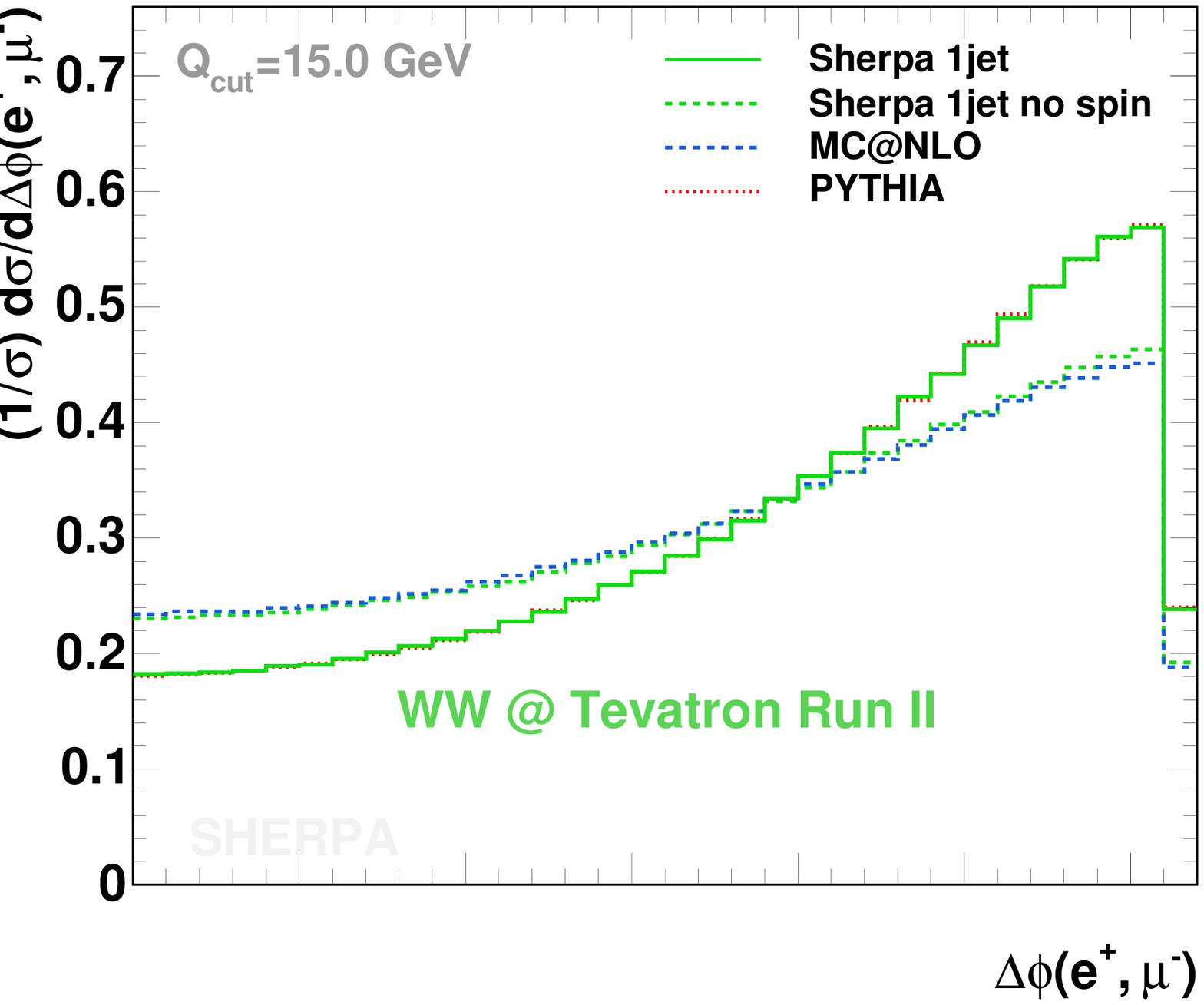}}
    \put(0,0){
      \includegraphics[width=70mm]{%
        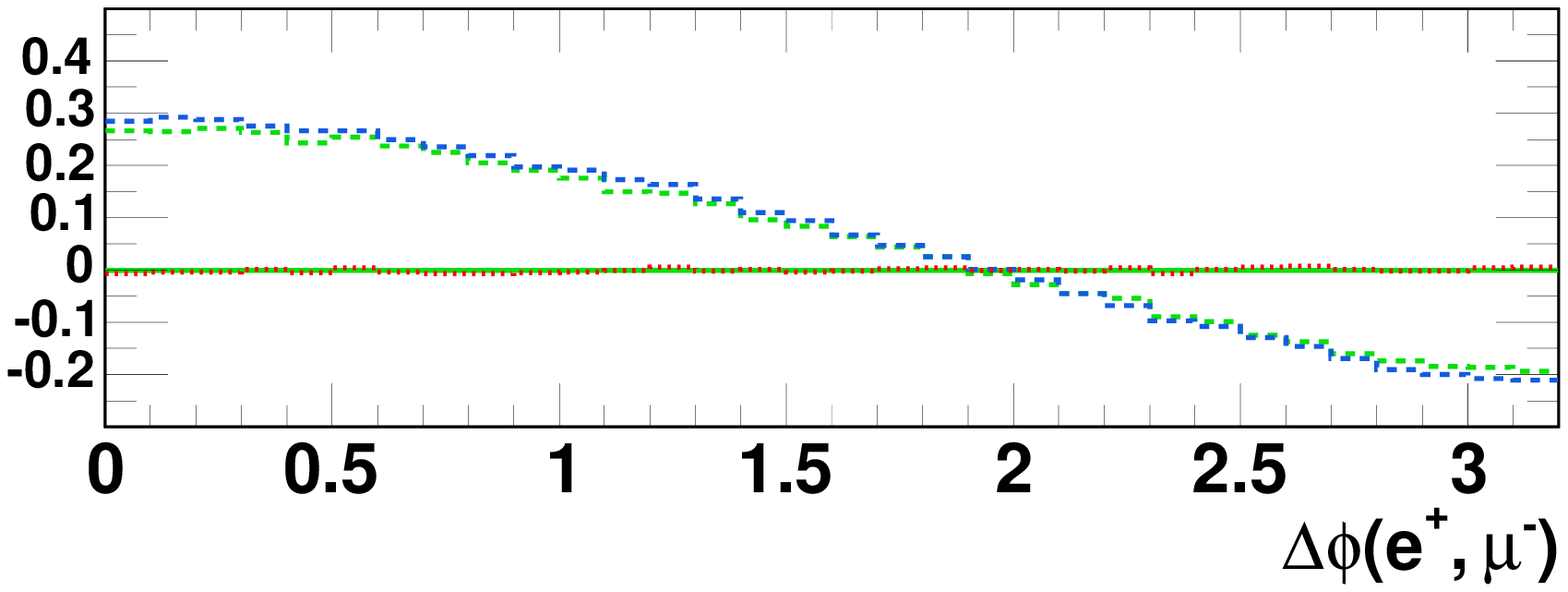}}
  \end{picture}
  \vspace{0mm}
  \caption{Normalized $\Delta\phi_{e\mu}$ distribution. Results of
    {\tt PYTHIA} (red dotted line) and \she\ (green solid line)
    including spin correlations are compared with those obtained from
    {\tt MC@NLO} (blue dashed line) and with results from \she, where
    spin correlations have been switched off (green dashed line). All
    predictions are obtained without the use of cuts. For input
    parameters, see App.\ \ref{app_input}. The lower part of the plot
    shows the normalized differences with respect to the \she\
    prediction including spin correlations.}
  \label{dphiemu_spini_mocas}
  \vspace{0mm}
\end{figure}
\begin{figure}[b!]
  \vspace{0mm}
  \begin{picture}(188,204)
    \put(0,0){
      \includegraphics[width=70mm]{%
        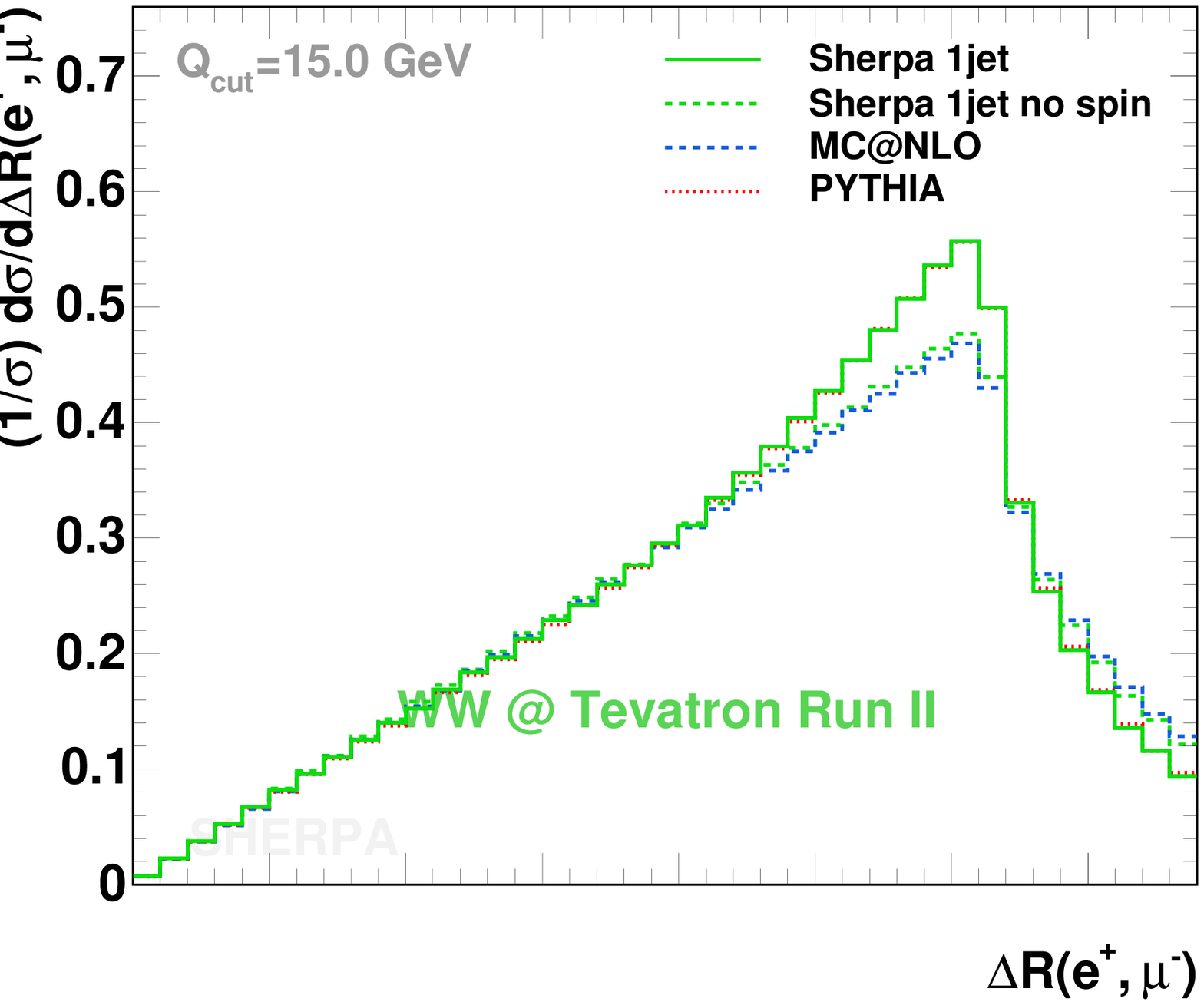}}
    \put(0,0){
      \includegraphics[width=70mm]{%
        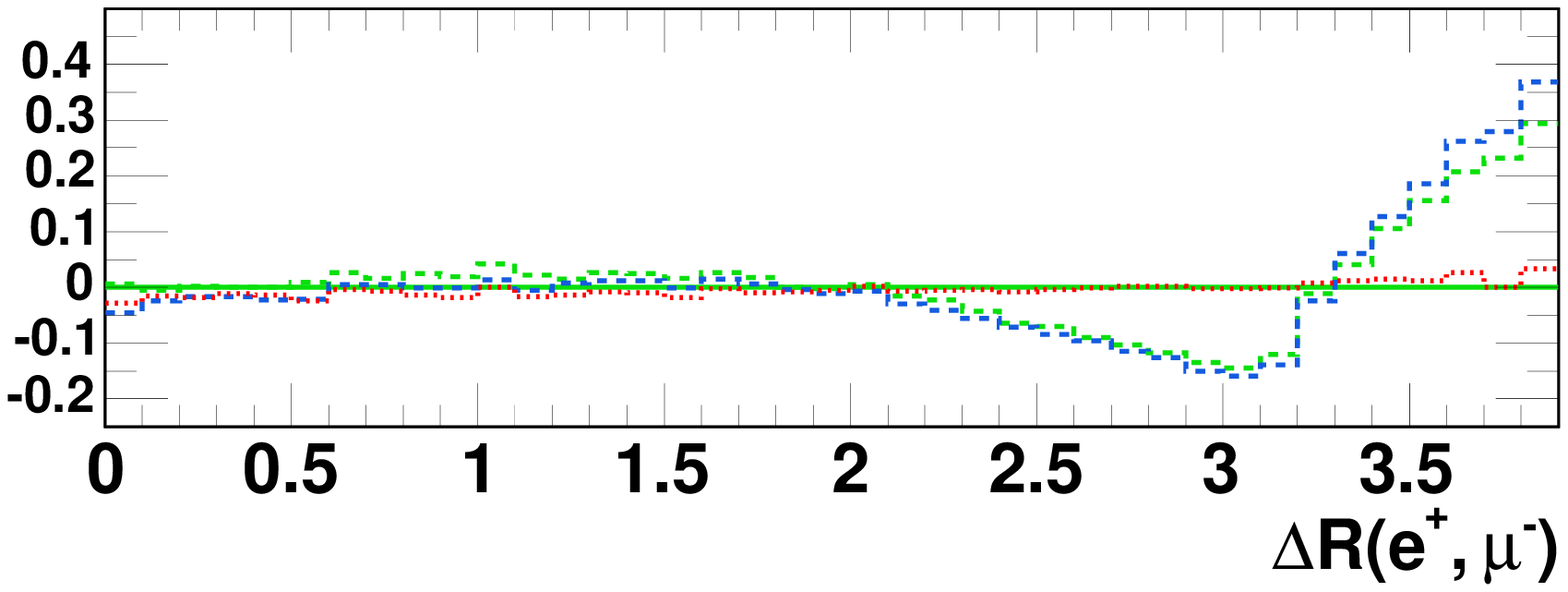}}
  \end{picture}
  \vspace{0mm}
  \caption{Normalized $\Delta R_{e\mu}$ distribution. Results of {\tt
    PYTHIA} (red dotted line) and \she\ (green solid line) including
    spin correlations are compared with those obtained from {\tt
    MC@NLO} (blue dashed line) and with results from \she, where spin
    correlations have been eliminated (green dashed line). All
    predictions are obtained without the use of cuts. For input
    parameters, see App.\ \ref{app_input}. The lower part of the plot
    shows the normalized differences with respect to the \she\
    prediction including spin correlations.}
  \label{dremu_spini_mocas}
  \vspace{0mm}
\end{figure}
\begin{figure*}[t!]
  \vspace{0mm}
  \begin{picture}(444,160)
    \put(300,0){
      \includegraphics[width=54mm]{%
       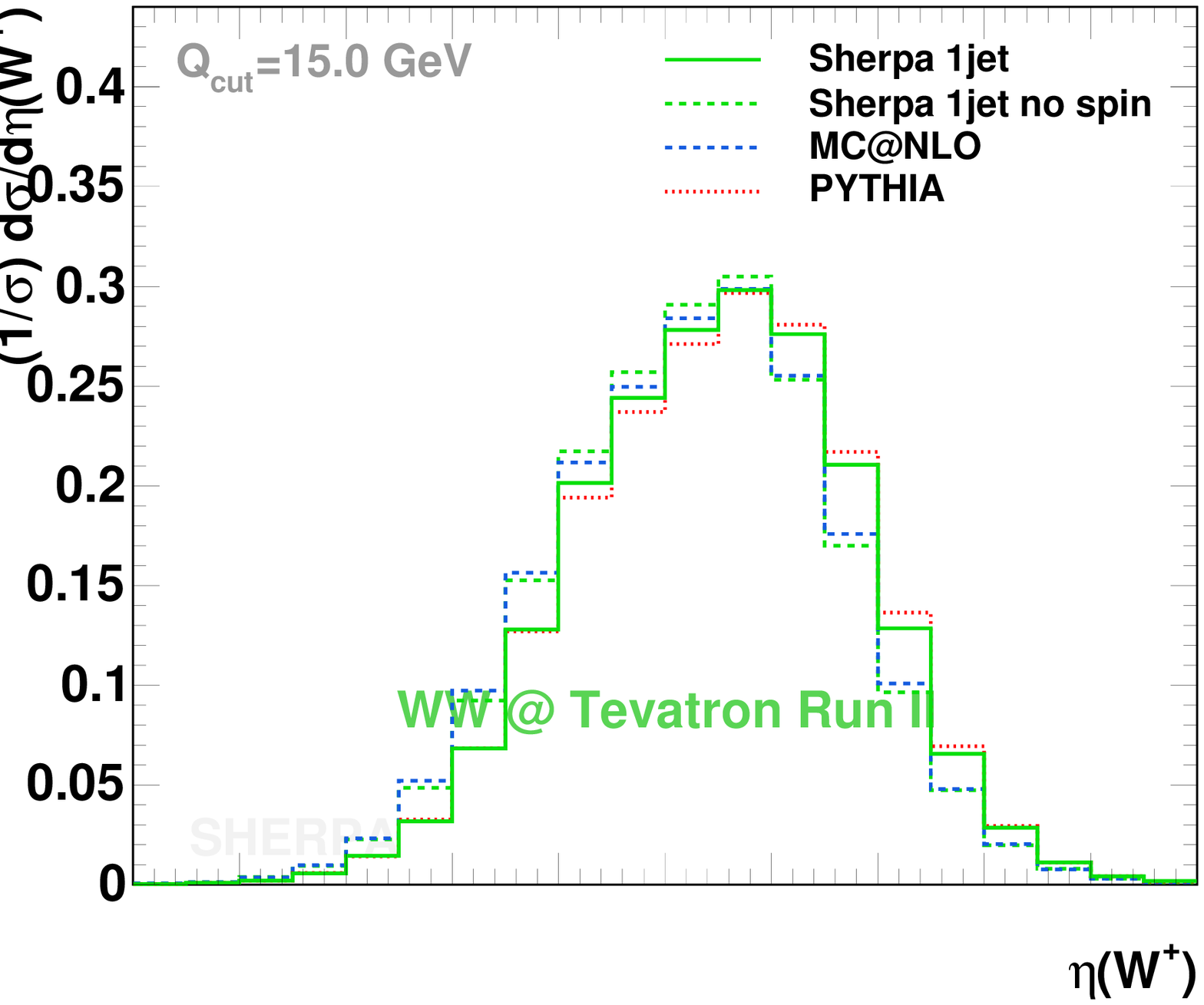}}
    \put(300,0){
      \includegraphics[width=54mm]{%
       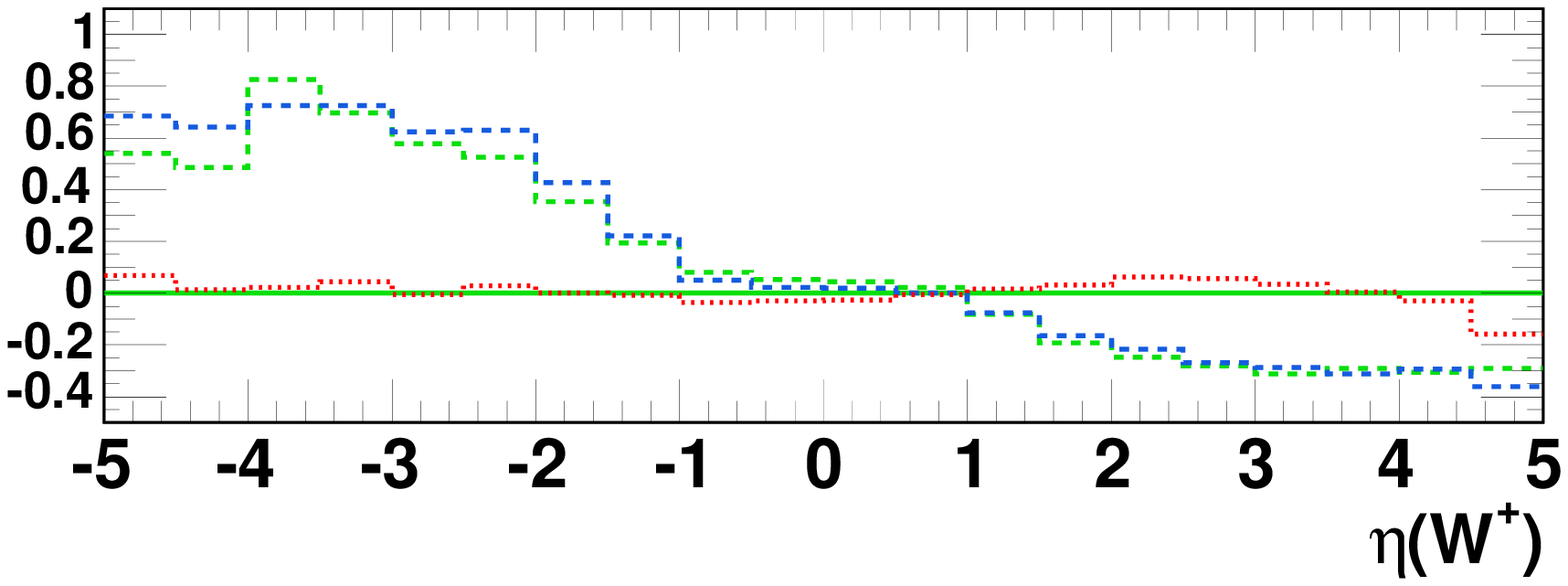}}
    \put(150,0){
      \includegraphics[width=54mm]{%
       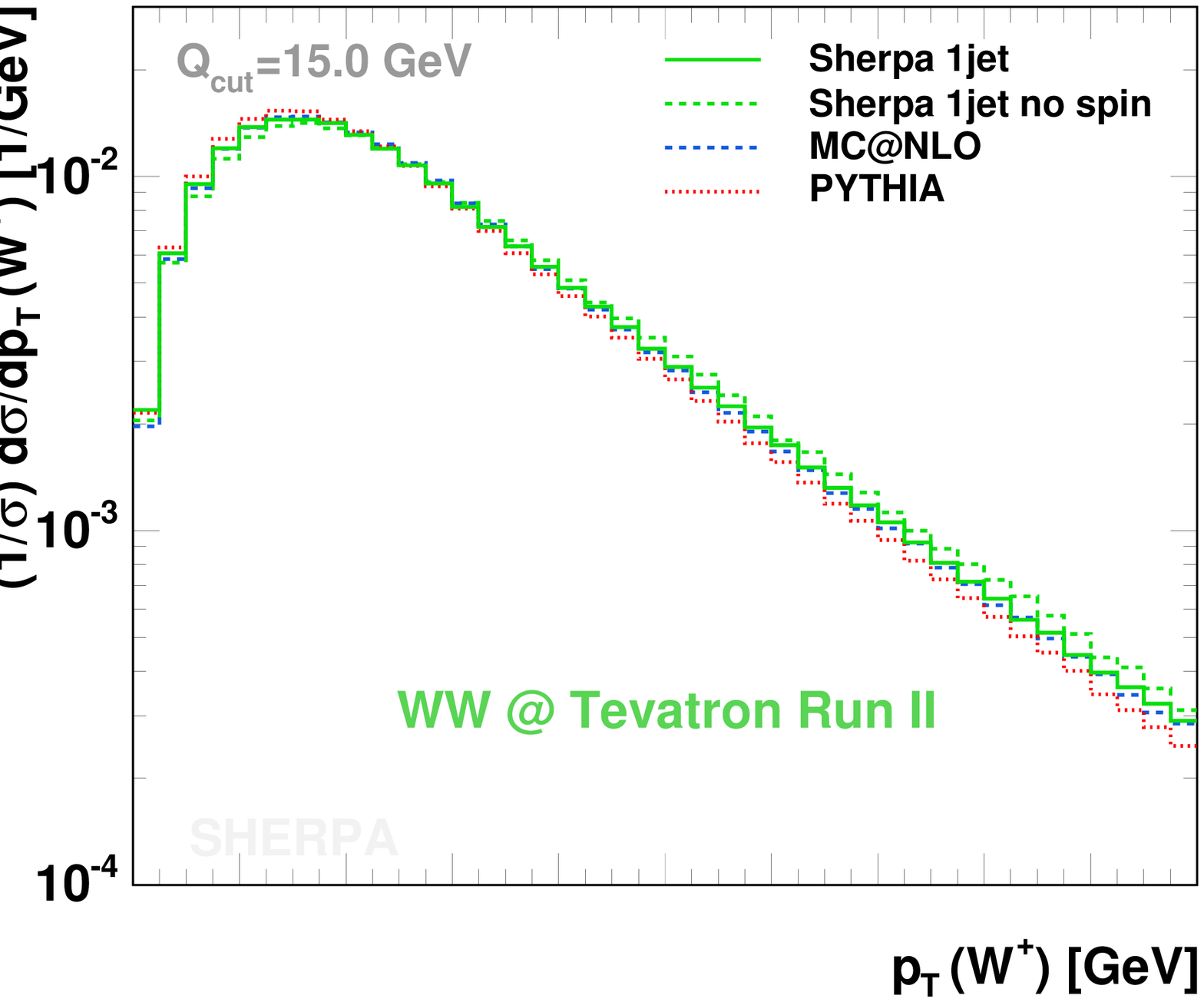}}
    \put(150,0){
      \includegraphics[width=54mm]{%
       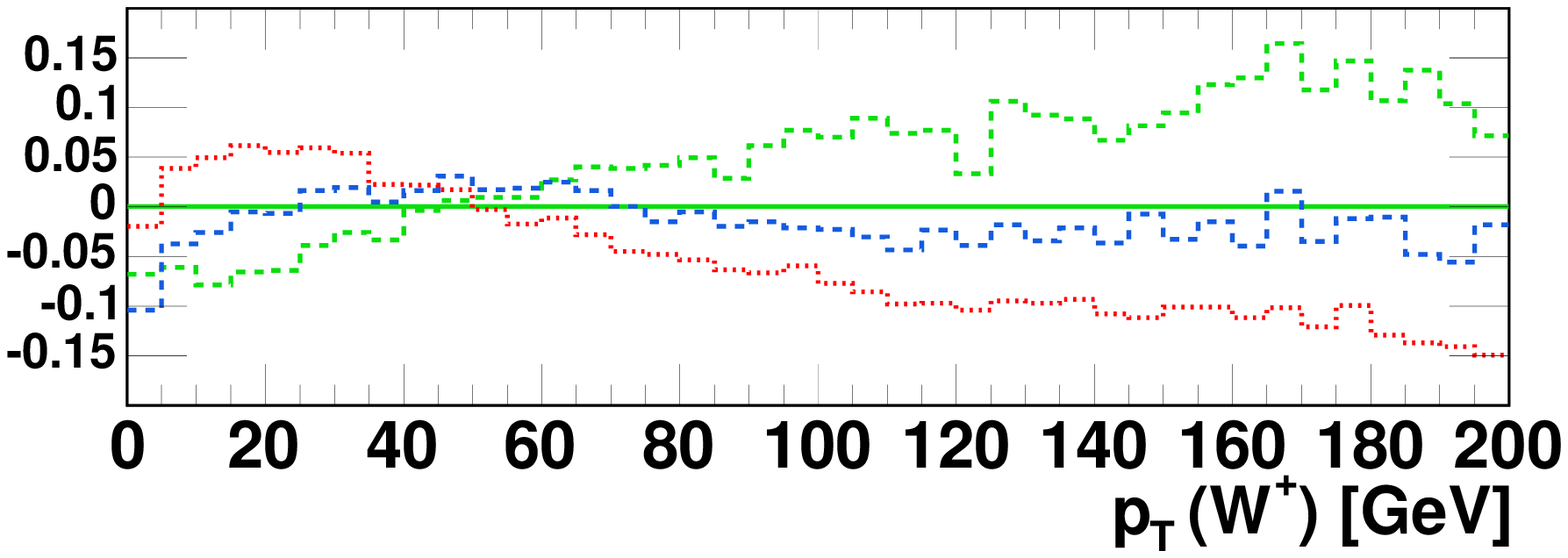}}
    \put(0,0){
      \includegraphics[width=54mm]{%
       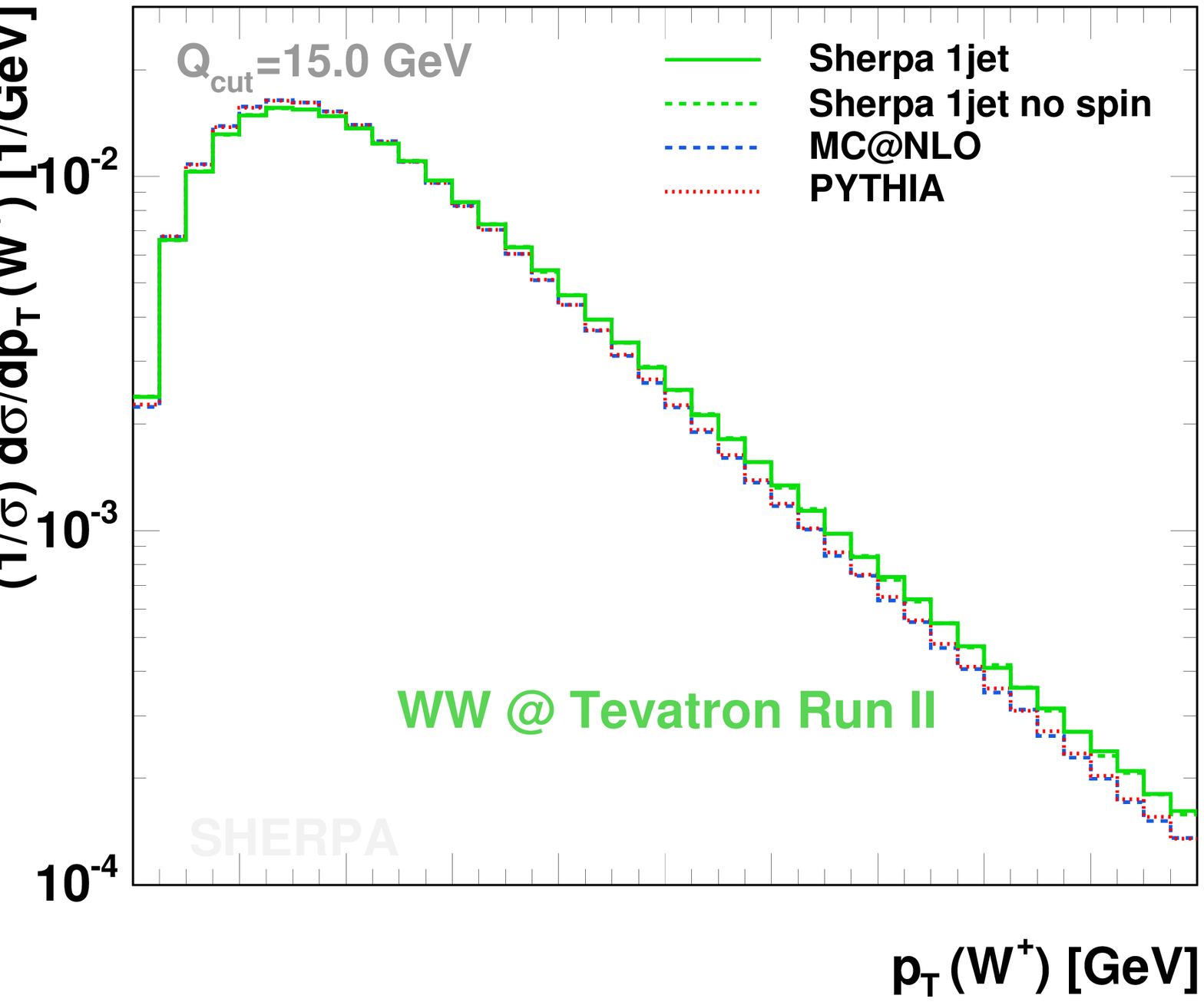}}
    \put(0,0){
      \includegraphics[width=54mm]{%
       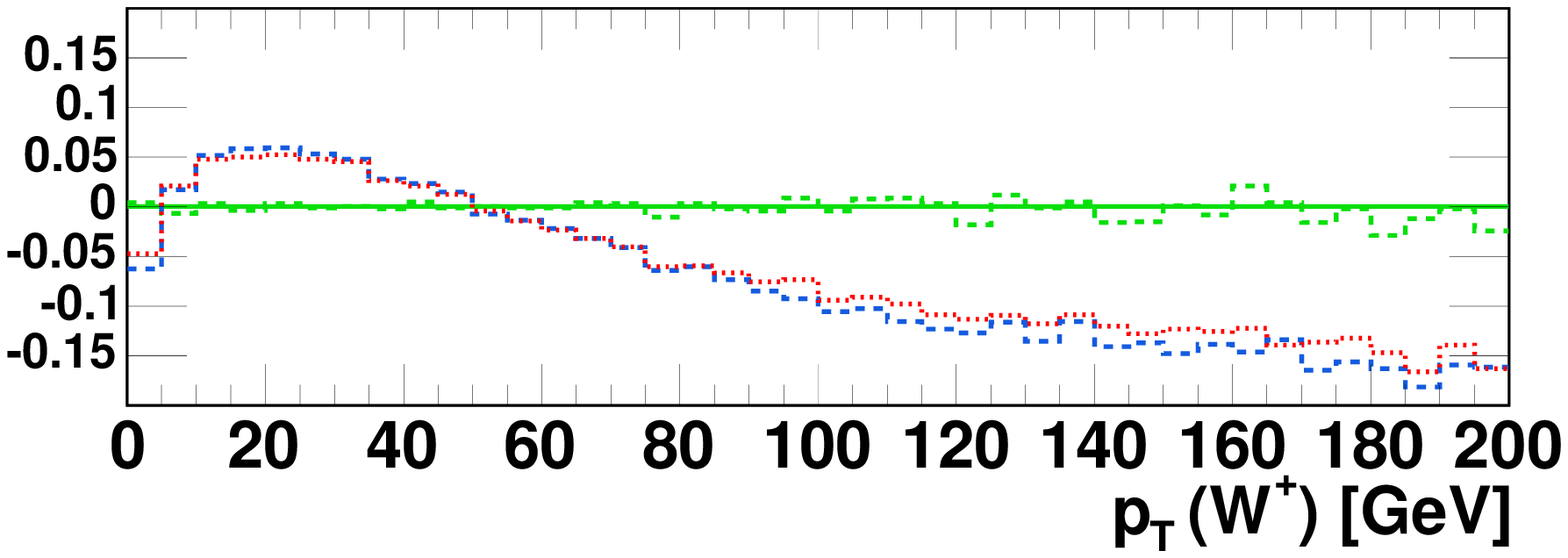}}
  \end{picture}
  \vspace{0mm}
  \caption{In the left and middle panel the $p_T$ spectrum of
    the $W^+$ before and after the application of cuts is depicted,
    respectively. The right panel exhibits the $\eta$\/ distribution
    of the $W^+$ under the influence of these cuts. The predictions
    compared are: {\tt PYTHIA} (red dotted line), \she\ (green solid
    line), {\tt MC@NLO} (blue dashed line) and \she\ without
    correlations in the boson decays (green dashed line).
    For input parameters, see App.\ \ref{app_input}. The lower part of
    the plots shows the normalized differences with respect to the
    \she\ prediction including spin correlations.}
  \label{cuts_mocas}
  \vspace{0mm}
\end{figure*}
\\
The impact of the lack of spin correlations already becomes visible
in one-particle observables, such as the $p_T$ or the $\eta$\/
spectrum of the positron produced in the $W^+$ decay. These are shown
in Figs.\ \ref{pte_spini_mocas} and \ref{etae_spini_mocas},
respectively. Confronting the two methods with each other, which
correctly respect spin correlations, for the transverse momentum
distribution of the $e^+$, the following pattern is revealed. Due to
the consistent inclusion of higher order tree-level matrix elements,
the \she\ $n_{\rm max}=1$ setup produces a considerably harder spectrum
than {\tt PYTHIA}. In contrast, the distributions with no spin
correlations both result in an even harder high-$p_T$ tail. They agree
quite well up to $p_T=60$ GeV, hence, this coincidence may be assigned
to the lack of spin correlations in the gauge boson decays. Above that
region, the {\tt MC@NLO} spectrum again becomes softer with respect to
the \she\ prediction where the spin correlations have been eliminated.
The fact that all four predicted distributions alter in their shape is
not solely triggered by the different spin correlation treatments,
again, the different descriptions of QCD radiation clearly contribute
to the deviations found.
\\
In contrast, a simpler pattern is found for the aforementioned
$\eta$\/ distribution of the $e^+$. The results of {\tt PYTHIA} and
\she\ with spin correlations on the one hand and of {\tt MC@NLO} and
\she\ without spin correlations on the other hand show perfect
agreement. Differences between the two spin correlation treatments
may, thereby, easily reach up to $40\%$.
\\
The influence of spin correlations can also be seen in observables
based on two particle correlations. As two illustrative examples take
the $\Delta\phi$\/ and the $\Delta R$\/ distribution of the
$e^+$ and the $\mu^-$ produced in the decay of the two $W$\/ bosons.
Again, the corresponding spectra, which have been exhibited in Figs.\
\ref{dphiemu_spini_mocas} and \ref{dremu_spini_mocas}, differ
significantly in shape depending on whether spin correlations are
taken into account or not.
\\
The discussion of the impact of spin correlations is completed by
exploring the influence of the application of experimental cuts
(cf.\ App.\ \ref{app_cuts}) on the shape of certain spectra. It is
clear that superimposing specific jet and lepton cuts strongly affects
the event sample. Here, the cuts are mainly on the $\eta$\/ and the
$p_T$ of the leptons. In turn their distributions alter. The
characteristics found for the cutfree case are not substantially
changed by the applied cuts and by the renormalization of the spectra
according to these given cuts indicated by the vertical lines in the
Figs.\ \ref{pte_spini_mocas} and \ref{etae_spini_mocas}. More
interestingly, however, these distributions drive alterations to
secondary observables. In the two-particle correlations mentioned
before, the effects already present without applying cuts are
enforced. The slopes of the $\Delta\phi$\/ distributions increase,
amplifying the difference between both sets of predictions, the
ones with and without spin correlations. The main change in the
$\Delta R$\/ spectrum is an additional deviation between $0.2$ (the
cut) and $2.0$, such that now the no-spin-correlation results are
roughly $20\%$ above the other ones. The case is different for the
pseudo-rapidity distribution of the $W^+$ boson. Without the
application of cuts one starts off distributions that agree on the
$10\%$ level. This is severely changed by the introduction of the
cuts, see the rightmost panel of Fig.\ \ref{cuts_mocas}. In contrast
to the aforementioned two-particle correlations, here the predictions
without spin correlations are well separated from the other ones only
after the application of the cuts. As a last example, consider the
transverse momentum distribution of the $W^+$ boson. Both types of
predictions stemming from uncutted (left panel) and from samples
analysed with cuts (middle panel) are pictured in Fig.\
\ref{cuts_mocas}. The inclusion of cuts apparently brings {\tt MC@NLO}
and \she\ including the full correlations into good agreement, but
this clearly happened accidentally.
\\
To summarize, the examples shown here, clearly hint that the
superposition of spin correlations (or their absence) together with
cuts triggers sizeable effects in both types of observables, such that
have already shown deviations in the absence of cuts and, more
crucially, such that have not. In specific cases, such as the $p_T$
spectrum of the $W^+$, this may possibly lead to misinterpretations of
the results.

\section{Conclusion}

\noindent
In this work, the merging procedure for multiparticle tree-level
matrix elements and the parton shower implemented in \she\ has been
further validated; this time, the case of $W$\/ pair production at the
Fermilab Tevatron has been considered. First, it has been shown that
the results obtained with \she\ are widely independent of specific
merging procedure details such as the choice of the merging scale and,
for sufficiently inclusive observables, the number of extra jets
covered by the tree-level matrix elements. In addition, it has been
shown that the specific form of the spectra produced by \she\ is
nearly independent -- with deviations less than $20\%$ -- of the
choice of the factorization scale and the renormalization scale.
\\
Having established the self-consistency of the \she\ results, they
have been compared to those from an NLO calculation provided through
{\tt MCFM}. There, good agreement of the two codes has been found,
again on the $20\%$ level. Thus it is fair to state
that the \she\ results for the shapes are within theoretical errors
consistent with an NLO calculation. The inclusion of the parton shower
connected with specific scale choices in \she, however, produces a
surplus of QCD radiation with respect to the single parton emission in
the real part of the NLO correction in {\tt MCFM}. 
\\
Finally, the results of \she\ have been compared with those of other
hadron-level event generators, namely with {\tt PYTHIA} and {\tt
MC@NLO}. In this comparison it turned out that \she\ predicts a
significant increase of QCD radiation with respect to the other two
codes. For the $p_T$ spectra of jets accompanying the two $W$\/
bosons, the differences are dramatic in the high-$p_T$ tails. In
addition, the impact of spin correlations has been quantified. In the
observables considered here, it reaches $20\ldots50\%$. This may be
even larger than the impact of higher order corrections.

\begin{acknowledgments}
  \noindent
  The authors would like to thank Stefan H\"oche for valuable
  collaboration on the development of \she. Furthermore, they would
  like to thank Marc Hohlfeld (D\O) for pleasant conversation on the
  experimental aspects of this work. The authors are also indebted to
  Torbj\"orn Sj\"ostrand, John Campbell, Tim Stelzer, and Stefano
  Frixione for helpful advise. Financial support by BMBF, DESY, and
  GSI is gratefully acknowledged.
\end{acknowledgments}

\begin{appendix}

\section{Input parameters of SHERPA\label{app_input}}

\noindent
All \she\ studies have been carried out with the cteq6l PDF set
\cite{Pumplin:2002vw}. The value of $\alpha_s$ has been chosen
according to the corresponding value of the selected PDF, namely
$\alpha_s=0.118$.
The running of the strong coupling constant is determined by the
corresponding two-loop equation, except for the \she\ {\tt MCFM}
comparison. There an one-loop running has been employed for
$\alpha_s$. Jets or initial partons are defined by gluons and all
quarks but the top quark; this one is allowed to appear within the
matrix elements only through the coupling of the $W$\/ boson with the
$b$\/ quark. In the \she\ {\tt MCFM} comparison \she\ runs, however
are restricted to the light-flavour sector, \ie the
$g,\,d,\,u,\,s,\,c$ sector. In the matrix element calculation the
quarks are taken massless, only the shower will attach current masses
to them. The shower cut-offs applied are $2$ GeV and $1$ GeV for the
initial and the final state emissions, respectively. If explicitly
stated a primordial $k_{\perp}$ Gaussian smearing has been employed
with both, mean and standard deviation being equal to $0.8$ GeV.
The Standard Model input parameters are:
\bea
&&m_W = 80.419\;{\rm GeV}\,,\quad \Gamma_W = 2.06\;{\rm GeV,}\nonumber\\
&&m_Z = 91.188\;{\rm GeV}\,,\quad \Gamma_Z = 2.49\; {\rm GeV,}\nonumber\\
&&G_{\mu} = 1.16639 \times 10^{-5}\; {\rm GeV}^{-2},\nonumber\\
&&\sin^2\theta_W = 1 - m^2_W/m^2_Z,\nonumber\\
&&\alpha_s = 0.118.
\eea
The electromagnetic coupling is derived from the Fermi constant $G_{\mu}$
according to
\bea
&&\alpha_{\rm em} = \frac{\sqrt{2}\,G_{\mu}\,M^2_W\,\sin^2\theta_W}{\pi}\,.
\eea
The constant widths of the electroweak gauge bosons are introduced
through the fixed-width scheme. The CKM matrix has been always taken
diagonal.

\section{Setups for MCFM, MC@NLO and PYTHIA\label{app_sets}}

\noindent
\underline{\tt MCFM}
\\[1mm]
The program version employed is {\tt MCFM v4.0}. The process chosen is
{\tt nproc=61}. The investigations have been restricted to the
$d,\,u,\,s,\,c$\/ quark sector. The PDF set used is cteq6l. The
default scheme for defining the electroweak couplings has been used
and their input values have been adjusted with the corresponding
parameter settings given for \she. The renormalization scale and the
factorization scale are fixed and set to $\mu_R=\mu_F=M_W$.
\\\\
\underline{\tt MC@NLO}
\\[1mm]
The program version used is {\tt MC@NLO 2.31}. The process number is
taken as {\tt IPROC=-12850}, so that the underlying event has not
been taken into consideration. The two $W$\/ boson decays into leptons
are steered by the two {\tt MODBOS} variables being set to {\tt 2} and
{\tt 3} for the first and the second choice, respectively. The lepton
pairs have been generated in a mass window of
\be
  M_W-40\,\Gamma_W < m_{l\nu} < M_W+40\,\Gamma_W\,.
\ee
Again, the cteq6l PDF set as provided by {\tt MC@NLO}'s own PDF
library is used. The weak gauge boson masses and widths are aligned to
the settings used for the previous codes. All other parameters have
been left unchanged with respect to their defaults.
\\\\
\underline{\tt PYTHIA}
\\[1mm]
The {\tt PYTHIA} version used is {\tt 6.214}. The process $p\bar p\to
W^+W^-+X$ is selected through {\tt MSUB(25)=1}. The specific decay
modes of the two $W$'s are picked by putting {\tt MDME(206,1)=2} and
{\tt MDME(207,1)=3}, where all other available modes are set to zero.
The possibility of parton shower emissions right up to the limit,
which has been proven to be more convenient for jet production
\cite{Miu:1998ju}, is achieved with {\tt MSTP(68)=2}. This increases
the IS shower start scale in {\tt PYTHIA} to $\sqrt{s}=1960$ GeV and
accounts for a reasonably higher amount of hard QCD radiation. For all
comparisons here, the underlying event is switched off, other
parameters are left to their default.

\section{Phase space cuts\label{app_cuts}}

\noindent
Two different analyses are used for the comparisons of the results
obtained throughout this publication. A simple analysis has been taken
to verify the pure behaviour of the considered programs. For
this case, only jets are analysed utilizing the Run II $k_{\perp}$
clustering algorithm defined in \cite{Blazey:2000qt} with a
pseudo-cone size of $R=1$. The jet transverse momentum has to be
greater than $15$ GeV.
\\
For more realistic experimental scenarios, an analysis applying
jet and lepton cuts has been availed. Then, the pseudo-cone
size of the jet algorithm has been set to $R=0.7$, and the jets have
to fulfil the following constraints on the pseudo-rapidity and the
transverse momentum,
\be
  |\eta^{\rm jet}|<2.0\,,\quad p^{\rm jet}_T>15\ {\rm GeV}\,.
\ee
For the charged leptons the cuts on these observables are given by
\be
  |\eta^{\rm lep}|<1.0\,,\quad p^{\rm lep}_T>20\ {\rm GeV}\,,
\ee
however, a cut on the missing transverse energy has not been
introduced. There is a final selection criteria corresponding to the
separation of the leptons from each other and from the jets,
\be
  \Delta R_{\rm ll}>0.2\,,\quad\Delta R_{\rm lj}>0.4\quad\,.
\ee

\end{appendix}


\end{document}